\documentclass[11pt]{article}
\pdfoutput=1
\usepackage{jheppub}
\usepackage{braket} 
\usepackage{amssymb,amsxtra}
\usepackage{bm}
\usepackage[dvipsnames]{xcolor}
\edef\restoreparindent{\parindent=\the\parindent\relax}
\usepackage{parskip}
\restoreparindent

\def\d{{\rm d}}
\def\i{{\rm i}}
\def\rn{{\rm{n}}}
\def\rD{{\rm{D}}}
\def\rE{{\rm{E}}}
\def\rL{{\rm{L}}}

\def\CB{{\cal B}}

\def\CD{{\cal D}}

\def\CI{{\cal I}}
\def\CJ{{\cal J}}

\def\CM{{\cal M}}
\def\CN{{\cal N}}
\def\CO{{\cal O}}

\def\BH{\mathbb{H}}
\def\BR{\mathbb{R}}
\def\BS{\mathbb{S}}

\def\SO{\mathrm{SO}}
\def\GL{\mathrm{GL}}

\def\Del{\Delta}
\def\tP{\tilde{P}}
\def\tX{\tilde{X}}
\def\tZ{\tilde{Z}}

\def\tP{\tilde{P}}

\def\tCO{\tilde{\cal{O}}}

\def\del{\delta}
\def\bdel{\bar{\delta}}
\newcommand{\nn}{\nonumber}

\newcommand{\be}{\begin{align}}
\newcommand{\ee}{\end{align}}
\newcommand{\bea}{\begin{align}}
\newcommand{\eea}{\end{align}}
\def\bDel{\bar{\Delta}}
\def\btau{\bar{\tau}}

\def\rB{{\rm{B}}}

\title{The gravity dual of Lorentzian OPE blocks}

\author[a]{Heng-Yu Chen,}
\author[a]{Lung-Chuan Chen,}
\author[b,c]{Nozomu Kobayashi}
\author[b]{and Tatsuma Nishioka}

\affiliation[a]{Department of Physics, National Taiwan University, Taipei 10617, Taiwan}
\affiliation[b]{Department of Physics, Faculty of Science,
The University of Tokyo,\\
Bunkyo-ku, Tokyo 113-0033, Japan}
\affiliation[c]{Kavli Institute for the Physics and Mathematics of the Universe (WPI), \\
The University of Tokyo Institutes for Advanced Study, The University of Tokyo, \\
Kashiwa, Chiba 277-8583, Japan}

\abstract{
We consider the operator product expansion (OPE) structure of scalar primary operators in a generic Lorentzian CFT and its dual description in a gravitational theory with one extra dimension.
The OPE can be decomposed into certain bi-local operators transforming as the irreducible representations under conformal group, called the OPE blocks.
We show the OPE block is given by integrating a higher spin field along a geodesic in the Lorentzian AdS space-time when the two operators are space-like separated.
When the two operators are time-like separated however, we find the OPE block has a peculiar representation where the dual gravitational theory is not defined on the AdS space-time but on a hyperboloid with an additional time coordinate and Minkowski space-time on its boundary.
This differs from the surface Witten diagram proposal for the time-like OPE block, but in two dimensions we reproduce it consistently using a kinematical duality between a pair of time-like separated points and space-like ones.
}

\preprint{UT-19-27, IPMU19-0174}

\begin{document}
\maketitle

\section{Introduction and summary}

The last decade has seen a revival of interests in Conformal Field Theory (CFT) in higher dimensions with recent advances of numerical techniques applied to conformal bootstrap programme \cite{Rattazzi:2008pe} (see e.g. \cite{Rychkov:2016iqz,Simmons-Duffin:2016gjk,Poland:2018epd} for reviews).
Correlation functions are severely constrained by conformal symmetry and completely fixed up to three-point functions.
For the higher-point functions, they can be reduced to the lower-point ones by using the operator product expansion (OPE) \cite{Wilson:1969zs,Wilson:1972ee}:
\begin{align*}
        \CO_i (x)\, \CO_j (0) = \sum_k\,C_{ijk}(x, \partial_x)\,\CO_k(0) \ ,
\end{align*}
where $\CO_i(x)$ is a primary operator labeled by a unitary irreducible representation $i$ of the conformal group and $C_{ijk}(x, \partial_x)$ is a composite derivative operator which generates the contributions of the descendant operators in the conformal multiplet of $\CO_k$ and whose form is fixed by conformal symmetry up to the OPE coefficient $c_{ijk}$.
Hence a CFT can be, in principle, characterized by the spectrum of primary operators and the OPE data.

Most works so far have been devoted to Euclidean CFT's partly due to their relevance to critical phenomena in statistical systems and the AdS/CFT correspondence.
In Euclidean theories correlation functions are free from the intricate causal ordering issues and there can only be one type of correlation functions.
Constructions of conformal correlation functions were undertaken in \cite{Polyakov:1970xd,Schreier:1971um,Dobrev:1976vr,Osborn:1993cr,Erdmenger:1996yc} and a systematic method of the classification was developed by \cite{Costa:2011mg} using the embedding space formalism \cite{Dirac:1936fq,Mack:1969rr}.
Among others, four-point functions have been extensively studied by decomposing them as a linear combination of universal building elements called conformal blocks \cite{Dolan:2000ut,Dolan:2003hv,Dolan:2011dv,Costa:2011dw}.
They form the basis of numerical conformal bootstrap and have an intuitive holographic description known as the geodesic Witten diagram \cite{Hijano:2015zsa}.

More recently, Lorentzian CFT's have attracted renewed interest in light of the novel development in analytic bootstrap in lightcone \cite{Fitzpatrick:2012yx,Komargodski:2012ek,Alday:2015ewa}, the Regge limits of correlation functions \cite{Cornalba:2006xm,Cornalba:2006xk,Cornalba:2007fs,Cornalba:2007zb,Costa:2014kfa}, and the remarkable applications in deriving averaged null energy condition in flat space \cite{Hartman:2016lgu,Afkhami-Jeddi:2017rmx} and the Lorentzian inversion formula \cite{Caron-Huot:2017vep,Simmons-Duffin:2017nub}.
Conformal blocks play a vital role also there as a probe of causal relationships between operators in Minkowski space, whose structures have been deeply tied to a class of quantum integrable systems in the past few years \cite{Isachenkov:2016gim,Isachenkov:2017qgn,Isachenkov:2018pef}.

However in contrast with the somewhat simplified Euclidean cases, in Lorentzian theories, there are various types of correlation functions such as time-ordered and Wightman correlators due to the causal ordering of the primary operators.
Despite the variety, any Lorentzian correlation function can in principle be recovered by analytic continuation of the Euclidean correlation function with a proper $\i\,\epsilon$-prescription \cite{Osterwalder:1973dx,Osterwalder:1974tc,Luscher:1974ez}.
In this sense Lorentzian conformal field theories appear to be subsidiaries of their Euclidean counterpart but there are intrinsic properties associated with the causal structure of space-time that become manifest only in Lorentzian settings.
For instance, a recent study shows that there exist continuous spin operators in any Lorentzian CFT \cite{Kravchuk:2018htv}.\footnote{They are non-local operators called light-ray operators and make an appearance in QCD \cite{Braun:2003rp}.}
The key to the construction of such an operator is the light transform that integrates a local operator along a null direction, which is obviously possible only in Lorentzian signature.
Lorentzian CFT's have just begun to be re-explored with a fresh eye and there remains the question of what constraint we can draw on the OPE data from the causal structure.

In this paper, we revisit the OPE structure in Lorentzian CFT and derive its holographic representation. In particular we focus on the so-called ``OPE blocks'' introduced in \cite{Ferrara:1971vh} during the nascent stage of conformal field theory studies, they are bi-local operators which package the individual primary operator and its infinite descendants in the OPE:
\begin{align}\label{Def:OPE-block}
 \CO_i (x_1)\, \CO_j (x_2) = \sum_k c_{ijk}\,\CB_{k}(x_{12})\ .
\end{align}
The precise definition of $\CB_{k}(x_{12})$ is sensitive to the nature of the separation $x_{12}\equiv x_1 - x_2$, and differs for space-like and time-like cases.
We will construct the gravitational/holographic dual of OPE block for various cases.
In this work we focus on the OPE of two scalar primaries $i = [\Delta_1, 0],\,  j = [\Delta_2, 0]$ with general conformal dimensions $\Delta_1, \Delta_2$ for simplicity,
thus only the OPE blocks in spin-$J$ representations ($k=[\Delta, J]$) contribute to the right hand side of \eqref{Def:OPE-block}.
It is pointed out in \cite{daCunha:2016crm} (see also \cite{Czech:2016xec}) that the scalar OPE block $(J=0)$ for a pair of space-like separated points takes the form of ``half" a geodesic Witten diagram:
\begin{align}\label{OPEB_intro}
    \CB_{\Delta}(x_{12}) \sim \frac{1}{(x_{12}^2)^\frac{\Delta_1 + \Delta_2}{2}}\, \int_{\gamma_{12}}\,\d \lambda\, e^{\lambda (\Delta_1 - \Delta_2) }\,\Phi_\Delta\left( y(\lambda)\right) \ ,
\end{align}
where $y(\lambda)$ parametrizes the geodesic $\gamma_{12}$ connecting $x_1$ and $x_2$ in the Poincar\'e AdS$_{d+1}$ space-time with parameter $\lambda$ and $\Phi_\Delta$ is the so-called HKLL representation of an AdS scalar field  \cite{Hamilton:2005ju,Hamilton:2006az}.
In two dimensions the space-like spinning OPE block in the conserved current representation ($\Delta = d-2 + J$) is also considered and shown to take a similar form with $\Phi_\Delta$ replaced with a massless higher spin field on AdS$_3$ \cite{Das:2018ajg}.
However, there have been no general results for spin-$J$ case in higher dimensions (and non-conserved case in two dimensions), especially for the OPE blocks of a pair of time-like separated points.
Deriving the holographic description of the time-like OPE block explicitly is one of the main results of this paper.

We start section \ref{sec:euclidean_OPE_block} with the Euclidean OPE block to set the stage for the Lorentzian case in the subsequent sections.
We use the embedding space formalism to give an integral representation of the OPE block using the shadow projector \cite{Ferrara:1971vh,Ferrara:1973eg} (see also \cite{Dobrev:1976vr,Dobrev:1977qv}) implemented in the embedding space \cite{SimmonsDuffin:2012uy}.
The Euclidean OPE block $\CB_{\Delta, J}^{(\rE)}$ turns out to be an integral transform of a primary $\CO_{\Delta, J}$:\footnote{We suppress the contracted spin indices for simplicity.}
\begin{align}
    \CB_{\Delta, J}^{(\rE)}(x_{12}) \sim \int [\d^d x_0]_\rE\, K_{\Delta_1, \Delta_2, [\bar\Delta, J]}(x_1, x_2, x_0)\, \CO_{\Delta, J}(x_0) \ ,
\end{align}
with an integral kernel $K_{\Delta_1, \Delta_2, [\bar\Delta, J]}(x_1, x_2, x_0)$ transforming as the three-point function of $\CO_{\Delta_1}(x_1)$, $\CO_{\Delta_2}(x_2)$ and a primary operator of shadow dimension $\bar\Delta \equiv d-\Delta$ and spin $J$ at $\CO_{\Del, J}(x_0)$.
This expression will be our starting point of deriving the holographic form of the OPE block.
A Feynman parametrization of the $K$-kernel naturally introduces the parameter $\lambda$ appearing in \eqref{OPEB_intro}, and by exchanging the order of integration we derive an analogous expression to \eqref{OPEB_intro} for spinning OPE blocks with the ``bulk" field $\Phi^{(\rE)}_{\Delta, J}$ satisfying the equation of motion for a spin-$J$ field on the Euclidean AdS space.
The correlator of two OPE blocks correctly reproduces the geodesic Witten diagram with spinning exchange \cite{Chen:2017yia, Castro:2017hpx, Dyer:2017zef,Sleight:2017fpc} as expected.
We also derive an alternative representation of the spinning OPE block using a differential operator, which will be used in comparing with the Lorentzian result in the later section.

In section \ref{sec:Momentum_shadow} we set up the shadow formalism adapted to Minkowski space-time to write down an integral representation of the Lorentzian OPE block in parallel with the Euclidean case.
The Lorentzian shadow projector can be obtained by analytically continuing the Euclidean projector but it is more transparent to construct it in momentum space \cite{Gillioz:2016jnn,Gillioz:2018mto}.
Applying the momentum shadow projector to a pair of operators we derive the Lorentzian OPE block $\CB_{\Delta, J}^{(\rL)}$ as an integral transform of $\CO_{\Delta, J}$ in momentum space:
\begin{align}\label{Lorentzian_OPE_intro}
    \CB_{\Delta, J}^{(\rL)}(x_{12}) \sim \int [\rD^d p]_\rL\, Q_{\Delta_1, \Delta_2, [\bar\Delta, J]}(x_1, x_2, p)\, \CO_{\Delta, J}(p) \ .
\end{align}
We should stress here that this expression is not new, but a well-established result derived in \cite{Ferrara:1971vh,Ferrara:1972ay,Ferrara:1973vz,Dobrev:1975ru,Dobrev:1977qv} about half a century ago, and we reproduce it here in a slightly different way using the momentum shadow projector.
Thus this section may be regarded as a self-contained review of the known facts about the Lorentzian OPE block.
The reader not interested in the derivation can safely skip to the next section after skimming the definition of the $Q$-kernel \eqref{Q-kernel} and the final result \eqref{Lorentzian_OPE_block_momentum}.

Armed with the integral representation \eqref{Lorentzian_OPE_intro} we consider the space-like OPE block and derive the holographic representation in section \ref{sec:space-like_OPE_block}.
When $x_1$ and $x_2$ are space-like separated, the result shares many similarities with the Euclidean case.
We obtain a generalization of \eqref{OPEB_intro} to arbitrary spin-$J$ representation.
In the special case when $\CO_{\Delta,J}$ is a conserved current, we find that the space-like OPE block has an illuminating gravity dual description, given by integrating a massless higher spin field \cite{Sarkar:2014dma} over the geodesic $\gamma_{12}$.

We then turn our attention to the time-like OPE block in section \ref{sec:time-like_OPE_block}.
By analytically continuing the integral representation \eqref{Lorentzian_OPE_intro} to the time-like configuration and performing similar manipulations to the space-like case, we derive a new representation of the time-like OPE block.
Contrary to one's naive expectation, the time-like OPE block has a quite different structure and interpretations from the space-like one.
It takes the form of half geodesic Witten diagram not on the Lorentzian AdS space-time, but on a hyperboloid with two time directions in $\BR^{2,d}$.
This new structure is the manifestation of the causal structure in Lorentzian CFT, which is inaccessible from the Euclidean perspective and can be understood as follows:
In Euclidean signature, the Euclidean AdS$_{d+1}$ space is the unique extension of a flat space $\BR^d$ CFT's live on in the embedding space $\BR^{1,d+1}$.
On the other hand, there are two different extensions of Minkowski space $\BR^{1,d-1}$ to a higher-dimensional hyper-surface inside the Lorentzian embedding space $\BR^{2,d}$, each of which precisely corresponds to the space-like and time-like case respectively.

The time-like OPE block has been studied in relation to entanglement entropy in \cite{Czech:2016xec,deBoer:2016pqk} and is proposed to be holographically described by the surface Witten diagram that integrates an AdS field over a codimension-two space-like hyper-surface in the AdS space-time.
Our result based on the analytic continuation is different from the proposal.
Motivated by this discrepancy, we present another derivation of the time-like OPE block using a kinematical duality exchanging a pair of time-like separated operators with a space-like codimension-two defect \cite{Gadde:2016fbj,Fukuda:2017cup}.
While we are not able to address this issue in full generality, we show that the duality method reduces the time-like configuration to a space-like one, ending up with the surface Witten diagram in two dimensions.

We discuss the implications of our results and few open problems in section \ref{sec:Discussion}.
Appendix \ref{sec:Miscellaneous} includes useful formulas used in the main text and a brief review on the embedding space formalism in both Euclidean and Lorentzian signatures.
Appendix \ref{Appendix:Wightman_and_Feynman} summarizes our conventions for Euclidean and Wightman two-point functions.
In appendix \ref{Appendix:Integral_reps_of_Bessel} integral representations of various types of Bessel functions used in deriving the holographic OPE blocks are collected.
Appendix \ref{app:SurfaceWitten} gives the computation of a three-point function with the surface Witten diagram as a modest check of the validity.

\section{Euclidean OPE block}\label{sec:euclidean_OPE_block}

Let us start by constructing the holographic dual configuration in $(d+1)$-dimensional Euclidean AdS-space  for the spinning OPE block in a $d$-dimensional Euclidean CFT. 
This somewhat more familiar warm up case will help us to fix various notations and serve as clear comparison with the Lorentzian cases we will consider next.
In this section we will work in the so-called ``embedding space'' formalism, 
our conventions for the embedding space coordinates in so-called Poincar\'{e} section are given in appendix \ref{app:Embedding}.

\subsection{Shadow formalism in the embedding space}

Consider the operator product expansion (OPE) between two scalar primary operators $\CO_{\Del_1}(P_1)$ and $\CO_{\Del_2}(P_2)$,
which can be decomposed into the summation of bi-local functions $\CB_{\Delta, J}^{(\rE)}(P_1, P_2)$ called the OPE blocks:
\begin{align}\label{Def:E-OPE-block}
    \CO_{\Del_1} (P_1)\, \CO_{\Del_2}(P_2) = \sum_{[\Delta, J]}\,c_{\Delta_1, \Delta_2, [\Delta, J]}\, \CB_{\Delta, J}^{(\rE)} (P_1, P_2)\ .
\end{align}
$\CB_{\Delta, J}(P_1, P_2)$ are labeled by the scaling dimension $\Delta$ and spin $J$ of the conformal group $\SO(1, d+1)$, 
which contains the contribution of the exchange primary operator $\CO_{\Del, J}$ and its infinite descendants, i.\,e., the entire conformal family.
The expansion coefficient $c_{\Delta_1, \Delta_2, [\Delta, J]}$ is the OPE coefficient we define from the unnormalized Euclidean three-point function with subscript $\rE$,
\begin{align}
    \langle\, \CO_{\Delta_1} (P_1)\, \CO_{\Delta_2}(P_2)\,\CO_{\Delta, J}(P_3, Z_3)\, \rangle_{\rE} = c_{\Delta_1, \Delta_2, [\Delta, J]}\, 
        E_{\Delta_1, \Delta_2, [\Delta, J]}(P_1, P_2, P_3; Z_3)\ ,
\end{align}
and the normalized three-point function,
\begin{align}\label{Normalized_3pt}
       E_{\Delta_1, \Delta_2, [\Delta, J]}(P_1, P_2, P_3; Z_3)
            &= \frac{\left[-2 P_1\cdot C_{3}\cdot P_2\right]^J }{P_{12}^\frac{\Delta^+_{12} - \Delta + J}{2} \, P_{13}^\frac{\Delta + \Delta^-_{12} +J}{2}\,P_{23}^\frac{\Delta -\Delta^-_{12} + J}{2}} \ , \qquad \Del_{ij}^{\pm}=\Del_i\pm \Del_j\ ,
\end{align}
where $Z_3$ is the polarization vector for spin.
The conformally invariant separation $P_{ij}$ and the gauge invariant anti-symmetric tensor $C_i^{AB}$ are given by:
\begin{align}
P_{ij} \equiv -2P_i \cdot P_j\ , \qquad C_{i}^{AB} \equiv Z_i^A P_i^B - P_i^A Z_i^B\ , \qquad i, j=1,2,3\ .
\end{align}
As we will discuss momentarily, the OPE block $\CB_{\Delta, J}(P_1, P_2)$ acting on the vacuum state is fixed by conformal symmetry \cite{Ferrara:1971vh,Ferrara:1972uq,Ferrara:1973vz,Dobrev:1977qv,Mack:1976pa}.\footnote{The determination of the OPE block would be more complicated on a general state than on the vacuum \cite{Schroer:1974de,Dobrev:1975ru}, {as it involves four-point functions that cannot be fixed uniquely by conformal symmetry}.
}

By using the operator-state correspondence, we can consider the so-called ``Shadow Projector'' \cite{Ferrara:1972uq,SimmonsDuffin:2012uy}:
\begin{align}\label{Spinning-Shadow-Projector}
    \begin{aligned}
    |\CO_{\Delta,J}| &= \frac{1}{\alpha_{\Del, J}\,\alpha_{\bDel, J}}\,\frac{1}{J!(h-1)_J} \int [ \rD^{d} P]_{\rE}\, |\,\tCO_{\bDel, J}(P, D_Z)\,\rangle\, \langle\, \CO_{\Delta, J} (P, Z)\,|\ ,\\
    \bDel &= d-\Del\ ,\qquad h=\frac{d}{2} \ ,
    \end{aligned}
    \end{align}
where  the subscript $``\rE"$ again denotes ``Euclidean'', and $D_{Z}$ explicitly given in \eqref{Def:DZ} is the embedding space lift of Todorov operator which generates the projector for symmetric trace-less transverse (STT) tensors,
and the integration measure in the embedding space is defined to be:
\begin{align}\label{Measure-P}
[\rD^{d}P]_{\rE} \equiv \frac{\d P^+ \d P^- \d^d P^i}{{\rm Vol}(\GL(1, \BR)^+)}\, \delta(-P^+ P^- +P^i P^i)\,\Theta(P^0)\ .
\end{align}
In \eqref{Spinning-Shadow-Projector}, we have also introduced the shadow operator $\tCO_{\bDel}(P, Z)$ defined to be:
\begin{align}\label{Scalar-Shadow}
    \begin{aligned}
        \tCO_{\bDel, J}^{}(P, Z) &\equiv \frac{1}{J!(h-1)_J}\int [\rD^{d}P']_\rE \, E_{[\bDel, J]}(P, Z; P', D_{Z'})\, \CO_{\Del}(P', Z') \\ 
            &=\int [\rD^{d}P']_\rE \, \frac{1}{(-2P\cdot P')^{\bDel}} \,\CO_{\Del, J}(P', Z\cdot \CI(P', P))\ ,
    \end{aligned}
\end{align}
where we introduced the rank-two tensor:
\begin{align}\label{Def:IAB-PP}
\CI_{AB}(P', P) \equiv \delta_{AB} + \frac{2P'_A P_B}{(-2P\cdot P')} \ ,
\end{align}
satisfies the properties:
\begin{align}
  P^A\,\CI_{AB}(P', P) = \CI_{AB}(P', P)\, P'^B = 0\ , \qquad \CI_{AC}(P',P)\,\CI^{C}_{\ \, B} (P', P) = \CI_{AB}(P', P) \ .
\end{align}
Finally in \eqref{Scalar-Shadow}, we have also introduced the normalized two-point function for the symmetric trace-less tensor primary fields:
\begin{align}\label{2pt-Norm}
    \begin{aligned}
        E_{[\Del, J]}(P, Z; \tP, \tZ)
                &= \frac{[(Z\cdot \tZ)(-2P\cdot \tP)+2(P\cdot \tilde Z)(\tilde P\cdot Z)]^J}{(-2P\cdot \tP)^{\Del+J}} \\
                &= \frac{[Z\cdot \CI(\tP, P) \cdot \tZ]^J}{(-2P\cdot \tP)^\Del}\ .
    \end{aligned}
\end{align}
In \eqref{Spinning-Shadow-Projector}, the symmetric factor $J!(h-1)_J$ comes from the tensor contraction convention in \cite{Costa:2011mg}, while $\alpha_{\Del, J}$ (and $\alpha_{\bDel, J}$) are fixed by the invariance under repeated shadow transformations to be:\footnote{Our normalization differs from the one in \cite{Dolan:2011dv} but will be simpler when we consider momentum space shadow projector in the next section.}
\begin{align}\label{Def:alpha}
    \alpha_{\Delta, J} \equiv  2^{d-2\Delta}\,\pi^h\,\frac{(\Delta - 1)_J\,\Gamma (h-\Delta)}{\Gamma(\Delta + J)} \ .
\end{align}
For later purpose, let us also write down here the spinning bulk to boundary propagator in AdS space:
\begin{align}\label{boundary-bulk-Norm}
    \begin{aligned}
        K_{[\Del, J]}(X, Z; \tP, \tZ) 
            &= \frac{[(W\cdot \tZ)(-2X\cdot \tP)+2(X\cdot \tZ)(\tP\cdot W)]^J}{(-2X\cdot \tP)^{\Del+J}}\\
            &= \frac{[W \cdot \CJ(\tP, X) \cdot \tZ]^J}{(-2X\cdot \tP)^\Del}\ ,
    \end{aligned}
\end{align}
where 
\begin{align}\label{Def:JAB-XP}
\CJ_{AB}(\tP, X) \equiv \delta_{AB} + \frac{2\tP_A X_B}{(-2X\cdot \tP)} \ ,
\end{align}
satisfies the properties:
\begin{align}
X^A\CJ_{AB}(\tP, X)=\CJ_{AB}(\tP, X) \tP^B =0\ , \qquad \CJ_{AC} (\tP, X)\, \CJ^C_{\ \, B} (\tP, X)=\CJ_{AB}(\tP, X)\ .
\end{align}
Moreover $\CI_{AB}(P', \tP)$ and $\CJ_{AB}(P', X)$ satisfy the interesting properties:
\begin{align}\label{IJ-identity}
\CI_{AC}(P', \tP)\,\CJ^{C}_{~\, B}(P', X)=\CI_{AB}(P', \tP)\ , \qquad \CJ_{AC}(P', X)\,\CI^{C}_{~\, B}(P', \tP)=\CJ_{AB}(P', X)\ .
\end{align}
Notice that we can introduce the following differential operators
\begin{align}\label{Def:Dp-PX}
    \begin{aligned}
    D^A_{P} &= Z^A \left(Z\cdot \frac{\partial}{\partial Z}\right)-{C}^{AB}\frac{\partial}{\partial P^B}\ , &\quad C^{AB} &= Z^A P^B-P^A Z^B\ , \\
    D^A_{X} &= W^A \left(W\cdot \frac{\partial}{\partial W}\right)-{B}^{AB}\frac{\partial}{\partial X^B}\ , &\quad B^{AB} &= W^A X^B-X^A W^B,
    \end{aligned}
\end{align}
to relate both spinning two-point function \eqref{2pt-Norm} and bulk to boundary propagator \eqref{boundary-bulk-Norm} to the scalar ones:
\begin{align}
    \begin{aligned}
        E_{[\Del, J]}(P, Z; \tP, \tZ)
            &=
            \frac{1}{(\Del)_J}[\tZ \cdot {D}_P ]^J E_{[\Del, 0]}(P; \tP)
            =
            \frac{1}{(\Del)_J}[Z \cdot {D}_{\tP} ]^J E_{[\Del, 0]}(P; \tP)\ , \\
        K_{[\Del, J]}(X, W; \tP, \tZ)
            &= \frac{1}{(\Del)_J}[\tZ \cdot D_{X} ]^J K_{[\Del, 0]}(X; \tP)=\frac{1}{(\Del)_J}[W \cdot {D}_{\tP}]^J K_{[\Del, 0]}(X; \tP) 
            \ .
    \end{aligned}
\end{align}
We will use these spin differential operators to give an alternative derivation of integral representation of the bulk tensor field momentarily.

\subsection{Holographic dual of Euclidean OPE blocks}

To obtain the holographic configuration of the spinning OPE block \eqref{Def:E-OPE-block}, let us first contract it with the shadow projector \eqref{Spinning-Shadow-Projector}, and we can express it as: 
\begin{align}\label{OPE_Block_formal}
    \begin{aligned}
    c_{\Del_1, \Del_2, [\bDel, J]}\,&\CB_{\Delta, J}^{(\rE)} (P_1, P_2)\\
        &= \frac{1}{\alpha_{\Del, J}\,\alpha_{\bDel, J}}\,\frac{1}{J! (h-1)_J}\int [\rD^{d}P_0]_\rE \, \langle\, \CO_{\Del_1} (P_1)\, \CO_{\Del_2}(P_2)\,\tCO_{\bar\Delta, J}(P_0, D_{Z_0})\,\rangle_{\rE}\, \CO_{\Delta, J}(P_0, Z_0)\ ,
    \end{aligned}
\end{align}
where we have introduced the Euclidean scalar-scalar-shadow three-point function \cite{Dobrev:1975ru}: 
\begin{align}
        \langle\, \CO_{\Delta_1} (P_1)\, \CO_{\Delta_2}(P_2)\,\tCO_{\bar\Delta, J}(P_0, Z_0)\,\rangle_{\rE} 
            &= c_{\Del_1, \Del_2, [\bDel, J]}\,\gamma_{\Delta, J}\,E_{\Delta_1, \Delta_2, [\bar\Delta, J]}(P_1, P_2, P_0; Z_0) \ , 
\end{align}
and the coefficient $\gamma_{\Delta, J}$
is given by\footnote{We use the shorthand notation, $\Gamma \left(x \pm y \right) \equiv \Gamma \left(x + y \right)\,\Gamma \left(x - y \right)$.}
\begin{align}
    \gamma_{\Delta, J} = \frac{1}{\kappa_{\bar\Delta, J}}\,\frac{\Gamma \left( \bdel_{12}^{\pm} +  J\right)}{\Gamma \left( \del_{12}^{\pm} + J\right)} \ , \qquad \kappa_{\Del, J} = \frac{\Gamma(\Del+J)}{\pi^h\,(d-\Del-1)_J\,\Gamma(h-\Del)}\ ,
\end{align}
with the parameters
\begin{align}\label{Def:delpm}
    \del_{12}^{\pm}=\frac{\tau\pm\Del_{12}^-}{2}\ ,\qquad  \bdel_{12}^{\pm}=\frac{\btau\pm\Del_{12}^-}{2}\ ,\qquad \tau =\Del-J \ ,\qquad \btau = \bDel-J \ .
\end{align}
Note that $\kappa_{\Delta, J}$ and $\kappa_{\bDel, J}$ satisfy the relation:\footnote{Note that $\kappa_{\Del, J}$ here can be related to the normalization constant $k_{\Del, J}$ in the shadow transformation in \cite{Dolan:2011dv} by $\kappa_{\Del, J} = \frac{k_{\Del, J}}{\pi^h}$.}
\begin{align}\label{kappa_to_alpha}
        \kappa_{\Delta, J}\, \kappa_{\bar\Delta, J} 
        =
        \frac{1}{\alpha_{\Delta, J}\,\alpha_{\bar\Delta, J}}
        \ .
\end{align}
The normalization constant $\gamma_{\Delta, J}$ appeared in the Euclidean three-point function involving shadow operator agrees with (D.20) in \cite{Liu:2018jhs} where the coefficient is denoted by $S^{\Delta_1, \Delta_2}_{[\Delta, J]}$.
We obtain the expression for the spinning OPE block:
\begin{align}\label{OPE_Block-Norm}
    \CB_{\Delta, J}^{(\rE)} (P_1, P_2) = \frac{\gamma_{\Delta, J}}{\alpha_{\Del, J}\,\alpha_{\bDel, J}}\,\frac{1}{J! (h-1)_J}\int [\rD^{d}P_0]_{\rE} \, E_{\Delta_1, \Delta_2, [\bar\Delta, J]}(P_1, P_2, P_0; D_{Z_0})\, \CO_{\Delta, J}(P_0, Z_0)\ ,
\end{align}
which is fixed kinematically and is our starting point for its holographic interpretation.
 
Applying the standard Feynman parametrization to the $P_0$-dependent part in the denominator of the three-point function, we find
\begin{align}\label{3pt_Feynman}
    \begin{aligned}
        \frac{ 1 }{P_{10}^\frac{\bar\Delta + \Delta^-_{12} + J}{2}\,P_{20}^\frac{\bar\Delta -\Delta_{12}^- + J}{2}}
            &= \rB^{-1}\left( \bdel_{12}^{+} +  J, \bdel_{12}^{-} +  J\right)\,\int_0^1 \frac{\d u}{u(1-u)}\frac{ u^{\bdel_{12}^{+} +  J}\,(1-u)^{\bdel_{12}^{-} +  J}}{(u\,P_{10} + (1-u)\, P_{20})^{\bDel + J}} \\
            &= 2\, \rB^{-1}\left( \bdel_{12}^{+} +  J, \bdel_{12}^{-} +  J\right)\,\frac{1}{(P_{12})^{\frac{\bar\Delta + J}{2}}}\int_{-\infty}^\infty \d \lambda\, e^{\lambda\Delta^-_{12}}\frac{1}{(-2X(\lambda)\cdot P_0)^{\bar\Delta + J}} \ ,
    \end{aligned}
\end{align}
where $\rB(x, y)$ is the Beta function and $\rB^{-1}(x,y) = 1/\rB(x,y)$
and we make a change of variable 
\begin{align}\label{u_to_lambda}
    u = \frac{1}{1 + e^{-2\lambda}} \ ,
\end{align}
in the second equality.
Interestingly, the combined coordinate above:
\begin{align}\label{geodesic_coordinates}
\gamma_{12}~:~X^A(\lambda) \equiv \frac{e^{\lambda} P_1^A + e^{-\lambda}P_2^A}{P_{12}^{\frac{1}{2}}} \ ,\qquad X(\lambda)^2=-1 \ ,
\end{align}
naturally exists in $(d+1)$-dimensional AdS space.
More precisely, we can regard $\lambda$ as the line parameter for the AdS-geodesic $\gamma_{12}$ connecting boundary points $P_1$ and $P_2$ \cite{Hijano:2015zsa}.\footnote{In physical space, $P_{12}= (x_1 - x_2)^2 \ge 0$ so $P_{12}^{1/2} = |x_1 - x_2|$.}
Substituting \eqref{3pt_Feynman} back into \eqref{OPE_Block_formal}, the spinning OPE block becomes
\begin{align}\label{Euclidean_OPEB_Direct_Derivation}
    \begin{aligned}    
    \CB_{\Delta, J}^{(\rE)}(P_1, P_2) 
            &= 
             \frac{2\,\gamma_{\Delta, J}}{\alpha_{\Del, J}\,\alpha_{\bDel, J}}\,\frac{1}{J!(h-1)_J}\, \rB^{-1}\left( \bdel_{12}^{+} +  J, \bdel_{12}^{-} +  J\right)\\
             &\quad\times \int_{-\infty}^\infty \d \lambda\,\frac{e^{\lambda \Delta^-_{12}}}{ P_{12}^{\frac{\Delta^+_{12}}{2}+ J}}
            \int [\rD^{d}P_{0}]_\rE \,\frac{\left[-2 P_1\cdot C_{0}\cdot P_2\right]^J|_{Z_0 \to D_{Z_0}}}{(-2X(\lambda)\cdot P_0)^{\bar\Delta + J}}\,\CO_{\Delta, J}(P_0, Z_0)\\
            &= 
            2\,\kappa_{\Del, J}\,\CN_{12,[\Delta, J]}\,\Gamma(\bar\Delta + J) 
            \,\frac{1}{J!(h-1)_J}\,
            \int [\rD^{d}P_0]_{\rE} 
            \\
            & \quad\times \int_{-\infty}^\infty \d \lambda \,\frac{1}{(-2P_1\cdot X(\lambda))^{\Delta_1} (-2P_2\cdot X(\lambda))^{\Delta_2}}\frac{\left[ V_{0,12}\right]^J
            |_{Z_0\to D_{Z_0}}}{(-2P_0\cdot X(\lambda))^{\bar\Delta + J}}\,\CO_{\Delta, J}(P_0, Z_0)\ ,
    \end{aligned}
\end{align}
where we simplify the coefficient in the second equality by introducing a frequently appearing constant in the following sections by
\begin{align}\label{Def:N12Delta}
    \CN_{12,[\Del, J]}
    \equiv 
    \frac{1}{\Gamma\left( \delta_{12}^\pm + J\right) } \ .
\end{align}
In \eqref{Euclidean_OPEB_Direct_Derivation}, we have also introduced a gauge invariant tensor structure \cite{Costa:2011mg}:
\begin{align}
    V_{0,12} \equiv \frac{P_1\cdot C_0 \cdot P_2}{P_1\cdot P_2}= \frac{-2P_1\cdot C_{0}\cdot P_2 }{P_{12}}\ ,
\end{align}
which is the unique combination for scalar-scalar-tensor three-point function.
In the limit of $J=0$, if we further identify:
\begin{align}\label{HKLL-scalar}
    \Phi^{(\text{E})}_{\Del, 0}(X) = \int [\rD^{d}P_0]_\rE \, K_{[\bDel, 0]}(X, P_0)\, \CO_{\Del, 0}(P_0)\ , 
\end{align}
as the so-called HKLL integral representation of the bulk scalar field given in \cite{Hamilton:2006fh}, 
we can rewrite
\eqref{Euclidean_OPEB_Direct_Derivation} as:
\begin{align}\label{Euclidean_scalar_OPEB}
    \CB_{\Del, 0}^{(\rE)}(P_1, P_2)= 2\,\kappa_{\Del, 0}\,\CN_{12,[\Delta, 0]}\,\Gamma(\bar\Delta) \,
             \frac{1}{P_{12}^{\frac{\Del_{12}^+}{2}}}\int_{-\infty}^\infty \d \lambda\, e^{\lambda \Del_{12}^-}\,\Phi^{(\text{E})}_{\Del, 0}(X(\lambda))\ .
\end{align}
This reduces to the euclidean version of the proposal for the holographic scalar OPE block given in \cite{daCunha:2016crm, Guica:2016pid} up to an overall normalization constant, and we can interpret it as restricting a scalar HKLL bulk field to move along the AdS-geodesic $\gamma_{12}$.
Moreover in the last equality of \eqref{Euclidean_OPEB_Direct_Derivation}, we also made it clear that the three-point scalar correlation function can be reproduced by a three-point scalar geodesic Witten diagram.  
 
\begin{figure}[t]
    \centering
    \includegraphics[width=13cm]{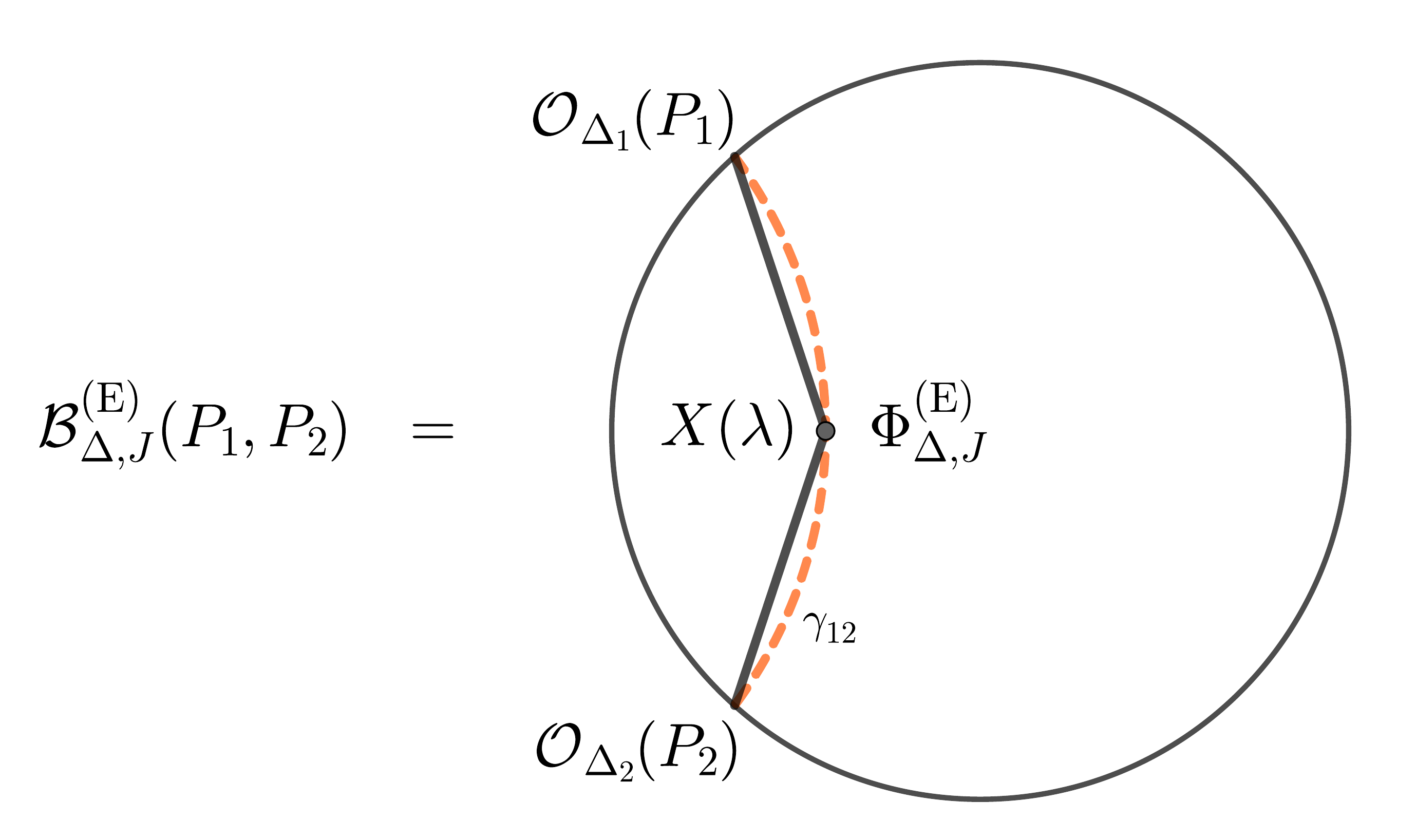}
    \caption{The holographic description of the OPE block. The orange dashed curve represents the bulk geodesic $\gamma_{12}$ connecting the boundary points $P_1$ and $P_2$, on which the spinning bulk field $\Phi_{\Delta, J}^{(\rE)}$ is smeared.
    The black lines are the scalar bulk to boundary propagators $K_{[\Delta, 0]}$ to the point $X(\lambda)$ on the geodesic.
    }
    \label{fig:AdS_OPE_block}
\end{figure}

For arbitrary $J$, we can consider the identity along the geodesic $\gamma_{12}$:
\begin{align}
V_{0,12} = -\frac{\d X(\lambda)}{\d\lambda}\cdot C_{0}\cdot X(\lambda) \ ,
\end{align}
and express \eqref{Euclidean_OPEB_Direct_Derivation} as:
\begin{align}\label{Spinning-OPE block}
    \begin{aligned}
        \CB_{\Delta, J}^{(\rE)} (P_1, P_2)
            &= 
            2\,\kappa_{\Del, J}\,\CN_{12,[\Delta, J]}\,\Gamma(\bar\Delta + J) 
            \,
    \int^{\infty}_{-\infty}\, \d\lambda\, 
     \frac{\Phi^{(\text{E})}_{\Del, J}\left(X(\lambda), \frac{\d X(\lambda)}{\d\lambda}\right)}{(-2P_1\cdot X(\lambda))^{\Del_1}(-2P_2\cdot X(\lambda))^{\Del_2}} \\        
            &= 
            2\,\kappa_{\Del, J}\,\CN_{12,[\Delta, J]}\,\Gamma(\bar\Delta + J) 
            \,\frac{1}{J!(h-1)_J}\, \int^{\infty}_{-\infty} \d\lambda\, e^{\lambda \Del_{12}^-} \left[\frac{\d X(\lambda)}{\d\lambda} \cdot K\right]^J \Phi^{(\text{E})}_{\Del, J}(X(\lambda), W) \\
            &= 
            2\,\kappa_{\Del, J}\,\CN_{12,[\Delta, J]}\,\Gamma(\bar\Delta + J) 
            \,
        \frac{1}{P_{12}^{\frac{\Del_{12}^+}{2}}}
        \int^{\infty}_{-\infty} \d\lambda\, e^{\lambda \Del_{12}^-}\, \Phi^{(\text{E})}_{\Del, J}\left(X(\lambda), \frac{\d X(\lambda)}{\d\lambda}\right)\ .
    \end{aligned}
\end{align}
In above we have introduced the AdS spin projector $K_A$ \eqref{Def:AdS-Todorov-op} and the spinning generalization of the HKLL bulk field \cite{Hamilton:2005ju, Hamilton:2006fh, Hamilton:2006az}:
\begin{align} \label{Euclidean_higher_spin_HKLL_in_Embedding}
    \begin{aligned}
    \Phi_{\Del, J}^{(\rE)}(X, W) 
        &\equiv \frac{1}{2^J J!(h-1)_J}\int [\rD^{d}P_0]_{\rE}\, K_{[\bar\Delta,J]}(X, P_0; W, D_{Z_0})\,\CO_{\Delta, J}(P_0, Z_0)\\
        &=\frac{1}{2^J}\int [\rD^{d}P_0]_{\rE} \, \frac{1}{(-2P_0\cdot X)^{\bDel}}\,\CO_{\Del, J}(P_0, W\cdot \CJ(P_0, X) ) \\
        &=\frac{1}{2^J}\int [\rD^{d}P_0]_{\rE} \, \frac{1}{(-2P_0\cdot X)^{\bDel}}\,(W\cdot \CJ(P_0, X))_{A_1} \cdots (W\cdot\CJ(P_0, X))_{A_J}\CO_{\Del}^{A_1 \cdots A_J}(P_0)\ , 
    \end{aligned}
\end{align}
where we use $\CJ_{AB}(P,X)$ defined in \eqref{Def:JAB-XP}.
Notice that the effective boundary polarization vector
\begin{align}
    \left(W\cdot \CJ(P_0, X)\right)_A =W_A+\frac{2W\cdot P_0}{(-2X \cdot P_0)}\,X_A
\end{align}
is manifestly invariant under bulk gauge transformation $W^A \to W^A+\alpha X^A$. Moreover this expression \eqref{Euclidean_higher_spin_HKLL_in_Embedding}
clearly satisfies the equation of motion for a spin-$J$ field in the AdS bulk. The natural interpretation of \eqref{Spinning-OPE block} is therefore that we are considering the pull-back of the rank-$J$ AdS-tensor field $\Phi^{(\text{E})}_{\Del, J}(X, W)$ moving along the geodesic $\gamma_{12}$, which is in completely agreement with the scalar case (see Fig.\,\ref{fig:AdS_OPE_block}).

Here we note that an alternative way of deriving the integral representation \eqref{Euclidean_higher_spin_HKLL_in_Embedding} is to 
start with the differential operators constructed in \eqref{Def:Dp-PX} and the following well-known conformal integral:
\begin{align}\label{Conf-Int-Scalar}
\frac{\Gamma(\bDel)}{\pi^h \Gamma(\bDel-h)}\int [\rD^{d}P_0]_{\rE} \,
\frac{1}{(-2P_0\cdot {X})^{\bar{\Delta}}}
\frac{1}{(-2\tP\cdot P_0)^\Delta}  =  \frac{(-X^2)^{h-\bDel}}{(-2\tP\cdot {X})^\Delta}\ , 
\end{align}
where $X^A$ is an arbitrary vector with non-vanishing norm in the embedding space $\BR^{1, d+1}$ \cite{SimmonsDuffin:2012uy}.
This implies that by setting $X^A$ to be AdS embedding coordinate in \eqref{Conf-Int-Scalar} and repeatedly acting with $D_{P}^A$, we can obtain the following integral relation between the normalized spinning bulk to boundary propagator and two-point function:
\begin{align}\label{HKLL-Derivation1}
    \begin{aligned}
    \frac{1}{(\Del)_J} &{D}^{A_1}_{\tP} \dots {D}^{A_J}_{\tP}  K_{[\Del,0]}(\tP, X)
     \\
    &=\frac{1}{(\Del)_J}{D}^{A_1}_{\tP} \dots {D}^{A_J}_{\tP} \frac{\Gamma(\bDel)}{\pi^h \Gamma(\bDel-h)}\int [\rD^{d}P_0]_\rE\, \frac{1}{(-2P_0\cdot X)^{\bar{\Del}}}\, E_{[\Del, 0]}(\tP, P_0 )\\
    &=\frac{\Gamma(\bDel)}{\pi^h \Gamma(\bDel-h)}\int [\rD^{d}P_0]_\rE\, \frac{1}{(-2P_0 \cdot X)^{\bar{\Del}}}\frac{(\tZ\cdot \CI(P_0, \tP))^{A_1}\cdots (\tZ\cdot \CI(P_0, \tP ))^{A_J}}{(-2P_0\cdot \tP)^{\Del}}\\
    &=\frac{\Gamma(\bDel)}{\pi^h \Gamma(\bDel-h)}\int [\rD^{d}P_0]_\rE\, \frac{1}{(-2P_0 \cdot X)^{\bar{\Del}}}\frac{(\tZ\cdot \CI(P_0, \tP)\cdot \CJ(P_0, X))^{A_1}\cdots (\tZ\cdot \CI(P_0, \tP )\cdot \CJ(P_0, X))^{A_J}}{(-2P_0\cdot \tP)^{\Del}}\ .
    \end{aligned}
\end{align}
Here in the last line we have used the identity \eqref{IJ-identity}, i.\,e., $\CI^{AC}(P_0, \tP)\,\CJ_C^{~B}(P_0, X) = \CI^{AB}(P_0, \tP)$,
this allows us to directly covariantize the primary operator $\CO_{\Del, B_1, \dots, B_J}(P_0)$ and make the following identification:
\begin{align}\label{HKLL-Spin-2}
\Phi_{\Del}^{(\rE) A_1\dots A_J}(X)\propto \int [\rD^{d}P_0]_\rE \,\frac{1}{(-2P_0 \cdot X)^{\bar{\Del}}}\, \CJ^{A_1 B_1}(P_0, X)\dots \CJ^{A_J B_J}( P_0, X)\,\CO_{\Del, B_1\dots B_J}(P_0)\ .
\end{align}
We can then recover \eqref{Euclidean_higher_spin_HKLL_in_Embedding} by contracting with the bulk polarization vector $W^A$.
We notice that the additional tensor terms generated in \eqref{HKLL-Spin-2} all contain explicit $P_0^A$ dependences  due to covariantization, which all vanish when projected onto physical space and can be regarded as the equivalence class of the embedding space tensor, however these additional terms are necessary to preserve the invariance under the shift $W^A \to W^A +\alpha X^A$.

\begin{figure}[t]
    \centering
    \includegraphics[width=16cm]{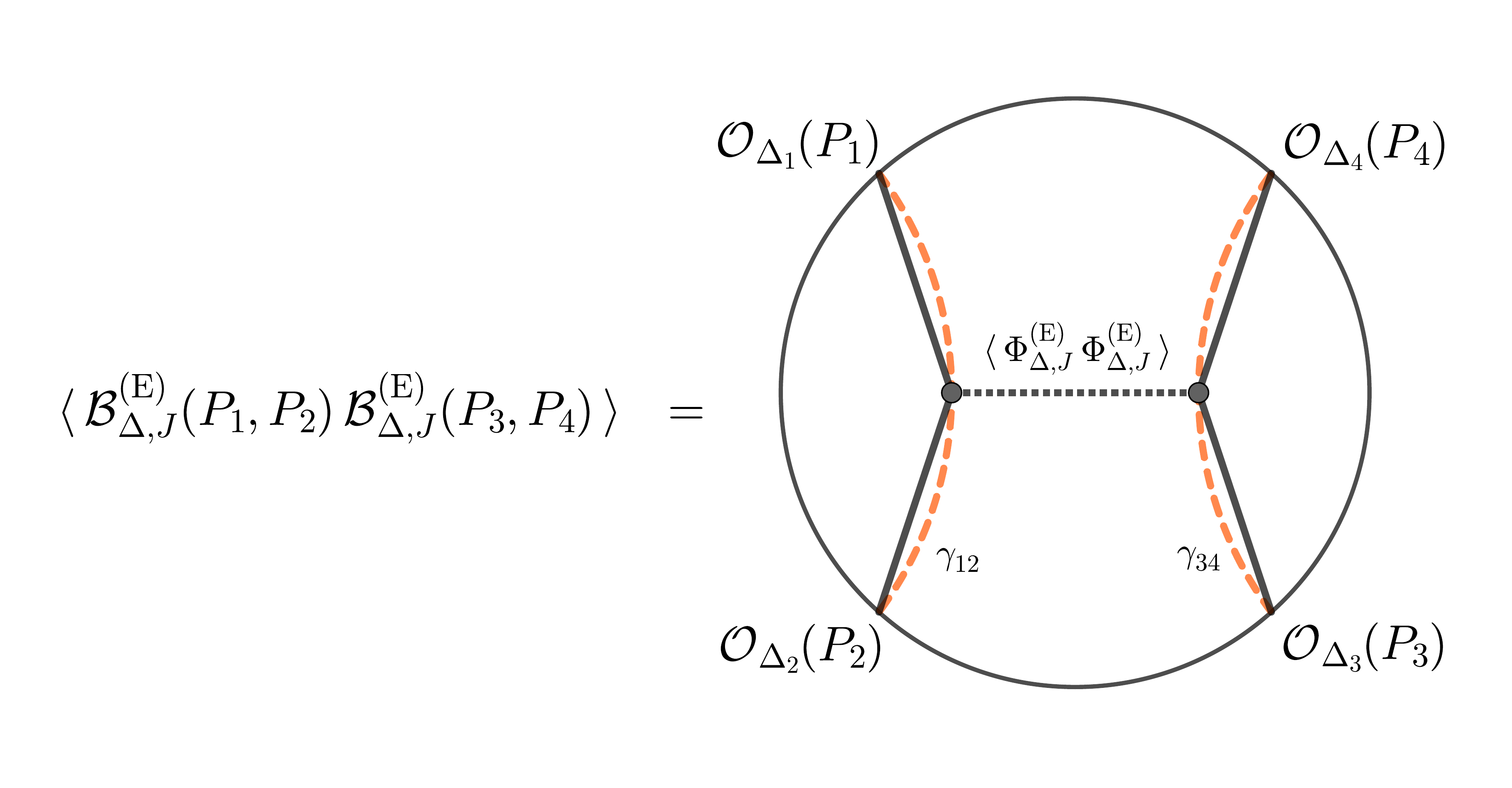}
    \caption{The geodesic Witten diagram as two-point function of the OPE blocks.
    The dashed black line represents the bulk to bulk propagator of the spinning bulk field from a point on $\gamma_{12}$ to another on $\gamma_{34}$. 
    }
    \label{fig:Geodesic_Witten}
\end{figure}

This interpretation of spinning OPE block \eqref{Spinning-OPE block} as the pullback of the bulk tensor field along the geodesic also implies we can use the last line of \eqref{Euclidean_higher_spin_HKLL_in_Embedding} to express the correlation function among two of them as:
\begin{align}\label{2pt-OPE block}
    \begin{aligned}
    &\langle\, \CB^{(\rE)}_{\Delta, J} (P_1, P_2)\, \CB^{(\rE)}_{\Delta, J} (P_3, P_4)\,\rangle\\
        &= \left( 2\,\kappa_{\Del, J}\,\Gamma(\bar\Delta + J) 
            \,\right)^2\, \CN_{12,[\Delta, J]}\,\CN_{34,[\Delta, J]}\,\\
    &\quad\times  \int^{\infty}_{-\infty}\, \d\lambda \int^{\infty}_{-\infty}\, \d\lambda'
    \frac{\left\langle
        \Phi^{(\text{E})}_{\Del, J}\left(X(\lambda), \frac{\d X(\lambda)}{\d\lambda}\right)\,\Phi^{(\text{E})}_{\Del, J}\left(\tX(\lambda'), \frac{\d\tX(\lambda')}{\d\lambda'} \right)\right\rangle_{\rE}}{(-2P_1\cdot X(\lambda))^{\Del_1}(-2P_2\cdot X(\lambda))^{\Del_2}(-2P_3\cdot \tX(\lambda'))^{\Del_3} (-2P_4\cdot \tX(\lambda'))^{\Del_4}}\ .
    \end{aligned}
\end{align}
Identifying the pull-back of the spinning bulk to bulk propagator onto the geodesics $\gamma_{12}$ and $\gamma_{34}$, we precisely obtain the four-point spinning geodesic Witten diagram up to overall constants \cite{Hijano:2015zsa, Chen:2017yia} (see Fig.\,\ref{fig:Geodesic_Witten}).\footnote{{Note that the bulk propagator is of Wightman type \cite{Satoh:2002bc} in Lorentzian case.}}
This is again consistent with the proposal \cite{Czech:2016xec} for the scalar case for constructing scalar conformal block from two-point function of the scalar OPE blocks.

\subsection{Differential representation of three-point function}

In this subsection, we derive an alternative representation of the holographic spinning OPE block based on the spin differential operators. This representation will be particularly useful for performing the Fourier transformation, which in turn allows us to consider the Lorentzian OPE block in momentum space in the subsequent sections.
One should note that its physical space projection already appeared in the early CFT literature such as \cite{Dobrev:1977qv}.

Let us first expand the numerator of the normalized three-point function \eqref{Normalized_3pt} into the following summation:
\begin{align}
        E_{\Delta_1, \Delta_2, [\bar\Delta, J]}(P_1, P_2, P_3; Z_3)
        =\frac{1}{[(P_1-P_2)^2]^{\frac{\Del_{12}^+-\bar\tau}{2}}}\sum_{r=0}^{J} \binom{J}{r}\frac{(Z_3\cdot P_1)^r}{[(P_1-P_3)^2]^{\bdel_{12}^+ +r} }\frac{(-Z_3\cdot P_2)^r}{[(P_2-P_3)^2]^{\bdel_{12}^- +J-r} }\ ,
\end{align}
where the parameters $\bar\delta_{12}^\pm$ are defined by \eqref{Def:delpm}
and we have deliberately taken $P_i$ off the lightcone as we will differentiate with respect to them, and only impose the lightcone condition, the fact that $(P_i-P_j)^2= -2P_i\cdot P_j$, at the end.
This allows us to use the simple differential identity:
\begin{align}
\frac{1}{(a)_r}\left(Z_3 \cdot \frac{\partial}{\partial P_i}\right)^r \frac{1}{[(P_i-P_3)^2]^a} = \frac{(-2Z_3\cdot P_i)^r}{[(P_i-P_3)^2]^{a+r}}\ ,\qquad i=1,2   \ , 
\end{align}
to rewrite the summation into:
\begin{align}\label{3pt-DiffRep-embed}
    \begin{aligned}
         E_{\Delta_1, \Delta_2, [\bar\Delta, J]}&(P_1, P_2, P_3; Z_3)\\
         &=\frac{1}{[(P_1-P_2)^2]^{\frac{\Del_{12}^+-\btau}{2}}}\frac{\CD_J\left(\bdel_{12}^{+}, Z_3\cdot \frac{\partial}{\partial P_1}; \bdel_{12}^-, Z_3\cdot \frac{\partial}{\partial P_2} \right)}{2^J (\bdel_{12}^+)_J (\bdel_{12}^-)_J}\frac{1}{[(P_1-P_3)^2]^{\bdel_{12}^+} }\frac{1}{[(P_2-P_3)^2]^{\bdel^-_{12}}}\ ,
    \end{aligned}
\end{align}
where the combined differential operator is defined through the following summation:
\begin{align}\label{3pt_differential}
    \begin{aligned}
    \CD_J (a, \lambda; b, \mu) & \equiv J!\, (\lambda + \mu)^J\, P_J^{(a-1,b-1)} \left(\frac{\mu -\lambda}{\mu + \lambda} \right)\\ 
    &= \sum_{r=0}^J \binom{J}{r} (a + r)_{J-r}\, (b+J-r)_{r}\, (- \lambda)^r\, \mu^{J-r} \\
    &= \frac{(a)_J}{2^J}\,\sum_{r=0}^J\frac{(-J)_r\,(a+b+J-1)_r}{(a)_r\,r!}\,{}_2 F_1 (1-b-J, r-J; a+r; -1)\,(\lambda - \mu)^r\,(\lambda + \mu)^{J-r}\ ,
    \end{aligned}
\end{align}
where $P_J^{(a, b)}(x)$ is the degree-$J$ Jacobi polynomial and ${}_2 F_1(a, b; c ;z)$ is Gauss Hypergeometric function.

Notice that when we project \eqref{3pt-DiffRep-embed} into physical space, we recover:
\begin{align}\label{3-pt_differential_rep}
        E_{\Delta_1, \Delta_2, [\bar\Delta, J]}(x_1, x_2, x_3; z_3)
        =  \frac{1}{ (x_{12}^2)^{\frac{\Delta_{12}^+ - \btau}{2}} } 
        \frac{\CD_J \left(\bdel_{12}^+,  z_3 \cdot \frac{\partial}{\partial x_1}; \bdel_{12}^-,  z_3 \cdot \frac{\partial}{\partial x_2} \right)} { 2^J \left(\bdel_{12}^+\right)_J\, \left(\bdel_{12}^-\right)_J} \left[\frac{1}{(x_{13}^2)^{\bdel_{12}^+} (x_{23}^2)^{\bdel_{12}^- }  }\right]\ ,
\end{align}
which is precisely the differential representation of three-point function given in \cite{Dobrev:1977qv}, and we can thus regard  \eqref{3pt-DiffRep-embed} as its embedding space lift and our simple computation as its derivation.
Although the differential representation of three-point function given in \eqref{3pt-DiffRep-embed} and \eqref{3-pt_differential_rep} might seem unfamiliar, it is indeed useful when we consider the Fourier transform with respect to $x_3$.
To conclude, if we substitute \eqref{3pt-DiffRep-embed} into the OPE block, and perform the similar Feynman reparametrization as before we find:
\begin{align}\label{OPE-block-Diff}
    \begin{aligned}
        \CB_{\Del, J}^{(\rE)}(P_1, P_2) &= \frac{1}{J! (h-1)_J}\, \kappa_{\Del, J}\,\CN_{12,[\Delta,J]}\,\Gamma(\bar\tau)\\
        &\quad\times\frac{1}{P_{12}^{\frac{\Del_{12}^+}{2}}}\,\tilde{\CD}_J \left(\bdel_{12}^{+}, D_{Z_3}\cdot \frac{\partial}{\partial P_1}; \bdel_{12}^-, D_{Z_3}\cdot \frac{\partial}{\partial P_2} \right) \int^{\infty}_{-\infty} \d\lambda\,{e^{\lambda \Del_{12}^-}} \int [ \rD^d  P_3]_{\rE}\, \frac{\CO_{\Del, J}(P_3, Z_3) }{[-2P_3\cdot X(\lambda)]^{\bar{\tau}}}\ ,
    \end{aligned}
\end{align}
where the transformed differential operator is
\begin{align}
   \tilde{\CD}_J \left(\bdel_{12}^{+}, D_{Z_3}\cdot \frac{\partial}{\partial P_1}; \bdel_{12}^-, D_{Z_3}\cdot \frac{\partial}{\partial P_2} \right) = {P_{12}^{\frac{\btau}{2}}}\, \CD_J \left(\bdel_{12}^{+}, D_{Z_3}\cdot \frac{\partial}{\partial P_1}; \bdel_{12}^-, D_{Z_3}\cdot \frac{\partial}{\partial P_2} \right) P_{12}^{-\frac{ \btau}{2}}\ .
\end{align}
Notice that the $P_3$-dependent integral in \eqref{OPE-block-Diff} is no longer conformal, the conformality is only restored after the action of the transformed differential operator $\tilde{\CD}_J$.
Compared with \eqref{Spinning-OPE block},
we thus have an alternative representation of the pull-back of the HKLL tensor field $\Phi^{(\text{E})}_{\Del, J}(X, W)$ along the geodesic $\gamma_{12}$:
\begin{align}\label{Diff-HKLL-Spinning}
    \begin{aligned}
        \Phi^{(\text{E})}_{\Del, J}&\left(X(\lambda), \frac{\d X(\lambda)}{\d\lambda}\right) \\
            &=
            \frac{\Gamma(\bar\tau)}{2\,\Gamma(\bar\Delta + J)}\,\frac{1}{\,J! (h-1)_J} \tilde{\CD}_J \left(\bdel_{12}^{+}, D_{Z_3}\cdot \frac{\partial}{\partial P_1}; \bdel_{12}^-, D_{Z_3}\cdot \frac{\partial}{\partial P_2} \right)\, \int [ \rD^d  P_3]_{\rE}\, \frac{\CO_{\Del, J}(P_3, Z_3) }{[-2P_3\cdot X(\lambda)]^{\btau}}\ .  
    \end{aligned}
\end{align}
While the gauge invariance in the bulk embedding space is not manifest, this expression is equivalent to \eqref{Euclidean_higher_spin_HKLL_in_Embedding} and is better suited for comparing with the Lorentzian OPE block in section \ref{sec:space-like_OPE_block}.

\section{Lorentzian OPE block in momentum space}\label{sec:Momentum_shadow}
 
We have studied the OPE blocks in Euclidean CFT using the embedding space formalism and derived the unique holographic description as the pull-back of the appropriately defined integral representation of the bulk field along the geodesics. However, in Lorentzian CFT's, there are generally two types of the OPE blocks depending on whether the two external primary operators are space-like or time-like separated.
The Lorentzian OPE block should be an appropriate analytic continuation of the Euclidean OPE block \eqref{OPE_Block_formal}, so it would be natural to assume that it takes the same form as in the Euclidean case:
\begin{align}\label{Lorentzian_OPE_block_position_temp}
    \CB_{\Delta, J}^{(\rL)} (x_1, x_2) \underset{?}{=} \int [\d^d y]_{\rL}\, \langle\, \CO_1 (x_1)\, \CO_2(x_2)\,\tilde\CO_{\bar\Delta, J}(y, d_z)\,\rangle\, \CO_{\Delta, J}(y, z) \ .
\end{align}
While this expression appears to be a unique choice of conformal integral for reproducing the OPE structure, there remain several subtleties to make it well-defined.
Firstly, it is not clear which type of the correlation function should be used in \eqref{Lorentzian_OPE_block_position_temp} as there are a few types of three-point functions such as time-ordered and Wightman functions in Lorentzian case.
Secondly, there is no reason to have the same integration range in \eqref{Lorentzian_OPE_block_position_temp} as in the Euclidean case.  
Indeed there exists a natural space-time region for a pair of time-like separated operators, to which the Lorentzian OPE block can be associated \cite{Czech:2016xec,deBoer:2016pqk}.
Resolving these ambiguities carefully will also lead us to the generalizations of the HKLL integral representation of the spinning bulk field \eqref{Euclidean_higher_spin_HKLL_in_Embedding} in different Lorentzian settings.

To elucidate these subtleties we will give a complimentary derivation of the Lorentzian OPE blocks by starting with the momentum space shadow formalism.
In order to facilitate the momentum space formalism, in this section we will also represent various quantities in physical space coordinates wherever appropriate, their relations 
with the embedding space coordinates used in previous section are also summarized in \eqref{EmbeddingCoorPola}.
This section is written independently from the others and the main result is the reproduction of the well-known result \eqref{Lorentzian_OPE_block_momentum} in old literature \cite{Ferrara:1971vh,Ferrara:1972ay,Ferrara:1973vz,Dobrev:1975ru,Dobrev:1977qv}.
The busy reader can skip to the next section and come back if necessary.

\subsection{Momentum space projector}
 
In Euclidean CFT the OPE block was defined through the shadow projector \eqref{Spinning-Shadow-Projector} in the embedding space.
In Lorentzian CFT it is more pleasing to represent the projector in momentum space to make manifest the positivity of the Hilbert space \cite{Gillioz:2016jnn,Gillioz:2018mto,Erramilli:2019njx}. To construct the projector in momentum space, let us define the momentum eigenstate of symmetric traceless primary as follows,
\begin{align}\label{Def:Momentum-Estate}
    |\,\CO_{\Delta, J}(p, z)\,\rangle \equiv \int [\d^d x]_{\rL} \,e^{\i\,p\cdot x}\,\CO_{\Delta, J}(x^0 + \i\, \epsilon, x^i, z)\, |0\rangle \ ,
\end{align}
where we have integration measures over $d$-dimensional Euclidean $\BR^d$ and  Minkowski space-time $\BR^{1,d-1}$:
\begin{align}
    \begin{aligned}
            {} [\d^d x]_{\rE} &= \d^d x^{i}\ , &\qquad i&=1, \dots, d\ , \\
            {} [\d^d x]_{\rL} &= \d^d x^{\mu}\ , &\qquad \mu &=0,\dots, d-1 \ ,
    \end{aligned}
\end{align}
and \eqref{Def:Momentum-Estate} has support on the future lightcone. Here we denote the coordinate $x$ with subscript L to emphasize that we are now in Lorentzian space-time. These states spans an orthogonal basis and its norm is given by
\begin{align}\label{Momentum_basis_norm}
    \langle \, \CO_{\Delta,J} (p_1,z_1 )\, |\, \CO_{\Delta,J} (p_2 , z_2)\, \rangle = (2 \pi)^d \,\delta^{(d)}(p_1 + p_2 )\, W_{[\Delta,J]}(p_2; z_1,z_2)\ ,
\end{align}
where we take $ \langle\, \CO (p)\, | = |\, \CO(-p) \,  \rangle^\dagger$ and $W_{[\Delta,J]}(p)$ is the Fourier transform of the two-point Wightman function (see appendix \ref{Appendix:Wightman_and_Feynman} for the derivation and explicit form).\footnote{A general comment on the notations. In this section, we start introducing the notations such as $E_{\Del_1, \Del_2, [\Del, J]}(x_1, x_2, x_3)$ and $E_{\Del_1, \Del_2, [\Del, J]}(x_1, x_2, p)$ for the Euclidean correlation functions in both position and momentum spaces to highlight the subsequent analytic continuation into Lorentzian space-time. Similarly we use $W_{\Del_1, \Del_2, [\Del, J]}(x_1, x_2, x_3)$ and $W_{\Del_1, \Del_2, [\Del, J]}(x_1, x_2, p)$ to denote the position and momentum space Wightman functions. We also omit the inclusion of polarization vector $z$ when it is contracted.}
In momentum space the state of the shadow operator is defined by
\begin{align}\label{Momentum_shadow}
    |\,\tilde \CO_{\bar\Delta, J}(p)\,\rangle \equiv 
    W_{[\bar \Delta, J]}(p)\,|\,\CO_{\Delta, J}(p)\,\rangle  \ .
\end{align}
Here we use the shorthand notation for the spin indices, so \eqref{Momentum_shadow} should be read as 
\begin{align}
|\,\tilde \CO_{\bar\Delta, J}(p, z)\,\rangle \equiv 
\frac{1}{ J!\,(h-1)_J}\, W_{[\bar \Delta, J]}(p; z, d_{z'})\,|\,\CO_{\Delta, J}(p, z')\,\rangle \ .
\end{align}
with the Todorov operator $d_z$ in physical space \eqref{Def:Todorov-op}.
We will keep using this notation throughout this paper to suppress the dependence of the physical space polarization vector $z^\mu$.
Note that given the normalization for the shadow transformation in momentum space, acting on $\CO_{\Del, J}(p)$ with shadow transformation twice yields
\begin{align}
    \tilde{\tilde{\CO}}_{\Delta, J}(p) = C_{\Delta, J}\,C_{\bDel, J}\,\CO_{\Delta, J}(p) \ ,
\end{align}
where the normalization constant $C_{\Del, J}$ is closely related to $\alpha_{\Del, J}$ in previous section and is defined in \eqref{Wightman_constant}.
This is due to the relation $W_{[\Delta,J]}(p) \, W_{[\bar\Delta,J]}(p) = C_{\Delta, J}\,C_{\bDel, J}\,\Theta(p^0)\,\Theta(-p^2)$ derived in \eqref{Wightman_inverse}, which means we can regard the shadow Wightman function $W_{[\bar\Delta,J]}(p)$ as the inverse of $W_{[\Delta,J]}(p)$ in momentum space.
Moreover, the two-point function of the shadow operators has a different norm from \eqref{Momentum_basis_norm}:
\begin{align}\label{Momentum_basis_shadow_norm}
    \langle \, \tilde\CO_{\bar\Delta,J} (p_1,z_1 )\, |\, \tilde\CO_{\bar\Delta,J} (p_2 , z_2)\, \rangle = C_{\Delta,J}\,C_{\bar\Delta, J}\,(2 \pi)^d \,\delta^{(d)}(p_1 + p_2 )\, W_{[\bar\Delta,J]}(p_2; z_1,z_2)\ .
\end{align}

Given these states, the projector in momentum space can be written as
\begin{align}\label{Momentum_projector}
    {\bf 1} = |0\rangle \langle 0| + \sum_{\Delta, J}\,\frac{1}{C_{\Delta, J}\,C_{\bDel, J}}\int [\rD^d p]_{\rL}\,|\, \tCO_{\bar\Delta, J}(-p)\,\rangle\, \langle \,\CO_{\Delta, J}(p)\,| \ ,
\end{align}
where we again suppress the spin indices in $|\, \tCO_{\bar\Delta, J}(-p)\,\rangle\, \langle \,\CO_{\Delta, J}(p)\,|$ and have defined the momentum space measure:
\begin{align}\label{Lorentz-P measure}
[\rD^d p]_{\rL}  \equiv  \frac{\d^d p}{(2\pi)^d}\, \Theta(p^0)\,\Theta(-p^2) \ .
\end{align}
One can easily confirm that \eqref{Momentum_projector} gives the correct projector if we insert this completeness relation between the Wightman two-point function $W_{[\Delta,J]}(x,z_1; 0,z_2) = \langle\, \CO_{\Delta,J}(x,z_1) \, \CO_{\Delta,J}(0,z_2)\, \rangle$ and use \eqref{Wightman_inverse}.
We also emphasize that the momentum space shadow projector defined in \eqref{Momentum_projector} however does not have the shadow contribution as opposed to the Euclidean shadow projector in position space.\footnote{The Lorentzian momentum projector may be obtained by analytic continuation from the Euclidean shadow projector \eqref{Spinning-Shadow-Projector} written in momentum space,
\begin{align}
    |\CO_{\Delta, J}| = \frac{1}{\alpha_{\Delta, J}\,\alpha_{\bar\Delta, J}}\,\int [\rD^d p]_\rE\, |\,\tilde \CO_{\bar \Delta, J} (-p)\,\rangle\,\langle\, \CO_{\Delta, J} (p)\,| \ , \qquad |\,\tilde \CO_{\bar \Delta, J} (p)\,\rangle \equiv E_{\bar\Delta, J}(p)\, |\,\CO_{\Delta, J} (p)\,\rangle \ ,
\end{align}
with the states satisfying
\begin{align}
    \langle \, \CO_{\Delta,J} (p_1,z_1 )\, |\, \CO_{\Delta,J} (p_2 , z_2)\, \rangle = (2 \pi)^d \,\delta^{(d)}(p_1 + p_2 )\, E_{[\Delta,J]}(p_2; z_1,z_2)\ .
\end{align}
Taking into account the difference of the normalization between the Euclidean and Wightman two-point functions given in appendix \ref{Appendix:Wightman_and_Feynman}, the analytic continuation amounts to the following replacement:
\begin{align}
    [\rD^d p]_\rE \to [\rD^d p]_\rL \ , \qquad E_{[\Delta, J]} \to W_{[\Delta, J]} \ , \qquad \alpha_{\Delta, J} \to C_{\Delta, J} \ .
\end{align}}
The step function inside the momentum integral guarantees that correlation functions including the momentum projector are of the Wightman type.

\subsection{Lorentzian OPE block}

Adopting the projector \eqref{Momentum_projector} in momentum space, we obtain the following form of the OPE block,
\begin{align}\label{Lorentzian_OPE_block}
    \CB_{\Delta,J}^{(\rL)}(x_1, x_2) = \frac{1}{C_{\Delta, J}\,C_{\bDel, J}}\,\int [\rD^d p]_{\rL} \, W_{\Delta_1,\Delta_2, [\Delta,J]}(x_1, x_2, -p)\, W_{[\bar\Delta,J]}(-p)\, \CO_{\Delta,J}(p)\ .
\end{align}
Here $W_{\Delta_1,\Delta_2, [\Delta,J]}(x_1, x_2, -p) $ is the Wightman three-point function in given order obtained from the Euclidean correlator $E_{\Del_1, \Del_2, [\Del, J]}(x_1, x_2, -p)$ with the exchanged primary operator Fourier transformed.
We will now proceed to obtain an explicit expression of $W_{\Delta_1,\Delta_2, [\Delta,J]}(x_1, x_2, -p)$ by performing an analytic continuation from the Euclidean space carefully.
\paragraph{}
The Wightman three-point function in physical space is defined as follows:
\begin{align}
        W_{\Delta_1,\Delta_2, [\Delta,J]}(x_i ; z) = \langle 0 |\, \CO_{\Delta_1}(x_1)\, \CO_{\Delta_2} (x_2)\, \CO_{\Delta, J} (x_3,z)\, | 0 \rangle \equiv \lim_{\epsilon_i \rightarrow 0} E_{\Delta_1,\Delta_2, [\Delta,J] } (\mathbf{x}_i, \i\, x^0_i + \epsilon_i; z ) \, ,
\end{align}
with the condition $\epsilon_1 > \epsilon_2 > \epsilon_3>0$\ .
To obtain the partially Fourier transformed expression, we start with the Euclidean three-point function in momentum space and then analytically continue to the Lorentzian signature.
The partially Fourier transformed Euclidean three-point function can be derived from the differential representation of the normalized three-point function \eqref{3-pt_differential_rep} with the help of the integral identity \eqref{useful_formula}:
\begin{align}
    E_{\Delta_1,\Delta_2, [\Delta,J] } (x_1,x_2,-p; z_3) &= \int [\d^d x_3]_{\rE} \,  E_{\Delta_1,\Delta_2, [\Delta,J] } (x_1,x_2,x_3; z_3) \, e^{\i\,p\cdot x_3}\nn \\
    &=\frac{\pi^h}{2^{J-1}}\,\frac{\CN_{12,[\Del, J]}}{ (x_{12}^2)^{\frac{\Delta_{12}^+- \tau}{2}} }\, \CD_J \left(\delta_{12}^+, \, z_3 \cdot \partial_1; \, \delta_{12}^-, \, z_3 \cdot \partial_2\right)  \left(\frac{p^2}{4x_{12}^2} \right)^{\frac{\tau-h}{2}} \nonumber \\
    & \times \int_0^1 \d u \, u^{\frac{\Delta_{12}^- + h}{2}-1} (1-u)^{\frac{ h - \Delta_{12}^-}{2}-1}  e^{\i\,p\cdot\left( u x_1 + (1-u) x_2\right)}  K_{h-\tau}\left(\sqrt{u(1-u)\,p^2\, x_{12}^2}\right)  ,
\end{align}
where we use the parameters given in \eqref{Def:delpm} and \eqref{Def:N12Delta}.
$\CD_J$ is the $J$-th order differential operator defined in \eqref{3pt_differential}, and $K_\nu (z)$ is the Bessel-$K$ function.
The Schwinger parameter $u$ here is related to the parameter $\lambda$ introduced in \eqref{3pt_Feynman} in the previous section
by the relation \eqref{u_to_lambda}.
Employing the relation between Bessel functions,
\begin{align}
    K_\nu (z) = \frac{\pi}{2 \sin{\pi \nu}} \left( I_{-\nu} (z) - I_\nu (z)  \right) \, ,
\end{align}
the Fourier transformed three-point function $E_{\Delta_1,\Delta_2, [\Delta,J] } (x_1,x_2,-p; z_3)$ can be further decomposed into the following expression:
\begin{align}\label{Splitting_rep_of_3-pt}
    \begin{aligned}
    E_{\Delta_1,\Delta_2, [\Delta,J] } (x_1,x_2,-p) = \frac{\pi^{h+1}\,\CN_{12,[\Del, J]}}{2^J \sin\left(\pi(h - \tau) \right)}  &\bigg[Q_{\Delta_1,\Delta_2, [\Delta,J]} (x_1,x_2,-p)  \\
     & \qquad - \kappa_{\Delta, J}\,E_{[\Delta,J]}(-p)\, Q_{\Delta_1,\Delta_2, [\bar\Delta,J]} (x_1,x_2,-p) \bigg] \ ,
    \end{aligned}
\end{align}
where the additional $E_{[\Delta, J]}(-p)$ is the Euclidean two-point function in momentum space given in appendix \ref{Appendix:Wightman_and_Feynman}. 
Restoring the polarization vector, the $Q$-kernel is defined by:
\begin{align}\label{Q-kernel}
    \begin{aligned}
    Q_{\Delta_1,\Delta_2, [\Delta,J]}&(x_1,x_2,-p; z_3)  \\
        &= - \frac{1}{ (x_{12}^2)^{\frac{\Delta_{12}^+- \tau}{2}} }\, \CD_J \left(\delta_{12}^+, \, z_3 \cdot \partial_1; \, \delta_{12}^-, \, z_3 \cdot \partial_2\right)  \left(\frac{p^2}{4x_{12}^2} \right)^{\frac{\tau-h}{2}} \nonumber \\
        &\quad \times \int_0^1 \d u \, u^{\frac{\Delta_{12}^- + h}{2}-1} (1-u)^{\frac{ h - \Delta_{12}^-}{2}-1}  e^{\i\,p\cdot \left( u x_1 + (1-u) x_2\right)}  I_{h-\tau}\left(\sqrt{u(1-u)\,p^2\, x_{12}^2}\right)\ .
    \end{aligned}
\end{align}
In deriving \eqref{Splitting_rep_of_3-pt} we used a non-trivial identity (see (A.12) in \cite{Dobrev:1975ru} for the derivation):
\begin{align}
    \begin{aligned}
        (x_{12}^2)^\frac{\tau}{2}&\,\CD_J \left(\delta_{12}^+, z\cdot \partial_1; \delta_{12}^-, z\cdot \partial_2 \right)\,\left( \frac{p^2}{4x_{12}^2}\right)^\frac{\tau - h}{2}\\
        &\qquad\times\int_0^1\,\d u\, u^{\frac{\Delta_{12}^- + h}{2}-1} (1-u)^{\frac{ h - \Delta_{12}^-}{2}-1}  e^{\i\,p\cdot x(u)}  I_{\tau - h}\left(\sqrt{u(1-u)\,p^2\, x_{12}^2}\right)\\
        &= \kappa_{\Delta, J}\,(x_{12}^2)^\frac{\bar\tau}{2}\, \frac{1}{J!(h-1)_J}\,E_{[\Delta, J]}(p; z, d_{z'})\,\CD_J \left(\bar\delta_{12}^+, z'\cdot \partial_1; \bar\delta_{12}^-, z'\cdot \partial_2 \right)\,\left( \frac{p^2}{4x_{12}^2}\right)^\frac{\bar\tau - h}{2}\\
        &\qquad \times \int_0^1\,\d u\, u^{\frac{\Delta_{12}^- + h}{2}-1} (1-u)^{\frac{ h - \Delta_{12}^-}{2}-1}  e^{\i\,p\cdot x(u)}  I_{h-\bar\tau}\left(\sqrt{u(1-u)\,p^2\, x_{12}^2}\right) \ .
    \end{aligned}
\end{align}

Next let us move on to the Wightman function by analytically continuing \eqref{Splitting_rep_of_3-pt} into Lorentzian space-time.
Its Fourier transform in $x$ follows from the analytic continuation of the Euclidean correlation function:
\begin{align}
        W_{\Delta_1,\Delta_2, [\Delta,J]}(x_1, x_2, x ; z) = \lim_{\epsilon_i \rightarrow 0} \int \frac{\d^{d-1} {\bf p}}{(2\pi)^{d-1}}\,e^{\i {\bf p}\cdot {\bf x}}\, \int_{-\infty}^\infty\frac{\d p^d}{2\pi}\,e^{-p^d (x^0 - \i\,\epsilon_3)}\,  E_{\Delta_1,\Delta_2, [\Delta,J] } (x_1,x_2,p; z) \ .
\end{align}
Here we have singled out
the $p^d$-integral and as it decays exponentially in the upper half plane, we can deform the contour to wrap the cut along the positive imaginary axis, which gives us the discontinuity.
{Notice that the $Q$-kernel defined in \eqref{Q-kernel} is an entire analytic function of $p$ \cite{Dobrev:1975ru,Dobrev:1977qv},
such that $E_{\Delta_1,\Delta_2, [\Delta,J]}(x_1, x_2, p; z)$ inherits the same branch cut from the two-point function $E_{[\Delta,J]}(-p)$. We thus obtain:
}
\begin{align}\label{Wightman_3pt_disc}
    \begin{aligned}
     W_{\Delta_1,\Delta_2, [\Delta,J]}&(x_1, x_2, x ; z) \\
        &= - \i 
 \lim_{\epsilon_i \rightarrow 0} \int \frac{\d^{d-1} {\bf p}}{(2\pi)^{d-1}}\,e^{\i {\bf p}\cdot {\bf x}}\,  \int_{0}^\infty\frac{\d p^0}{2\pi}\,e^{- \i\,p^0 (x^0 - \i\,\epsilon_3)}\, \text{Disc}\, E_{\Delta_1,\Delta_2, [\Delta,J] } (x_1,x_2,p;z)|_{p^d\to \i\,p^0} \ ,
    \end{aligned}
\end{align}
where we denote the discontinuity as $\text{Disc} \, f(x) \equiv f(x + \i\, 0) - f(x-\i\, 0) $ and it is explicitly defined by 
\begin{align}
    \text{Disc} \, E_{\Delta_1,\Delta_2, [\Delta,J] } = E_{\Delta_1,\Delta_2, [\Delta,J] } (x_1,x_2, {\bf p}, p^0 + \i\,0; z) - E_{\Delta_1,\Delta_2, [\Delta,J] } (x_1,x_2, {\bf p}, p^0 - \i\,0; z) \ .
\end{align}
It is easy to see that $ \text{Disc}\, Q_{\Delta_1,\Delta_2, [\Delta,J]} (x_1,x_2,p;z_3) = 0$, i.\,e., it is analytic from the definition of the $Q$-kernel \eqref{Q-kernel} as the Bessel function has the series expansion
\begin{align}
    I_{\nu}(x)=\i^{-\nu} J_{\nu}(\i\, x)= \left(\frac{x}{2}\right)^\nu \sum_{m=0}^{\infty} \frac{1}{m !\, \Gamma(m+\nu+1)}\left(\frac{x}{2}\right)^{2 m} \ ,
\end{align}
and the $Q$-kernel has a series expansion with respect to $p^2$ with non-negative integer powers.
Hence from the split expression \eqref{Splitting_rep_of_3-pt} we have
\begin{align}
    \begin{aligned}
    \text{Disc}\, E_{\Delta_1,\Delta_2, [\Delta,J] } (x_1,x_2,p) & = - \frac{\pi^{h+1}\,\kappa_{\Delta, J}\,\CN_{12,[\Del, J]}}{2^J \sin\left(\pi(h - \tau) \right)}\, Q_{\Delta_1,\Delta_2, [\bar\Delta,J]} (x_1,x_2,p) \, \text{Disc}\, E_{[\Delta,J]}(p) \\
    & =  -\i\, \frac{\pi^{h+1}\, \kappa_{\Delta, J}\,\CN_{12,[\Del, J]}}{2^J \sin\left(\pi(h - \tau) \right)}\, Q_{\Delta_1,\Delta_2, [\bar\Delta,J]} (x_1,x_2,p) \, W_{[\Delta,J]} (p) \ ,
    \end{aligned}
\end{align}
where we use the fact that the Wightman two-point function can be given by $W_{\Delta,J}(p) = - \i \, \text{Disc}E_{\Delta,J}(p)$.
Plugging into \eqref{Wightman_3pt_disc} gives 
\begin{align}\label{Wightman-3pt}
    \begin{aligned}
     W_{\Delta_1,\Delta_2, [\Delta,J]}(x_1, x_2, x) &= \int \frac{[\d^d  p]_{\rL}}{(2\pi)^d}\,e^{\i\, p\cdot x}\, W_{\Delta_1,\Delta_2, [\Delta,J]}(x_1, x_2, p) \\
     &=\int \frac{[\d^d  p]_{\rL}}{(2\pi)^d}\,e^{\i\, p\cdot x} \left[ - \frac{\pi^{h+1}\,\kappa_{\Delta, J}\,\CN_{12,[\Del, J]}}{2^J \sin\left(\pi(h - \tau) \right)}\, Q_{\Delta_1,\Delta_2, [\bar\Delta,J]} (x_1,x_2,p)\, W_{\Delta,J} (p)  \right] \bigg|_{p^d \to \i\, p^0} \ ,
     \end{aligned}
\end{align}
which allows us to identify directly the momentum space three-point Wightman function $W_{\Del_1, \Del_2, [\Del, J]}(x_1, x_2, p)$.
Finally by substituting \eqref{Wightman-3pt} into \eqref{Lorentzian_OPE_block} and using \eqref{Wightman_inverse} to simplify, 
we obtain an explicit expression of the Lorentzian OPE block:
\begin{align}\label{Lorentzian_OPE_block_momentum}
    \CB_{\Delta,J}^{(\rL)}(x_1, x_2) = -\frac{\pi^{h+1}\, \kappa_{\Delta, J}\,\CN_{12,[\Del, J]}}{2^J \sin\left(\pi(h - \tau) \right)}\, \int [\rD^d p]_{\rL} \, Q_{\Delta_1,\Delta_2, [\bar\Delta,J]} (x_1,x_2,-p)\, \CO_{\Delta,J}(p)\big|_{p^d \to \i\, p^0} \ .
\end{align}
This reproduces the result derived long time ago for the momentum space representation of OPE block in \cite{Ferrara:1971vh,Ferrara:1972ay,Ferrara:1973vz,Dobrev:1975ru,Dobrev:1977qv}.
We will use this particular representation  in the subsequent sections to consider when the external points $x_{1,2 }$ are space-like and time-like respectively.

\section{Space-like OPE block}\label{sec:space-like_OPE_block}

Starting with the integral representation \eqref{Lorentzian_OPE_block_momentum} given in momentum space, we will derive a holographic description of the Lorentzian OPE block for a pair of space-like separated points $x_1$ and $x_2$ with $x_{12}^2> 0$.
The resulting expression is the analytic continuation of the Euclidean OPE block \eqref{Spinning-OPE block} with $\Phi^{(\text{E})}_{\Delta,J}$ replaced with an HKLL-type higher spin field $\Phi^{(\text{L})}_{\Delta,J}$ propagating along the geodesic in Lorentzian AdS space-time.

\subsection{Scalar OPE block}
 
When the two operators are space-like separated, $x_{12}^2>0$, one can use the same Feynman parametrization \eqref{3pt_Feynman} as in the Euclidean case since the geodesic coordinate $X(\lambda)$ defined by \eqref{geodesic_coordinates} is still well-defined as $P_{12}^\frac{1}{2} = (x_{12}^2)^\frac{1}{2}$ remains real.
It is located on the geodesic $\gamma_{12}$ connecting the two boundary points $P_1$ and $P_2$ in the Lorentzian AdS space-time as parametrized in \eqref{geodesic_coordinates}.
We will begin with the scalar case to illustrate how to rewrite the Lorentzian OPE block in momentum space into the form facilitating the holographic interpretation. 
 
Let us consider the scalar block, where the $Q$-kernel gets simplified to give
\begin{align}
    \CB_{\Delta}^{(\rL)}(x_1, x_2) &=  \frac{\pi^{h+1}\,\kappa_{\Delta, 0}\,\CN_{12, [\Del, 0]}}{\sin\left(\pi(h - \Delta) \right)} \, \int [\rD^d p]_{\rL} \,   \frac{1}{(x_{12}^2)^{\frac{\Delta_{12}^+ - \bar \Delta}{2}}} \left( \frac{-p^2}{4 x_{12}^2} \right)^{\frac{h-\Delta}{2}} \nonumber \\  
    & \times \int_0^1 \d u \, u^{\frac{\Delta_{12}^- + h}{2}-1} (1-u)^{\frac{h-\Delta_{12}^-}{2}-1} \, e^{\i\,p\cdot (u x_1 + (1-u)x_2)} J_{\Delta- h} \left( \sqrt{-u(1-u)\,p^2\, x_{12}^2} \right) \CO_{\Delta}(p) \ ,
\end{align}
where we use the identity 
\begin{align}\label{Inu_to_Jnu}
    I_\nu (x) = \i^{-\nu} J_{\nu} ( \i\, x) \ ,
\end{align}
to rewrite it.
As a consistency check, this is equivalent to the results of \cite{Ferrara:1971vh,Ferrara:1973vz}.
To make contact with the bulk AdS interpretation, we define the following coordinate:
\begin{align}\label{Def:x-eta}
     x^\mu (u) \equiv u\, x_1^\mu + (1-u)\, x_2^\mu \ ,\qquad \eta(u) \equiv \sqrt{u(1-u)\,x_{12}^2}\ , 
\end{align}
which can be identified with the spatial and radial coordinates of the Poincar\'e patch of the Lorentzian AdS$_{d+1}$ space-time.
Indeed by the change of variable \eqref{u_to_lambda}, these are uplifted to the geodesic in the AdS embedding space coordinates \eqref{geodesic_coordinates} in the Poincar\'e section:
\begin{align}
    (X^+(u), X^-(u), X^\mu(u)) = \frac{1}{\eta(u)}\, (1, \eta^2(u) + x^2(u), x^\mu(u)) \ .
\end{align}
Using these variables, the scalar OPE block becomes
\begin{align}    \label{scalar_block_in_momentum}
    \CB_{\Delta}^{(\rL)}(x_1, x_2) &=  \frac{\pi^{h+1}\,\kappa_{\Delta, 0}\,\CN_{12, [\Del, 0]}}{2^{h-\Delta} \sin\left( \pi(h - \Delta)\right)}\,\frac{1}{(x_{12}^2)^{\frac{\Delta_{12}^+}{2}}} \,\int_0^1 \d u \, u^{\frac{\Delta_{12}^-}{2}-1} (1-u)^{-\frac{\Delta_{12}^-}{2}-1}  \nonumber \\  
    &\quad \times \int [\rD^d p]_{\rL}  \, e^{ \i\,p \cdot x(u)} \left(\sqrt{-p^2} \right)^{h-\Delta} \eta(u)^{h}\, J_{\Delta- h} \left( \sqrt{-p^2 }\, \eta(u) \right) \CO_{\Delta}(p) \ .
\end{align}
The second line of the above equation is precisely the so-called HKLL representation of the AdS scalar operator in momentum space \cite{Hamilton:2005ju,Hamilton:2006az,Hamilton:2006fh}.

\begin{figure}
    \centering
    \includegraphics[width=9cm]{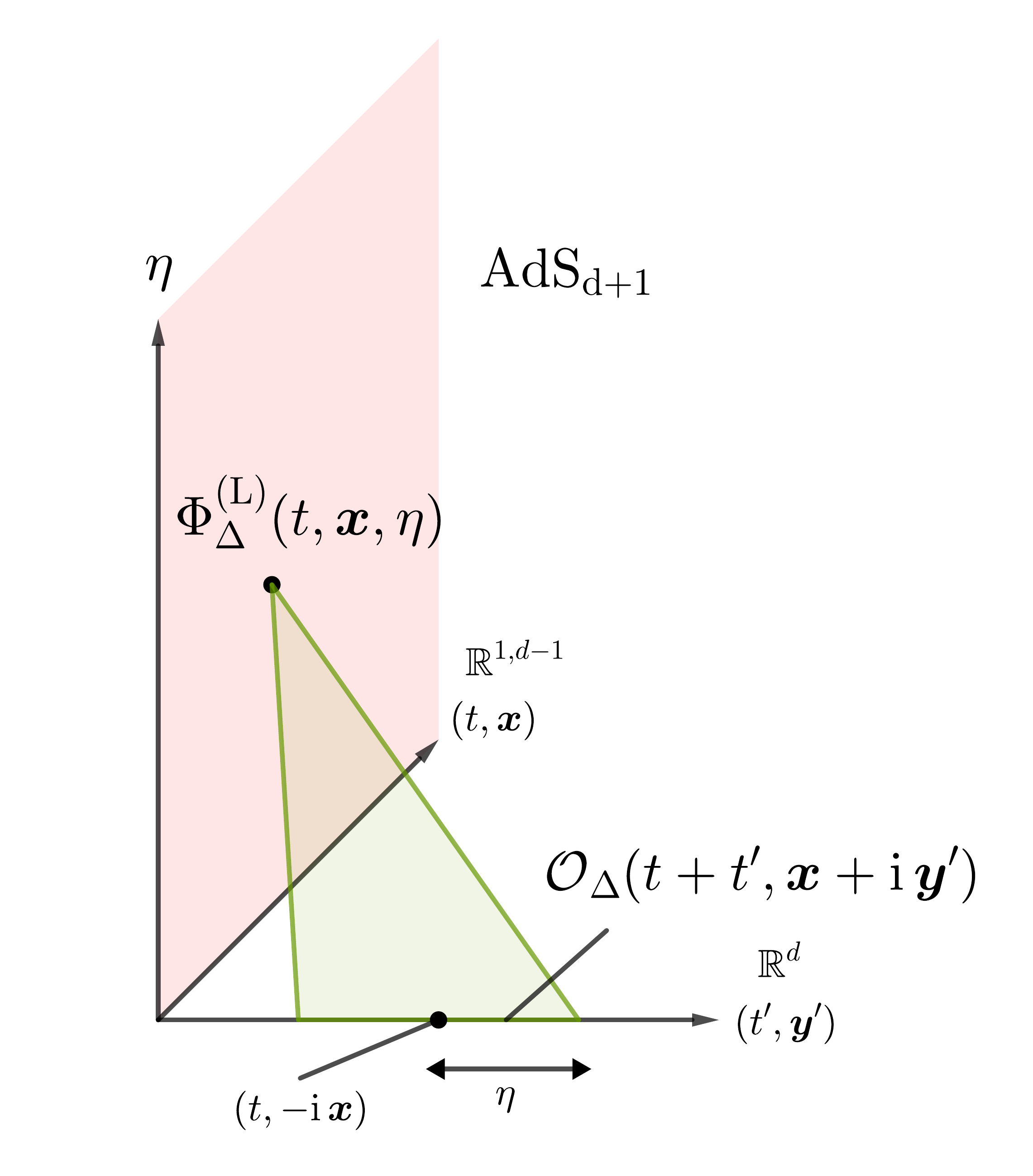}
    \caption{The HKLL representation of the AdS scalar field. The bulk field $\Phi^{(\rL)}_\Delta$ in Lorentzian AdS space-time is constructed by smearing a scalar primary $\CO_\Delta$ on a region in Euclidean space with the HKLL kernel.}
    \label{fig:HKLL_space-like}
\end{figure}

To make contact with the Euclidean result \eqref{Euclidean_scalar_OPEB},
we make the inverse Fourier transform of \eqref{scalar_block_in_momentum} to the position space expression of the scalar block.
To this end, we employ the following derived integral representation of $J_\nu (z)$ (see appendix \ref{ss:besselJ} for the derivation),
\begin{align} 
        J_{\nu} (|p_\text{E}| x) = \frac{1}{2^\nu \pi^h\Gamma(\nu - h + 1)}\left(\frac{|p_\text{E}|}{x}\right)^\nu\,\int_{|y| \le x}[\d^d  y]_{\rE}  \, (x^2 - |y|^2)^{\nu -h}\,e^{\i\,p_0 t + \i\, \bm{p}_\text{E} \cdot \bm{y}} \, ,
\end{align}
where $p_{\rE} = (p_0, \bm{p}_{\rE})$ and $y=(t, \bm{y})$ are $d$-dimensional vectors in $\BR^d$ with norms $|p_\text{E}| = \sqrt{p_{0}^2 + \bm{p}_\text{E}^2}$ and $|y|=\sqrt{t^2 + \bm{y}^2}$.
By analytically continuing $\bm{p}_\text{E} \to +\i\,\bm{p}$, we obtain the momentum vector $p = (p_0, \bm{p})$ in Lorentzian signature with the norm\footnote{Note that we use the mostly plus convention for the Lorentzian vector.} $\sqrt{-p^2} = \sqrt{p_0^2 - \bm{p}^2} \ge 0$.
With this parametrization the Bessel function takes the form:
\begin{align}\label{J_integral_rep}
    J_\nu \left(\sqrt{p_0^2 - \bm{p}^2}\, x \right) = \frac{1}{2^\nu \pi^h\Gamma(\nu - h + 1)}\left(\frac{\sqrt{p_0^2 - \bm{p}^2}}{x}\right)^\nu\,\int_{|y| \le x}[\d^d  y]_{\rE}  \, (x^2 - |y|^2)^{\nu -h}\,e^{\i(p_0 t + \i\, \bm{p}\cdot \bm{y})} \, .
\end{align}
Substituting this expression to \eqref{scalar_block_in_momentum} and performing the inverse Fourier transform
\begin{align}
        \int [\rD^d p]_{\rL} \, e^{ \i (p_0 t + \i\, \bm{p}\cdot \bm{y}) }\, \CO_\Delta (p) = \CO_\Delta \left(-t,\, \i\, \bm{y} \right) \ ,
\end{align}
we find the physical space representation of the scalar OPE block:

\begin{align}\label{Scalar_OPE_block_position}
    \begin{aligned}
    \CB_{\Delta}^{(\rL)}(x_1, x_2) &= \frac{\pi\,\kappa_{\Delta, 0}\,\CN_{12, [\Del, 0]}}{\sin\left( \pi(h - \Delta)\right)\,\Gamma(1-\bar\Delta)}\,\frac{1}{(x_{12}^2)^{\frac{\Delta_{12}^+}{2}}} \int_0^1 \d u \, u^{\frac{\Delta_{12}^- }{2}-1} (1-u)^{-\frac{\Delta_{12}^-}{2}-1}\\  
    & \quad\times    \int_{t^2 + \bm{y}^2 \le \eta(u)^2} [\d^d y]_{\rE}\, \left( \frac{\eta(u)}{\eta(u)^2 -t^2 - \bm{y}^2} \right)^{\bar\Delta}\, \CO_\Delta \left(t(u) + t, \, \bm{x}(u) + \i\, \bm{y} \right) \ ,
    \end{aligned}
\end{align}
where we denoted $x^\mu(u) = (t(u), \bm{x}(u))$ and redefined $t\to -t$.
Again the second line is exactly the same form as the HKLL representation of the AdS scalar field derived in \cite{Hamilton:2005ju,Hamilton:2006fh,Hamilton:2006az} (see figure \ref{fig:HKLL_space-like} for illustration):\footnote{Our normalization of the HKLL field is different from the one in \cite{Hamilton:2006fh}: 
$\displaystyle \Phi_\text{HKLL} = \frac{\Gamma(\Delta - h +1)}{\pi^h \Gamma(\Delta-d+1)}\,\Phi_\text{our}$.
The overall normalization constant does not matter in our discussion, and we make a simplest choice for our purpose.
}
\begin{align}\label{Scalar_HKLL_field}
    \Phi^{(\text{L})}_{\Del} \left( t, \bm{x}, \eta\right) = \int_{t'^2 + \bm{y}'^2 \le \eta^2} \d t'\d^{d-1}\bm{y}'\, \left(\frac{\eta}{\eta^2 - t'^2 - \bm{y}'^2}\right)^{\bar \Delta}\,\CO_\Delta \left(t + t', \, \bm{x} + \i\, \bm{y}' \right) \ .
\end{align}
The final result we obtained from combining the momentum space shadow projector and analytic continuation \eqref{Scalar_OPE_block_position} is a natural generalization of the work \cite{daCunha:2016crm} which derived the scalar OPE block in CFT$_2$ that decomposed into the holomorphic and anti-holomorphic parts, each of which can be regarded as CFT$_1$ respectively.
It is also easy to see that \eqref{Scalar_OPE_block_position} takes the same form as the Euclidean result \eqref{Euclidean_scalar_OPEB} given in the embedding space coordinates up to the signature of the metric and the integration region translates into the condition $(-2P\cdot X) \ge 0$, which was also valid in Euclidean case.

\subsection{OPE block for a spin-$J$ primary operator}
  
Let us next turn to the OPE block \eqref{Lorentzian_OPE_block_momentum} for a tensor field with spin $J$.
Using the integral representation of the $Q$-kernel \eqref{Q-kernel} and the analytic continuation we find the general expression for the spinning OPE block:
\begin{align}\label{Spinning_OPEB_momentum}
    \begin{aligned}
    &\CB_{\Delta,J}^{(\rL)}(x_1, x_2) =   \frac{\pi^{h+1}\,\kappa_{\Delta, J}\,\CN_{12, [\Del, J]}}{2^{\bar\Delta -h} \sin\left(\pi(h - \tau) \right)}\,\frac{1}{ (x_{12}^2)^{\frac{\Delta_{12}^+- \bar \tau}{2}} }\,\int_0^1 \d u \, u^{\frac{\bar\tau + \Delta_{12}^-}{2}-1} (1-u)^{\frac{ \bar\tau - \Delta_{12}^-}{2}-1} \, \int [\rD^d p]_{\rL}  \\
    & \times  \frac{1}{J!\,(h-1)_J}\,   \CD_J \left(\bdel_{12}^+, \, d_z \cdot \partial_1; \, \bdel_{12}^-, \, d_z \cdot \partial_2\right) \,e^{\i\,p\cdot x(u)}  \left(\frac{\sqrt{-p^2}}{\eta(u)} \right)^{\bar \tau-h} J_{h-\bar \tau}\left(\sqrt{-p^2}\,\eta(u)\right) \CO_{\Delta,J}(p,z) \ .
    \end{aligned}
\end{align}
In what follows, we will use the integral representation of the modified Bessel function \eqref{J_integral_rep} to recast the space-like OPE block into the holographic form in parallel with the scalar case.

\subsubsection{Conserved case}\label{ss:conserved}
 
Before examining general cases we will work out the case with a spin-$J$ conserved current satisfying the conservation law in momentum space:
\begin{align}
    p^{\mu_1}\, \CO_{\Delta, \mu_1 \mu_2\cdots \mu_J}(p) = 0 \ .
\end{align}
When $\CD_J$ acts on $e^{\i\,p\cdot x(u)}$ the resulting term always vanishes due to the conservation law above.
The remaining terms that $\CD_J$ acts on is a function depending on $x_{12}=x_{1}-x_2$, but not on $x_1 + x_2$ which can also be understood as the direct consequence of translational invariance.
Thus we introduce new coordinates by
\begin{align}
    x_+ \equiv x_1 + x_2 \ ,\qquad x_- \equiv x_1 - x_2 = x_{12} \ ,
\end{align}
and the derivative with respect to them by
\begin{align}
    \partial_\pm \equiv \frac{\partial_1 \pm \partial_2}{2} \ .
\end{align}
The expansion of $\CD_J$ in $\partial_\pm$ becomes
\begin{align}
    \CD_J = (-1)^J\,(\bar\Delta -1)_J\,(z\cdot \partial_-)^J + \cdots \ ,
\end{align}
where $\cdots$ are differential operators including $\partial_+^k~(k\ge 1)$, which vanish when acted on a function of $x_- = x_{12}$.

Using the Bessel function differential identity:
\begin{align}
    \frac{\partial (x^\nu J_\nu(a x))}{\partial x} = a\, x^\nu J_{\nu-1}(a x) \ ,
\end{align}
the second line of \eqref{Spinning_OPEB_momentum} takes the form:
\begin{align}
    \begin{aligned}
    &(-1)^J\,(\bar\Delta -1)_J\,\frac{1}{J!\,(h-1)_J}\, e^{\i\,p\cdot x(u)}\,\left( u(1-u)\right)^J\,(x_-\cdot d_z)^J \left(\frac{\sqrt{-p^2}}{\eta(u)} \right)^{\bar \Delta -h}\, J_{h-\bar \Delta}\left(\sqrt{-p^2}\,\eta(u)\right)  \, \CO_{\Delta,J}(p,z) \\
    &= \frac{(-1)^J(\bar\Delta -1)_J}{2^{\Delta -h}\pi^h \Gamma(1-\bar\Delta)}\,\frac{1}{J!\,(h-1)_J}\, \left( u(1-u)\right)^J\,(x_-\cdot d_z)^J\\
    &\qquad \times \int_{t^2 + \bm{y}^2 \le \eta(u)^2} [\d^d y]_\rE \,e^{-\i\,p_0 (t(u) - t) +\i\,\bm{p} \cdot (\bm{x}(u)+ \i\,\bm{y})}\,(\eta(u)^2 - t^2 - \bm{y}^2)^{-\bar\Delta} \, \CO_{\Delta,J}(p,z) \ .
    \end{aligned}
\end{align}
Putting all together we obtain the space-time representation of the spinning OPE block:
\begin{align}\label{Spinning-OPE block-L}
    \begin{aligned}
    \CB_{\Delta,J}^{(\rL)}(x_1, x_2) &= \frac{(-1)^{J}\,\pi\,(\bar\Delta -1)_J\,\kappa_{\Delta, J}\,\CN_{12, [\Del, J]}}{\sin\left( \pi(h - \Delta)\right)\,\Gamma(1-\bar\Delta)}\,\frac{1}{(x_{12}^2)^\frac{\Delta_{12}^+}{2} } \\
     & \quad\times \frac{1}{2^J}\int_0^1 \d u \, u^{\frac{\Delta_{12}^-}{2}-1} (1-u)^{-\frac{\Delta_{12}^-}{2} -1}\, \Phi^{(\text{L})}_{\text{con},\, \mu_1\cdots\mu_J} \left( x^\mu(u), \eta(u)\right) \, w^{\mu_1}(u)\cdots w^{\mu_J}(u)\ ,
    \end{aligned}
\end{align}
where we introduced a new vector:
\begin{align}
    w(u)^\mu \equiv 2 u(1-u)\,x_{12}^\mu = \frac{\d x^{\mu}(\lambda)}{\d\lambda} \ ,
\end{align}
and we have used the relation \eqref{u_to_lambda}
and a bulk spin-$J$ field constructed by\footnote{We redefined $t'\to -t'$ in deriving this expression.}
\begin{align}\label{Sarkar-Xiao_field}
    \Phi^{(\text{L})}_{\text{con},\, \mu_1\cdots\mu_J} \left( x^\mu, \eta\right)  \equiv \frac{1}{\eta^J}\int_{t'^2 + \bm{y}'^2 \le \eta^2} \d t'\,\d^{d-1} \bm{y}'\, \left( \frac{\eta}{\eta^2 - t'^2 - \bm{y}'^2} \right)^{\bar\Delta}\, \CO_{\Delta, \mu_1\cdots\mu_J} \left(t + t', \bm{x} + \i\,\bm{y}' \right) \ .
\end{align}
Notably this is the HKLL-type representation of massless higher spin fields derived in \cite{Sarkar:2014dma}.

The bulk field defined by \eqref{Sarkar-Xiao_field} takes a similar form to the Euclidean higher spin field given by \eqref{Euclidean_higher_spin_HKLL_in_Embedding}.
This is not a coincidence, and one can show the latter is the (Lorentzian version of) holographic field for a general CFT operator and reduces to the former when the dual CFT operator is conserved. We will return to explaining this point momentarily at the end of the next subsection.

\subsubsection{General spinning case}
 
One can generalize the previous argument to a non-conserved spinning operator.
To begin with, we consider the spin-$1$ case where the differential operator simplifies to
\begin{align}
    \CD_1 = z\cdot (\Delta_{12}^-\, \partial_+ - \bar \tau\, \partial_-) \ .
\end{align}
A straightforward calculation shows
\begin{align}\label{Spin1_OPE_Block}
    \begin{aligned}
    \CB_{\Delta,\,J=1}^{(\rL)}&(x_1, x_2) \propto \frac{1}{(x_{12}^2)^\frac{\Delta_{12}^+}{2} } \int_0^1 \d u \, u^{\frac{\Delta_{12}^-}{2}-1} (1-u)^{-\frac{\Delta_{12}^-}{2} -1}\\
    & \times\left[ - \bar\tau\, \frac{1}{\eta(u)}\int_{t^2 + \bm{y}^2 \le \eta(u)^2} [\d^d  y]_{\rE}  \left( \frac{\eta(u)}{\eta(u)^2 -t^2 - \bm{y}^2} \right)^{\bar\Delta}\,w(u)^\mu\, \CO_{\Delta,\mu} \left(t(u) + t, \bm{x}(u) + \i\,\bm{y}\right) \right. \\
    &\quad \left. + 2 \left(\Delta_{12}^- - (2u-1)\,\bar\tau\right)\,\int_{t^2 + \bm{y}^2 \le \eta(u)^2} [\d^d  y]_{\rE}  \left( \frac{\eta(u)}{\eta(u)^2 - t^2 - \bm{y}^2} \right)^{\bar\Delta-1}\,\partial^\mu \CO_{\Delta, \mu}\left(t(u) + t, \bm{x}(u) + \i\,\bm{y} \right) \right] \ .
    \end{aligned}
\end{align}
The terms in the square bracket looks similar to the reconstruction formula for a massive gauge field given by \cite{Kabat:2012hp} in (69), but the second term differs from theirs where $\partial^\mu \CO_\mu$ should roughly be replaced by $w_\nu\partial^\nu \partial^\mu \CO_\mu$.\footnote{While it is not immediately clear how to interpret the square bracket term inside \eqref{Spin1_OPE_Block} as a bulk field, we can show the second term does not contribute to the OPE block after the $u$-integration.
First we note that the OPE block is invariant under the exchange of the two scalar primary operators as they are space-like separated.
This exchange amounts to the following transformation in the right hand side of \eqref{Spin1_OPE_Block},
\begin{align}\label{Z2_symmetry}
    x_1 \leftrightarrow x_2 \ , \qquad \Delta_{12}^- \leftrightarrow - \Delta_{12}^- \ .
\end{align}
Making a change of variable $u\to 1-u$ together with the exchange does not change the integration measure for $u$ and the variables $\eta(u), x(u)$ and $w^\mu(u)$, but flips the sign in the second term in the square bracket.
Hence the second term does not contribute and the OPE block of a non-conserved spin-$1$ operator reduces to the same form as the one for conserved currents.
}
We will discuss this issue in section \ref{sec:Discussion}.

For higher spin cases, there are at most $J +1$ terms in the derivative operator $\CD_J$, hence the same number of terms in the OPE block \eqref{Spinning_OPEB_momentum}:
\begin{align}\label{space-like_OPEB_HS}
    \CB_{\Delta, J}^{(\rL)}(x_1, x_2) &= \frac{\pi\,\kappa_{\Delta, J}\,\CN_{12, [\Del, J]}}{\sin\left(\pi(h - \tau) \right)\,\Gamma(1-\bar\tau)}\, \frac{1}{(x_{12}^2)^\frac{\Delta_{12}^+}{2} } \int_0^1 \d u \, u^{\frac{\Delta_{12}^-}{2}-1} (1-u)^{-\frac{\Delta_{12}^-}{2} -1}\, \Phi^{(\text{L})}_{\Delta, J} \left( x^\mu(u), \eta(u) \right)\ ,
\end{align}
where we define the ``bulk higher spin field" $\Phi^{(\text{L})}_{\Delta, J}$ with the spin indices contracted with $w^\mu(u)$, which is schematically written as
\begin{align}\label{Lorentzian_higher_spin_HKLL}
    \begin{aligned}
    \Phi^{(\text{L})}_{\Delta, J} &(x^\mu(u), \eta(u)) \\
        &= \sum_{r=0}^{J}\, a_r(u)\, \frac{1}{\eta(u)^{J-r}}\,\int_{t^2 + \bm{y}^2 \le \eta(u)^2} [\d^d  y]_{\rE}  \left( \frac{\eta(u)}{\eta(u)^2 -t^2 - \bm{y}^2} \right)^{\bar\Delta - r}\, w(u)^{J-r}\, \partial^{r} \CO_{\Delta,J} \left(t(u) + t, \bm{x}(u) + \i\,\bm{y}\right) \ ,
    \end{aligned}
\end{align}
with some $a_r(u)$s which can be determined via explicit differentiation.
 
We have already seen the same type of the structure when we introduced the HKLL-type representation of a higher spin field \eqref{HKLL-Spin-2} in the Euclidean embedding space, where $J+1$ terms are generated by the binomial expansion of the tensor structure $\CJ^{AB}$ defined by \eqref{Def:JAB-XP}.
Indeed \eqref{Lorentzian_higher_spin_HKLL} should be a Lorentzian counterpart of the higher spin field \eqref{HKLL-Spin-2} as the Lorentzian OPE block \eqref{Spinning_OPEB_momentum} is obtained by the analytic continuation of the Euclidean one \eqref{Spinning-OPE block}.
To make their relation manifest, we use an alternative representation of $\Phi^{(\text{L})}_{\Delta, J}$ using the differential operator $\CD_J$, which directly follows from the definition of the OPE block \eqref{Spinning_OPEB_momentum}:
\begin{align}\label{Lorentzian_higher_spin_HKLL_diff}
    \begin{aligned}
    \Phi^{(\text{L})}_{\Delta, J}(x^\mu(u), \eta(u)) &= \frac{1}{2^J \,J! (h-1)_J}\,(x_{12}^2)^{\bar\tau}\,\CD_J \left(\bdel_{12}^+, \, d_z \cdot \partial_1; \, \bdel_{12}^-, \, d_z \cdot \partial_2\right)\,(x_{12}^2)^{-\bar\tau}\\
    &\quad\times \int_{t^2 + \bm{y}^2 \le \eta(u)^2} \d t\,\d^{d-1}\bm{y}\, \left(\frac{\eta(u)}{\eta(u)^2 - t^2 - \bm{y}^2}\right)^{\bar \tau }\,\CO_{\Delta, J} \left(t(u) + t, \, \bm{x}(u) + \i\, \bm{y} , z\right) \ .
    \end{aligned}
\end{align}
This should be compared with the differential representation of the Euclidean higher spin field $\Phi^{(\text{E})}_{\Delta, J}$ given in \eqref{Diff-HKLL-Spinning} and they are consistent to each other up to a constant as expected.

To summarize, we obtained two equivalent representations of higher spin fields $\Phi^{(\text{L})}_{\Delta, J}$ in Lorentzian AdS space-time, one \eqref{Lorentzian_higher_spin_HKLL} with $J+1$ terms and the other \eqref{Lorentzian_higher_spin_HKLL_diff} with the differential operator $\CD_J$.
The former was derived by first expanding $\CD_J$ and performing the Fourier transform afterwards, while the latter was given by pulling out $\CD_J$ outside the momentum integral and performing the Fourier transform.
We also showed that $\Phi^{(\text{L})}_{\Delta, J}$ is a counterpart of the Euclidean higher spin field $\Phi^{(\text{E})}_{\Delta, J}$ defined in \eqref{HKLL-Spin-2} by comparing their differential representations \eqref{Lorentzian_higher_spin_HKLL_diff} and \eqref{Diff-HKLL-Spinning}.

Before closing this section some comments are in order regarding the correspondence between the Lorentzian and Euclidean fields, $\Phi^{(\text{L})}_{\Delta, J}$ and $\Phi^{(\text{E})}_{\Delta, J}$:
\begin{itemize}
    \item 
        The Lorentzian expression \eqref{Lorentzian_higher_spin_HKLL} of $\Phi^{(\text{L})}_{\Delta, J}$ reduces to the massless higher spin field $\Phi_\text{con}^{(\rL)}$ of the form \eqref{Sarkar-Xiao_field} when $\CO_{\Delta, J}$ is a conserved current,\footnote{
        The two fields are related by $\displaystyle \Phi^{(\text{L})}_{\Delta, J} = \frac{(-1)^J\,(\bar\Delta -1)_J}{2^J}\,\frac{\sin\left(\pi (h-\tau)\right)\,\Gamma(1-\bar\tau)}{\sin\left(\pi (h-\Delta)\right)\,\Gamma(1-\bar\Delta)}\, \Phi^{(\text{L})}_{\text{con},\, \mu_1\cdots\mu_J} \, w^{\mu_1}\cdots w^{\mu_J}$.}
        which is not directly manifest in the general Euclidean expression \eqref{Euclidean_higher_spin_HKLL_in_Embedding}.
        We can however recover the corresponding expression from $\Phi_{\Del, J}^{(\rE)}$ for the conserved case by first expanding the numerator of the spinning bulk to boundary propagator in the integrand of \eqref{Euclidean_higher_spin_HKLL_in_Embedding} into a leading term which only contains $(W\cdot D_{Z_0})^J$ and independent of $P_0$ and the remaining $J$ terms which contain explicit $P_0$ dependences. 
        As discussed in \cite{Costa:2014kfa}, we can also package the remaining $J$ terms into a total derivative with respect to $P_0$. Upon substitution and integration by part, for the conserved boundary tensor field, i.\,e., $\partial_{P_0}^A \CO_{\Del, A B_2 \dots B_J}(P_0) = 0$ with $\Del  = d-2+J$, we can drop their contributions in the boundary $P_0$ integration, and the surviving leading term is directly projected into \eqref{Sarkar-Xiao_field}. This is simply a manifestation of the well-known AdS/CFT lore that the conservation of boundary operator implies the gauge invariance of the dual bulk field.
        
        \item More generally for non-conserved case, there are $J+1$ terms both in $\Phi^{(\text{L})}_{\Delta, J}$ and $\Phi^{(\text{E})}_{\Delta, J}$, which arise from the expansion of the differential operator $\CD_J$ and the product of $\CJ^{AB}$s in \eqref{Lorentzian_higher_spin_HKLL} and \eqref{HKLL-Spin-2} respectively.
        In the Euclidean expression \eqref{HKLL-Spin-2}, it is important to keep all the $J+1$ terms to maintain the invariance under the shift $W^A \to W^A+\alpha X^A$ while in the Lorentzian expression, the $J+1$ terms are necessary to construct a single bulk higher spin field dual to a general spin-$J$ CFT operator in \eqref{Lorentzian_higher_spin_HKLL}.
\end{itemize}

\section{Time-like OPE block}\label{sec:time-like_OPE_block}

As the discussion of section \ref{sec:space-like_OPE_block} relies on the fact that $x_1$ and $x_2$ are space-like separated, the result shares many similarities with the Euclidean case.
For the time-like separated case however, we need to modify the preceding discussion.
A straightforward way to achieve this is to analytically continue the space-like case to the time-like case with the $\i\, \epsilon$-prescription.
In section \ref{ss:Timelike_analytic_cont}, we perform this analytic continuation and derive an integral representation for the time-like OPE block, which is new to our best knowledge.
We further recast it into a holographic form in a similar manner to the space-like case, and find the resulting expression can be interpreted as a bulk field smeared over a geodesic not on the AdS space-time but on a de Sitter-like hyperboloid in $\BR^{2,d}$.

The time-like OPE block was also considered in a different context \cite{Czech:2016xec,deBoer:2016pqk} and proposed to have another holographic description known as the surface Witten diagram, which is quite different from our result in section \ref{ss:Timelike_analytic_cont} in general dimensions.
In order to make contact with the surface Witten diagram, we provide an alternative derivation based on the duality between conformal defects \cite{Gadde:2016fbj,Fukuda:2017cup} in section \ref{ss:Surface_Witten}.
Focusing on two-dimensional CFT's, the defect duality exchanges a pair of time-like separated points to a pair of space-like ones.
Then the time-like OPE block reduces to the space-like one, which agrees with the surface Witten diagram in two dimensions.

Section \ref{ss:OPE_block_position} is supplementary, where we discuss the space-time picture of the Lorentzian OPE block.
We show that the OPE block takes the same form as the Euclidean one with the integration range restricted to a natural causal space-time region associated with a pair of points.

\subsection{Analytic continuation from space-like separation}\label{ss:Timelike_analytic_cont}
 
To employ the $ \i\, \epsilon$-prescription appropriately, we first remark that there are two ways to analytic continue from an Euclidean theory to a Lorentzian one.
At the level of correlators, these correspond to Wightman correlators and time-ordered (Feynman) correlators
(see appendix \ref{Appendix:Wightman_and_Feynman}).
In considering the OPE block, the order of two operators is fixed, so we should use the Wightman-type analytic continuation for $x_{12}^2$.
To this end, without loss of generality we consider the OPE block of the order $\CO_1 \CO_2$.\footnote{We implicitly use the fact that these two operators act on vacuum in the sense that $\CO_1 \CO_2 \ket{0}$.}
In this ordering, starting with Euclidean time $\tau_{\text{E}, i} = \epsilon_i$, the desired analytic continuation is achieved by setting $\tau_{\text{E}, i} = \epsilon_i + \i\, t_i$ with $\epsilon_1 > \epsilon_2$.\footnote{This inequality is usually required for Wightman functions to be holomorphic in the upper half plane. Though we are now dealing with the OPE, we impose the condition so that our result should be consistent with the Wightman function.}
The distance $x_{12}^2$ becomes
\begin{align}
    x_{\text{E},12}^2 = ({\bm x}_1 - {\bm x}_2 )^2 + (\tau_{\text{E}, 1} - \tau_{\text{E}, 2})^2 \quad  \rightarrow \quad x_{\text{L},12}^2 = ({\bm x}_1 - {\bm x}_2 )^2 - (t_1 -t_2)^2 +  \i\, \epsilon\, (t_1 -t_2) \, ,
\end{align}
where $\epsilon = \epsilon_1 - \epsilon_2 >0$ and we take $\epsilon$ to zero in the end.
Here we again denote $x_{12}^2$ with the subscripts $\text{E}$ and $\text{L}$ to avoid confusion.
With this prescription, one can confirm that the OPE block for space-like separation is reduced to the one we obtain in the previous section.
 
\begin{figure}
    \centering
    \includegraphics[width=10cm]{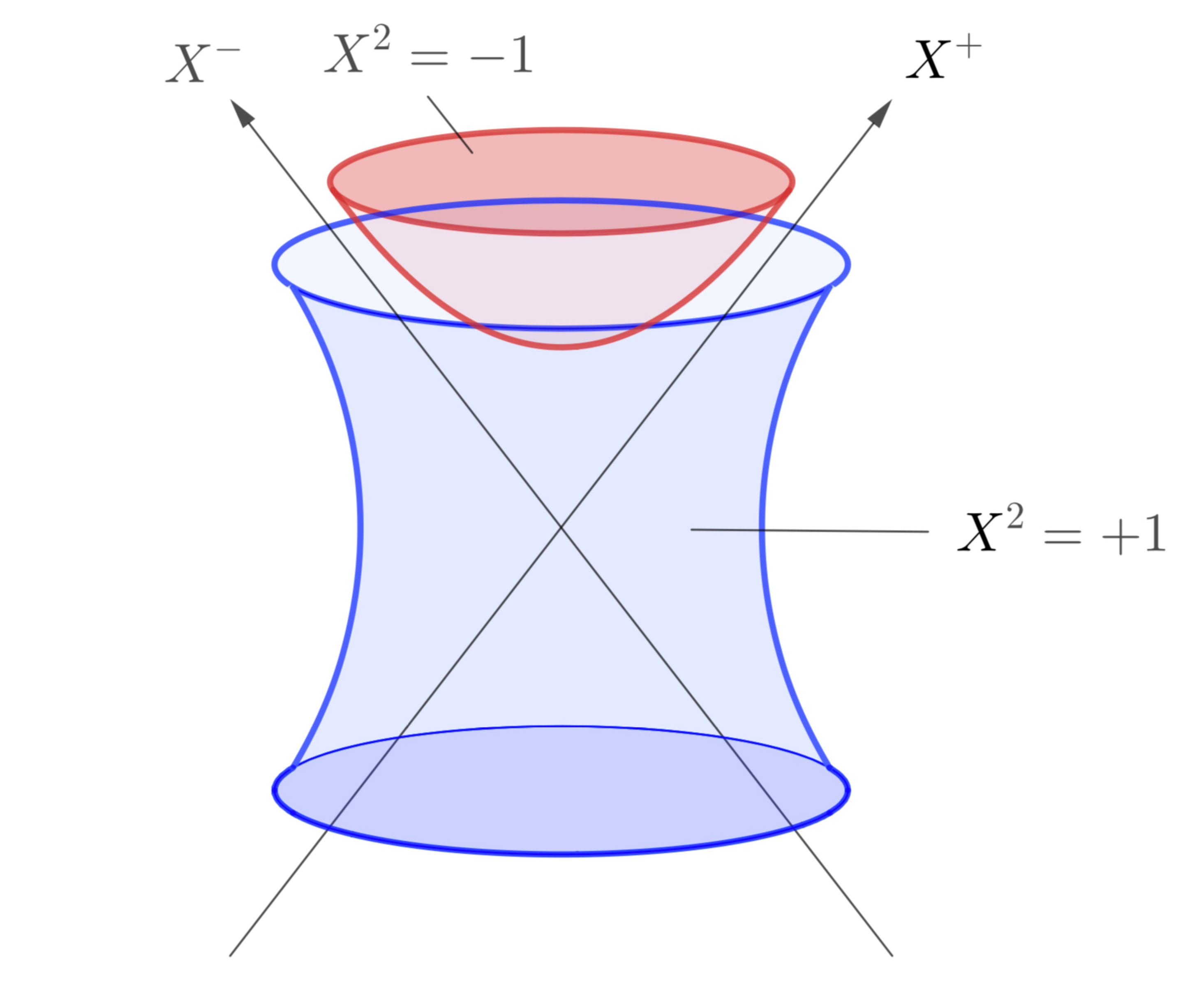}
    \caption{The embeddings of the AdS space-time ($X^2 = -1$) and the hyperboloid ($X^2 = 1$) in $\BR^{2,d}$.}
    \label{fig:deSitter_hyperboloid}
\end{figure}

Now we consider the time-like configuration such that $x_1$ sits in the forward lightcone of $x_2$, then the analytic continuation becomes\footnote{For the opposite configuration, $x_{12}^2 \rightarrow |x_{12}^2 |\, e^{-\i \pi}$ .}
\begin{align}\label{x12-analytic}
    \begin{aligned}
            x_{12}^2 &=  ({\bm x}_1 - {\bm x}_2 )^2 - (t_1 -t_2)^2 +  \i\, \epsilon\, (t_1 -t_2)  \\
    &= - \left\{ (t_1 -t_2)^2 - ({\bm x}_1 - {\bm x}_2 )^2 \right\} + \i\, \epsilon\, (t_1-t_2)   \\
    &\rightarrow |x_{12}^2|\, e^{\i \pi} \, .
    \end{aligned}
\end{align}
The OPE block for this time-like separation follows from \eqref{Lorentzian_OPE_block_momentum} with all $x_{12}^2$s replaced to $|x_{12}^2 |\, e^{\i \pi}$.
Let us focus on the scalar OPE block as the generalization to the spinning case is a straightforward exercise.
The scalar OPE block after the analytic continuation \eqref{x12-analytic} is given by
\begin{align}\label{time-like_OPEB_momentum}
    \CB_{\Delta}^{(\rL)}(x_1, x_2) &=  \frac{\pi^{h+1}\,\kappa_{\Delta, 0}\,\CN_{12, [\Del, 0]}\, e^{-\i \pi \frac{\Delta_{12}^+ - h}{2}}}{\sin\left(\pi(h - \Del) \right)} \, \int [ \rD^d  p]_{\rL}\,   \frac{1}{(|x_{12}^2|)^{\frac{\Delta_{12}^+ - \bar \Delta}{2}}} \left( \frac{-p^2}{4 |x_{12}^2|} \right)^{\frac{h-\Delta}{2}} \nonumber \\  
    & \times \int_0^1 \d u \, u^{\frac{\Delta_{12}^- + h}{2}-1} (1-u)^{\frac{h-\Delta_{12}^-}{2}-1} \, e^{\i\,p\cdot (u x_1 + (1-u)x_2)} I_{\Delta- h} \left( \sqrt{-u(1-u)p^2 |x_{12}^2|} \right) \CO_{\Delta}(p) \, .
\end{align} 
Similarly to the space-like case we can define a ($d+1$)-dimensional vector as
\begin{align} \label{time-like_x(u)}
    x^\mu (u) = u\, x_1^\mu + (1-u)\, x_2^\mu \ , \quad \chi(u) = \sqrt{u(1-u)\, |x_{12}^2|}  = \sqrt{-u(1-u)\, x_{12}^2} \ ,
\end{align}
where $\chi(u)$ will play the analogue of radial coordinate, but now in analytic continuation of de Sitter instead of the AdS space-time as we will shortly explain.
We can perform the inverse Fourier transform with respect to $p$ as in the case for the space-like OPE block.
The modified Bessel function $I_\nu (x)$ has a similar integral representation as the Bessel function $J_\nu (x)$ as shown in appendix \ref{ss:besselI}:
\begin{align}
        I_\nu (px) = \frac{1}{2^\nu \pi^h\,\Gamma(\nu-h+1)}\,\,\left( \frac{p}{x}\right)^\nu\,\int_{|y|\le x}[\d^d  y]_{\rE} \,(x^2 - y^2)^{\nu-h}\,e^{-p\cdot y} \ .
\end{align}
Following the same procedure as in the space-like case, we can rewrite the modified Bessel-$I$ function in \eqref{time-like_OPEB_momentum} as
\begin{align}
    \begin{aligned}
    I_{\Delta - h} \left (\sqrt{p_0^2 - \bm{p}^2}\, \chi(u)\right) &= \frac{1}{2^{\Delta-h} \pi^h\Gamma(1-\bar\Delta)}\,\left( \frac{\sqrt{p_0^2 - \bm{p}^2} }{\chi(u)} \right)^{\Delta - h}\\
        &\qquad \times\int_{t^2 + \bm{y}^2 \le \chi(u)^2} [\d^d  y]_{\rE}\,  \left( \chi(u)^2 -t^2 - \bm{y}^2 \right)^{-\bar\Delta} e^{- p_0 t + \i\,\bm{p} \cdot \bm{y}} \ .
    \end{aligned}
\end{align}
Substituting it into \eqref{time-like_OPEB_momentum} and performing the inverse Fourier transform gives us the scalar OPE block for time-like separation,
\begin{align}\label{time-like_OPE_block}
    \begin{aligned}
    \CB_{\Delta}^{(\rL)}(x_1, x_2) &=  \frac{\pi\,\kappa_{\Delta, 0}\,\CN_{12, [\Del, 0]} \, e^{-\i\pi \frac{\Delta_{12}^+ -h}{2}}}{\sin\left( \pi(h - \Delta)\right)\,\Gamma(1-\bar\Delta)}\,\frac{1}{|x_{12}^2|^{\frac{\Delta_{12}^+}{2}}} \int_0^1 \d u \, u^{\frac{\Delta_{12}^- }{2}-1} (1-u)^{-\frac{\Delta_{12}^-}{2}-1}\\  
     & \quad\times    \int_{t^2 + \bm{y}^2 \le \chi(u)^2} [\d^d  y]_{\rE}  \left( \frac{\chi(u)}{\chi(u)^2 -t^2 - \bm{y}^2} \right)^{\bar\Delta}\, \CO_\Delta \left(t(u) +\i\,t,  \bm{x}(u) + \bm{y} \right) \ .
    \end{aligned}
\end{align}
This expression closely resembles the space-like OPE block \eqref{Scalar_OPE_block_position} where we had the holographic interpretation as the bulk HKLL field smeared over the geodesic connecting the boundary points $x_1$ and $x_2$ in AdS space time. In the current time-like case, we wish to find a similar interpretation of \eqref{time-like_OPE_block} such that the second line represent a bulk field $\Phi_\Delta$ along the geodesic encoded by $(x^\mu(u), \chi(u))$.
To this end, we uplift the coordinates $(x^\mu(u), \chi(u))$ using \eqref{u_to_lambda} to the Lorentzian embedding space $\BR^{d, 2}$ as:
\begin{align}\label{Def:Y-geodesic}
    Y(\lambda) = \frac{e^{\lambda} P_1 + e^{-\lambda} P_2}{\sqrt{2 P_1 \cdot P_2}}\ , \quad Y^2(\lambda) = +1\ .
\end{align}
It is important to notice that $Y(\lambda)$ is no longer on the AdS space-time, but instead on a hyperboloid satisfying
\begin{align}\label{deSitter_embedding}
    X^2 = - X^+ X^- + \eta_{\mu\nu}\,X^\mu X^\nu = +1 \ ,
\end{align}
with the metric $\eta_{\mu\nu} = \text{diag}(-1, 1, \cdots, 1)$ (see Fig.\,\ref{fig:deSitter_hyperboloid}).
However as in the case of the AdS space-time we can parameterize points on this hyperboloid by
\begin{align}\label{deSitter_embedding_coord}
    (X^+, X^-, X^\mu) = \frac{1}{\chi} \left( 1,\, x^2 - \chi^2,\, x^\mu\right) \ ,
\end{align}
where $\chi$ indeed becomes the analogue of radial coordinate.
This parametrization makes it clear that the hyperboloid is an analytic continuation of de Sitter space-time in the flat slicing as seen from the metric:
\begin{align}\label{deSitter_hyperboloid}
    \d s^2 = \frac{-\d \chi^2 + \eta_{\mu\nu} \d x^\mu \d x^\nu}{\chi^2} \ .
\end{align}
It follows that $Y(\lambda)$ sits on the geodesic anchored on the points $P_1$ and $P_2$ at $\chi =0$ inside the hyperboloid.

\begin{figure}
    \centering
    \includegraphics[width=9cm]{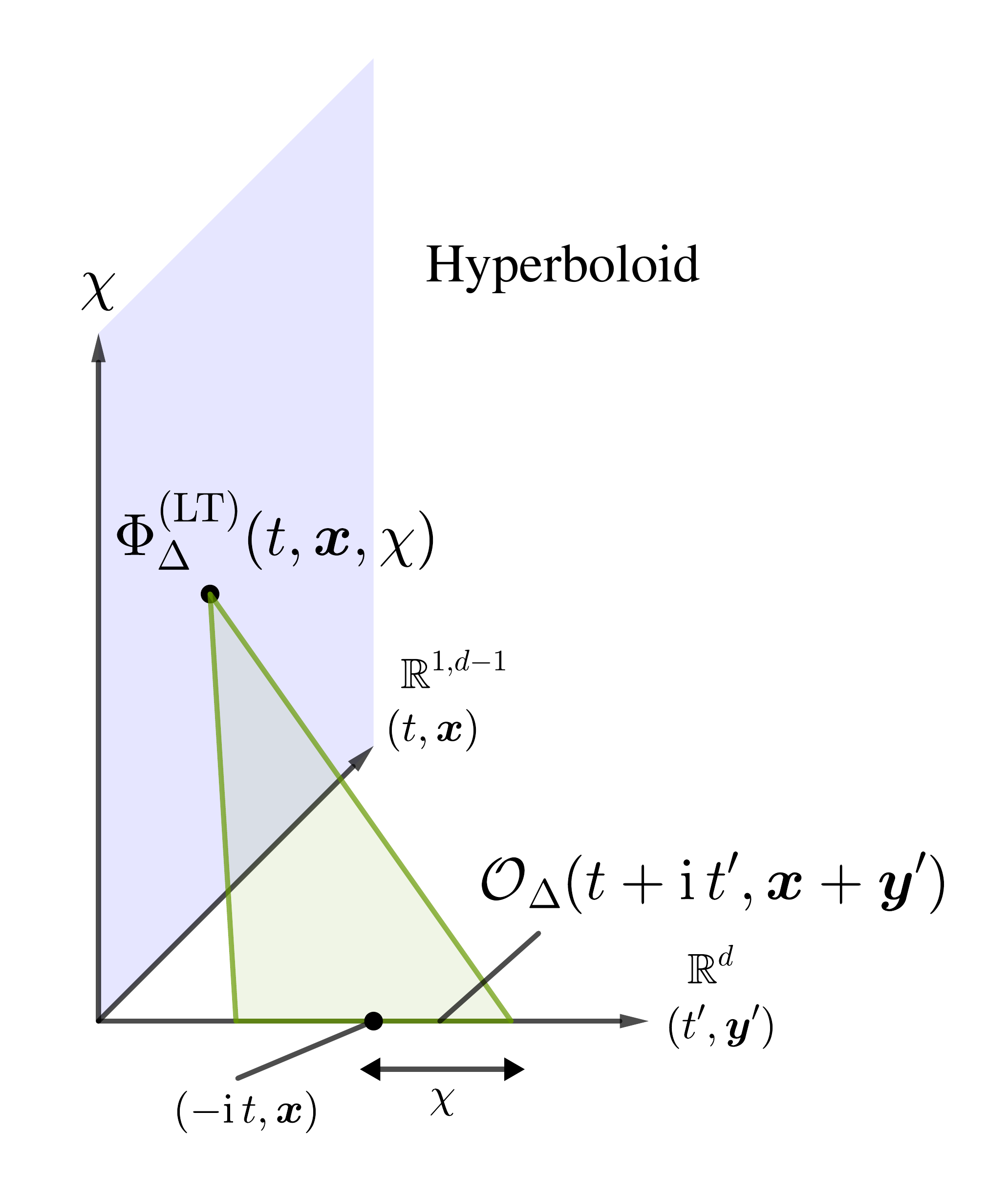}
    \caption{The HKLL-type representation of the scalar field propagating on the hyperboloid \eqref{deSitter_hyperboloid}.
    The bulk scalar field given by \eqref{Hyperboloid_scalar} is constructed by smearing a scalar primary $\CO_\Delta$ on the Euclidean space with the bulk to boundary propagator analogous to the HKLL kernel.
    }
    \label{fig:HKLL_time-like}
\end{figure}

With this in mind, we interpret the second line of \eqref{time-like_OPE_block} as a bulk scalar field constructed in a similar manner to the HKLL formula, but with support on the hyperboloid  \eqref{deSitter_embedding} instead:
\begin{align}\label{Hyperboloid_scalar}
    \Phi^{(\text{LT})}_\Delta\left( t, \bm{x}, \chi \right) \equiv \int_{t'^2 + \bm{y}'^2 \le \chi^2} \d t'\d^{d-1} \bm{y}' \left( \frac{\chi}{\chi^2 -t'^2 - \bm{y}'^2} \right)^{\bar\Delta}\, \CO_\Delta \left(t +\i\,t',  \bm{x} + \bm{y}' \right) \ .
\end{align}
Here the HKLL kernel is replaced with the bulk to boundary propagator in the hyperboloid from the Euclidean region to the Lorentzian bulk point $(x^\mu(u), \chi (u))$.
This means that we prepare the state $|\,\CO_\Delta (y)\,\rangle$ associated with the CFT operator $\CO_\Delta (y)$ using the Euclidean path integral and analytically continue it to the Lorentzian space-time, letting it propagate to the bulk point $(x^\mu(u), \chi(u))$ with the propagator given in the integration kernel above as in Fig.\,\ref{fig:HKLL_time-like}.
The time-like OPE block \eqref{time-like_OPE_block} is obtained by smearing the constructed bulk field over the geodesic in the hyperboloid \eqref{deSitter_embedding}.

\subsection{Dual defect picture and surface Witten diagram in two dimensions}\label{ss:Surface_Witten}
 
The key to the derivation of the time-like OPE block in the previous subsection was the analytic continuation of the three-point function with respect to $x_{12}^2$ from space-like separation to time-like separation.
Here we derive another representation of the time-like OPE block by rewriting the three-point function using the duality between a pair of time-like separated points and a codimension-two spacial defect \cite{Gadde:2016fbj,Fukuda:2017cup}.
As a disclaimer, in this subsection we will not be careful about the irrelevant overall normalization constants, as we are only interested in the kinematical structure to compare with a proposed form of the time-like OPE block by \cite{Czech:2016xec,deBoer:2016pqk}.

Let us recall the three-point function of scalar primaries
\begin{align}
    \begin{aligned}
    \langle\, \CO_1 (P_1)\, \CO_2(P_2)\,\tilde\CO_{\bar\Delta}(P_3)\,\rangle
    &\propto \frac{1}{P_{12}^\frac{\Delta_{12}^+ - \bar\Delta}{2}\,P_{23}^\frac{\bar\Delta -\Delta_{12}^-}{2}\,P_{31}^\frac{\Delta_{12}^- + \bar\Delta}{2}} \\
    & = \frac{1}{P_{12}^\frac{\Delta_{12}^+}{2}} \left( \frac{P_{23}}{P_{31}} \right)^\frac{\Delta_{12}^-}{2} \, \left( \frac{P_{12}}{P_{23} P_{31}} \right)^\frac{\bar \Delta}{2} \ .
    \end{aligned}
\end{align}
We notice that the last factor in the above expression can be identified with the three-point function of $\tilde \CO_{\bar \Delta}(P_3)$ with  scalar primaries $\varphi(P_1)$ and $\varphi(P_2)$ of dimension zero:
\begin{align}
    \left( \frac{P_{12}}{P_{23} P_{31}} \right)^\frac{\bar \Delta}{2} = \langle\, \varphi(P_1)\, \varphi(P_2)\, \tilde\CO_{\bar \Delta}(P_3)\, \rangle \ .
\end{align}
This three-point function can be viewed as the one-point function of $\tilde \CO_{\bDel}(P_3)$ in the presence of a codimension-$d$ scalar conformal defect $\CD^{(d)}$ representing a pair of operators $\varphi(P_1)$, $\varphi(P_2)$.
Given this interpretation, we are able to dualize the codimension-$d$ defect to a codimension-two defect\footnote{In a $d$-dimensional CFT with a conformal defect $\CD^{(p)}$ of codimension-$p$, the duality holds between $\CD^{(p)}$ and $\CD^{(d+2-p)}$ \cite{Gadde:2016fbj,Fukuda:2017cup}.}
\begin{align}
    \CD^{(d)}\quad \longleftrightarrow \quad \CD^{(2)} \ .
\end{align}

\begin{figure}[t]
\centering
    \includegraphics[width=13cm]{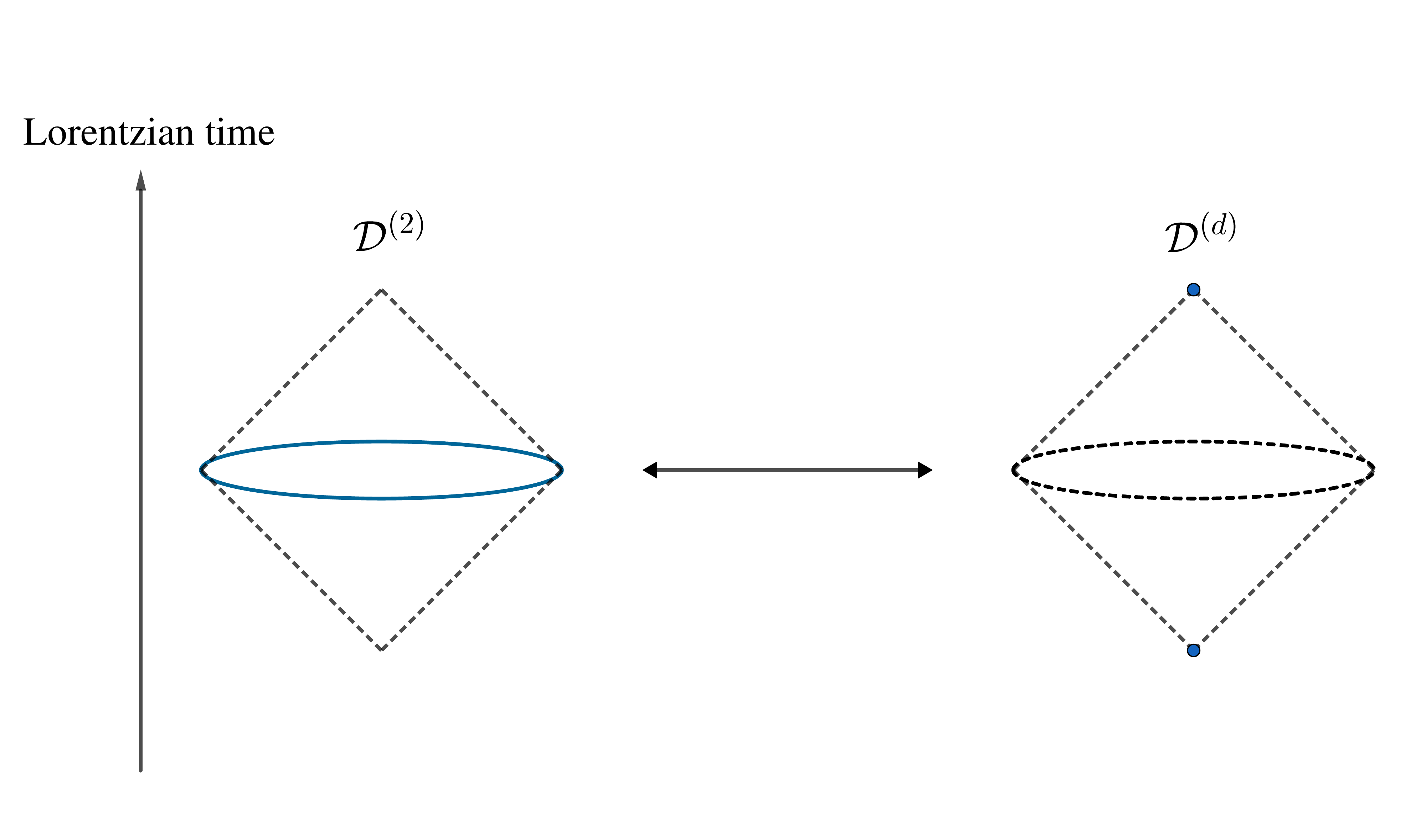}
    \caption{The duality between a codimension-two defect [Left] and a codimension-$d$ defect [Right].
    The codimension-$d$ defect is given by a pair of operators located at the tips of the causal diamond of the dual defect.}
    \label{fig:duality}
\end{figure}
More precisely, the kinematical part of a correlation function is invariant under the duality:
\begin{align}
    \langle\, \CD^{(d)}\, \tilde{\CO}_{\bar \Delta}(P_3)\, \rangle = \langle\, \CD^{(2)}\, \tilde{\CO}_{\bar \Delta}(P_3)\, \rangle \ .
\end{align}
When $x_{12}^2<0$, $\CD^{(d)}$ is time-like while the dual defect $\CD^{(2)}$ is a conformal defect supported on a space-like codimension-two surface $\Sigma_{12}$ that is uniquely determined by the pair of the time-like separated points $x_1$, $x_2$ as shown in  Fig.\,\ref{fig:duality}.
In the dual picture, the one-point function is given (up to a constant) by \cite{Gadde:2016fbj},
\begin{align}
    \langle\, \CD^{(2)}\, \tilde{\CO}_{\bar \Delta}(P_3)\, \rangle = \left( \frac{l_{\text{min}}\, l_{\text{max}}}{2r} \right)^{- \bar \Delta} \ , 
\end{align}
where $r$ is the radius of the spherical defect and $l_\text{min (max)}$ is the minimum (maximum) distance of $\tilde\CO_{\bar \Delta}(P_3)$ to the sphere.
If we put fictitious operators with vanishing dimension on the dual defect so that they are $l_\text{min (max)}$ separated from $P_3$, the one-point function in the presence of the defect can be regard as the three-point function of local operators,
\begin{align}
    \langle\, \CD^{(2)}\, \tilde{\CO}_{\bar \Delta}(P_3)\, \rangle =  \left( \frac{\tilde{P}_{12}}{\tilde{P}_{23} \tilde{P}_{31}} \right)^\frac{\bar \Delta}{2} \ , 
\end{align}
where $\tilde{P}_i$ are the coordinates of the fictitious operators on the space-like defect $\CD^{(2)}$. Notice that $\tilde{P_1}$ and $\tilde{P_2}$ depend on $P_3$ as well as $P_1$ and $P_2$.
With all this in mind, we can finally rewrite the three-point function with the dual coordinates,
\begin{align}\label{dual_3pt}
        \langle\, \CO_1 (P_1)\, \CO_2(P_2)\,\tilde\CO_{\bar\Delta}(P_3)\,\rangle  = \frac{1}{P_{12}^\frac{\Delta_{12}^+}{2}} \left( \frac{P_{23}}{P_{31}} \right)^\frac{\Delta_{12}^-}{2} \,  \left( \frac{\tilde{P}_{12}}{\tilde{P}_{23} \tilde{P}_{31}} \right)^\frac{\bar \Delta}{2} ,
\end{align}
up to some constants.
The main point is that in this expression $\tilde P_1$ and $\tilde P_2$ are space-like separated while $P_1$ and $P_2$ are time-like separated.

Given the duality one would expect that the time-like OPE block takes a similar form as the space-like one with $x_1, x_2$ replaced with the dual points $\tilde x_1, \tilde x_2$, but it  does not appear to work as $\tilde x_1$ and $\tilde x_2$ depend on $x_3$ in \eqref{dual_3pt} which makes the Fourier transform of the three-point function with respect to $x_3$ highly non-trivial.
It simplifies however in $d=2$ dimensions where the dual points are uniquely fixed by the original points $x_1$ and $x_2$ independent of $x_3$.
In this case, let us introduce the lightcone coordinates $x = (u,v)$ where they are related to the standard one by $u = (x^1 + x^0)/2$ and $v = (x^1 - x^0)/2$. 
When $x_1$ is located in the future lightcone of $x_2$, we can set $x_1 = (u_1,v_1)$ and $x_2 = (0,0)$ without loss of generality.
Since $\tilde{x}_1$ and $\tilde{x}_2$ are given by the intersection of the null rays of $x_1$ and $x_2$, one can choose $\tilde{x}_1 = (u_1,0)$ and $\tilde{x}_2 = (0,v_1)$.\footnote{The role of $\tilde{x}_1$ and $\tilde{x}_2$ can be swapped and the choice $\tilde{x}_1 = (0,v_1)$, $\tilde{x}_2 = (u_1,0)$ is also allowed.}
With this in mind, the OPE block becomes
\begin{align}\label{dual_time-like_OPE_block}
    \begin{aligned}
    \CB_{\Delta}^{(\rL)} (x_1, x_2 ) 
        &\propto 
        \frac{1}{|x_{12}^2|^\frac{\Delta_{12}^+}{2}}\, \int_{0}^1 \frac{\d u}{u(1-u)}\, \\
        &\quad \times \int_{t^2 + y^2 \le \tilde\eta (u)^2} \d t\,\d y \left( \frac{\tilde\eta (u)}{\tilde\eta (u)^2 -t^2 - y^2} \right)^{\tilde{\Delta} } \CO_\Delta \left(\tilde t(u) + t, \tilde{x}(u) + \i\,y \right) 
    \ ,
    \end{aligned}
\end{align}
where the ``holographic" coordinates $\tilde x^\mu(u)$ and $\tilde\eta (u)$ are
\begin{align}
    \tilde x^\mu (u) = u\, \tilde{x}_1^\mu (x_1, x_2 ) + (1-u)\, \tilde{x}_2^\mu (x_1, x_2) \ , \qquad \tilde\eta (u) = \sqrt{u(1-u)\, \tilde{x}^2_{12}} \ .
\end{align}
The second line can be interpreted as the HKLL field  $\Phi^{(\text{L})}_\Delta$ (defined by \eqref{Scalar_HKLL_field}) integrated over the bulk geodesic $\tilde\gamma_{12}$ between the space-like separated points $\tilde x_1$ and $\tilde x_2$:
\begin{align}
    \CB_{\Delta}^{(\rL)} (x_1, x_2 ) \propto
        \frac{1}{|x_{12}^2|^\frac{\Delta_{12}^+}{2}}\,\int_{\tilde\gamma_{12}}\d\lambda\, \Phi^{(\text{L})}_\Delta \left(\,\tilde t(\lambda),  \tilde x(\lambda) , \tilde \eta(\lambda)\,\right) \ ,
\end{align}
where $\lambda$ is the geodesic parameter related to $u$ by the relation \eqref{u_to_lambda}.

\begin{figure}[t]
    \centering
    \includegraphics[width=15cm]{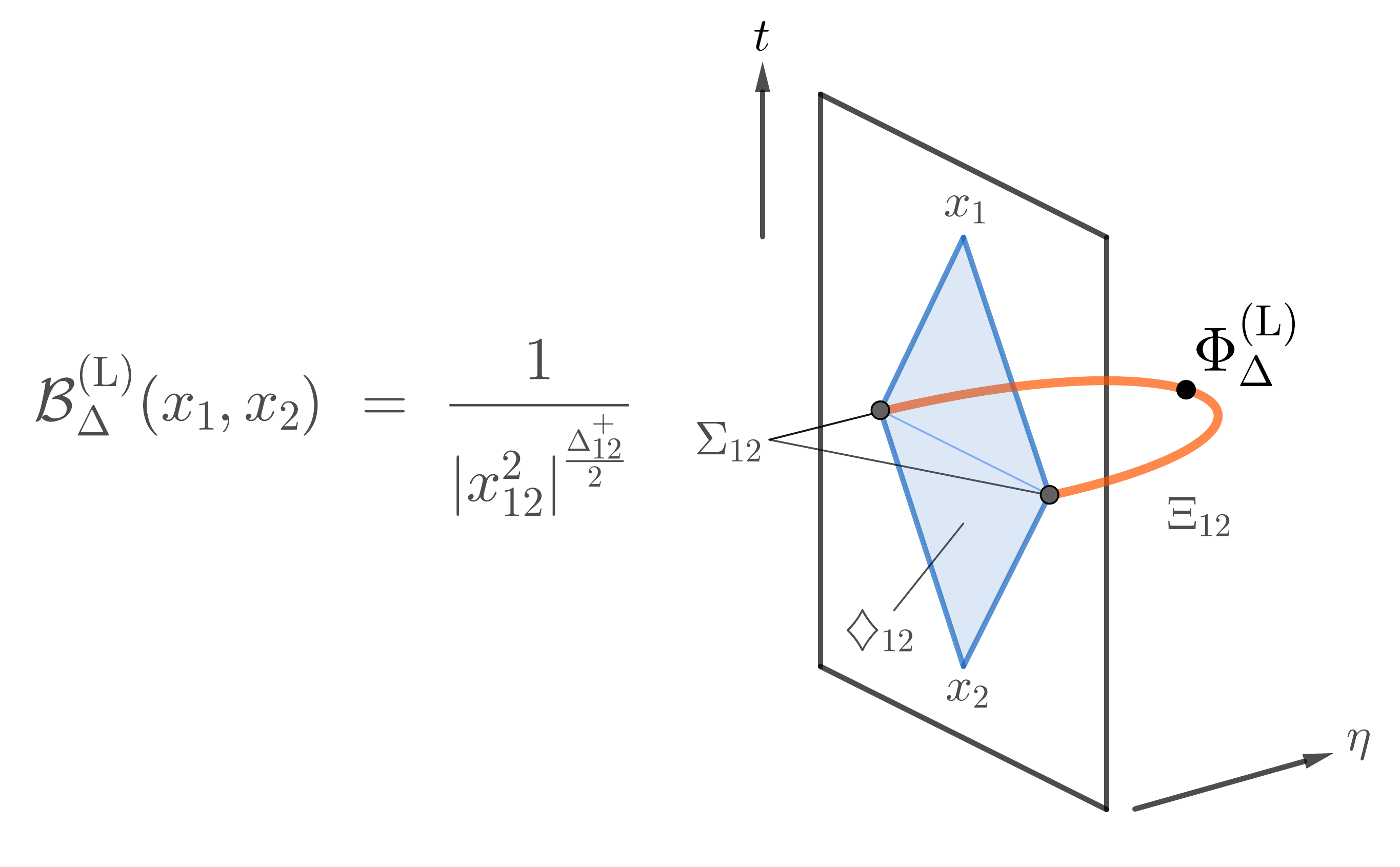}
    \caption{Time-like OPE block and surface Witten diagram.
    The blue shaded region is the causal diamond $\diamondsuit_{12}$ of the time-like separated points $x_1$ and $x_2$.
    The co-dimension-two surface $\Sigma_{12}$, which is the intersection of the past lightcone of $x_1$ and the future lightcone of $x_2$, is represented by two points in the figure.
    The bulk field is smeared over the codimension-two surface (the orange curve) in the bulk.
    }
    \label{fig:SurfaceWitten}
\end{figure}

It is worthwhile to compare the dual defect picture with the surface Witten diagram proposal \cite{Czech:2016xec,deBoer:2016pqk} as a holographic description of the time-like OPE block,
\begin{align}\label{Surface_Witten}
    \CB_{\Delta}^{(\rL)}(x_1, x_2) = \frac{1}{|x_{12}^2|^\frac{\Delta_{12}^+}{2}} \int_{\Xi_{12}}\d^{d-1}\xi\,\sqrt{h}\, \Phi^{(\text{L})}_\Delta\left( x^\mu(\xi), z(\xi)\right) \ ,
\end{align}
where $\Xi_{12}$ is the codimension-two surface bounding the dual defect $\Sigma_{12}$ on the boundary of the AdS space-time as shown in Fig.\,\ref{fig:SurfaceWitten}.
As a modest check one can show the two-point function of the surface Witten diagram and a scalar primary operator reproduces the correct structure of the three-point function (see appendix \ref{app:SurfaceWitten}). 
In two dimensions, the bulk surface $\Sigma_{12}$ becomes a one-dimensional curve that coincides with the bulk geodesic $\tilde\gamma_{12}$ between the dual points of the time-like separated points $x_1$ and $x_2$.
Hence the surface Witten diagram precisely agrees with our Lorentzian OPE block when $d=2$.

In higher dimensions, the surface Witten diagram associates the time-like OPE block to the bulk surface $\Xi_{12}$ anchored on the dual defect $\Sigma_{12}$ as in \eqref{Surface_Witten}, so one may still be tempted to extend the present method based on the dual defect picture to the higher-dimensional case.
Naively this would work if the surface Witten diagram reduced to the geodesic Witten diagram connecting the dual coordinates $\tilde x_1$ and $\tilde x_2$ on $\Sigma_{12}$ by regarding the surface as a collection of geodesics with different angles and integrating over the angular direction.
While we were not able to show it, it would be worthwhile to prove the equivalence between the surface Witten diagram and the time-like OPE block based on the approach outlined in this section.

\subsection{Lorentzian OPE block in position space}\label{ss:OPE_block_position}
 
We have derived the Lorentzian OPE block in the momentum space shadow formalism.
Comparing with the Euclidean case, the derivation in the Lorentzian case was more involved.
One may suspect the space-time derivation is much simpler than the momentum counterpart even in Lorentzian case, but it has to be done with extra care as we will show below.

Let us analytically continue the Euclidean OPE block \eqref{OPE_Block_formal} just by using $\i\epsilon$-prescription
\begin{align}
    \CB_{\Delta, J}^{(\rL)} (x_1, x_2) \underset{?}{=}
    \int [\d^d y]_{\rL} \,F_{\Delta_1, \Delta_2, [\bar \Delta, J]}(x_1, x_2, y; d_z)\, \CO_{\Delta,J}(y,z) \ ,
\end{align}
where $F_{\Delta_1, \Delta_2, [\bar \Delta, J]}(x_1, x_2, y; d_z)$ is the time-ordered three-point function and {the integration is performed over the entire space-time}.
This naive analytic continuation, however, does not agree with the correct result \eqref{Lorentzian_OPE_block}.
This is because the time-ordered correlator includes the shadow contribution proportional to the Wightman function $W_{\Delta_1, \Delta_2, [\bar \Delta, J]}$ 
\cite{Ferrara:1972ay} in addition to the direct one as seen from the decomposition
\begin{align}\label{F_W_relation}
    F_{\Delta_1, \Delta_2, [\bar \Delta, J]}(x_1, x_2, y; d_z) = c_1\,W_{\Delta_1, \Delta_2, [\bar \Delta, J]}(x_1, x_2, y; d_z) + c_2\,W_{\Delta_1, \Delta_2, [\Delta, J]}(x_1, x_2, y, d_{z'})\,W_{\bar \Delta, J}(y, z', d_z) \ ,
\end{align}
with some constants $c_1$ and $c_2$.\footnote{This expression follows from the analytic continuation of \eqref{Splitting_rep_of_3-pt} with \eqref{Wightman-3pt}.}
Hence subtracting the shadow contribution amounts to the Lorentzian OPE block in position space:
\begin{align}\label{Lorentzian_OPE_block_position}
    \CB_{\Delta, J}^{(\rL)} (x_1, x_2) =
    \int [\d^d y]_{\rL} \,W_{\Delta_1, \Delta_2, [\Delta, J]}(x_1, x_2, y; d_z)\, \tilde \CO_{\bar \Delta,J}(y,z) \ ,
\end{align}
which is Fourier transform of the momentum representation \eqref{Lorentzian_OPE_block}.\footnote{We suppress the overall constant for simplicity.}

Now let us compare our finding \eqref{Lorentzian_OPE_block_position} with the time-like OPE block proposed by \cite{Czech:2016xec,deBoer:2016pqk},
\begin{align}\label{time-like_OPE_block_proposal}
    \CB_{\Delta, J}^{(\rL)} (x_1, x_2) = \int_{\diamondsuit_{12}} [\d^d y]_{\rL} \, \langle\, \CO_1 (x_1)\, \CO_2(x_2)\,\tilde\CO_{\bar\Delta, J}(y, d_z)\,\rangle\, \CO_{\Delta, J}(y, z) \ ,
\end{align}
where the integration range $\diamondsuit_{12}$ is the causal diamond with the tips at $x_1$ and $x_2$.
The proposed expression \eqref{time-like_OPE_block_proposal} looks almost the same as ours \eqref{Lorentzian_OPE_block_position}, but they have different integration ranges.
Also it is not clear in the proposal which type of the three-point function should be used in \eqref{time-like_OPE_block_proposal}.
The problem of the integration range may be accounted for by using the conformal transformation to place the point $y$ inside the causal diamond $\diamondsuit_{12}$.
Under this map, the three-point functions in different causal regions takes the same form up to phase.
For instance, the time-ordered correlator is
\begin{align}
    F_{\Delta_1, \Delta_2, [\bar \Delta, J]}(x_1, x_2, y; d_z) = e^{\i\, \theta_F(x_1, x_2, y)}\,\langle\, \CO_1 (x_1)\, \CO_2(x_2)\,\tilde\CO_{\bar\Delta, J}(y, d_z)\,\rangle_{\rn} \ ,
\end{align}
where we define the kinematical three-point function by the analytic continuation of the Euclidean correlator without $\i\,\epsilon$-prescription,
\begin{align}\label{kinematical_3pt}
    \langle\, \CO_1 (x_1)\, \CO_2(x_2)\,\tilde\CO_{\bar\Delta, J}(y, d_z)\,\rangle_{\rn} \equiv E_{\Delta_1, \Delta_2, [\bar \Delta, J]}(x_1, x_2, y; d_z)|_{x_i^d \to \i\,x^0} \ ,
\end{align}
and the phase $\theta_F$ is constant taking different values depending on the causal relation between the three points $x_1, x_2$ and $y$.
The Wightman correlator also takes the same form with the phase $\theta_W$.
Hence splitting \eqref{Lorentzian_OPE_block_position} into the sum of integrals over different causal regions, rewriting with the time-ordered and Wightman correlators using the relation \eqref{F_W_relation} and mapping all the causal regions to $\diamondsuit_{12}$, the Lorentzian OPE block becomes
\begin{align}
    \CB_{\Delta, J}^{(\rL)} (x_1, x_2) = c\,\int_{\diamondsuit_{12}} [\d^d y]_{\rL} \, \langle\, \CO_1 (x_1)\, \CO_2(x_2)\,\tilde\CO_{\bar\Delta, J}(y, d_z)\,\rangle_{\rn}\, \CO_{\Delta, J}(y, z) \ ,
\end{align}
where $c$ is the sum of the phase factors associated with different casual regions.
So our Lorentzian OPE block \eqref{Lorentzian_OPE_block_position} is equivalent to the proposal by \cite{Czech:2016xec,deBoer:2016pqk} in the time-like case if the three-point function is understood as the kinematical one \eqref{kinematical_3pt}.

\section{Discussion}\label{sec:Discussion}

In this paper, we showed that the OPE block of two scalar primary operators in $d$-dimensional Lorentzian CFT's has a natural holographic description in $(d+1)$-dimensional space-time.
Depending on the causal relation of the two points, space-like or time-like, we obtained distinctly different pictures which would not have been realized simply by examining the Euclidean case and performing direct analytic continuation.
The space-like OPE block is given by integrating a higher spin field on the geodesic connecting the two CFT operators inside the AdS space-time.
This result extends the previous studies \cite{daCunha:2016crm,Das:2018ajg} to the most general case, and can be seen as half of the geodesic Witten diagram \cite{Hijano:2015zsa,Dyer:2017zef,Sleight:2017fpc}.
The time-like OPE block has a similar holographic description, but the dual gravity is not on the AdS space-time but on the hyperboloid with two time coordinates.

In contrast to the Euclidean case this new structure has some implications and raises additional subtleties for constructing the holographic description of conformal block in Lorentzian CFT's.
To see this, let us use the shorthand notations $x_1 \approx x_2$ when $x_1$ and $x_2$ are space-like separated and $x_1 > x_2$ when $x_1$ is in the future lightcone of $x_2$, and
consider three types of conformal block depending on the causal relations of four points:
\begin{itemize}
    \item When $x_1 \approx x_2$ and $x_3 \approx x_4$ the conformal block is given by the correlator of two space-like OPE blocks, hence it is described by the geodesic Witten diagram in the Lorentzian AdS space-time.
    The propagator between the two geodesics is of Wightman type.
    \item When $x_1 > x_2$ and $x_3 > x_4$ the conformal block is given by the correlator of two time-like OPE blocks, hence it is described by the analogue of the geodesic Witten diagram in the hyperboloid.
    \item When $x_1 \approx x_2$ and $x_3 > x_4$ the conformal block is given by the correlator of space-like and time-like OPE blocks.
    In this case, each block is holographically realized in a different space-time, so it is not possible to describe the conformal block in the same way as the geodesic Witten diagram.
\end{itemize}
In the third case, it is more natural to use the surface Witten diagram for the time-like OPE block as suggested in \cite{Czech:2016xec}, then the conformal block is given by a bulk to bulk propagator anchored on the codimension-two surface $\Xi_{12}$ and the geodesic $\gamma_{34}$ in the Lorentzian AdS space-time.
Only in two dimensions we were able to derive the surface Witten diagram from the time-like OPE block, and the problem of deriving the surface Witten diagram still remains open in higher dimensions.

Starting from the space-like OPE block \eqref{Spinning_OPEB_momentum} we arrived at the holographic description \eqref{space-like_OPEB_HS} as the HKLL-type spinning field \eqref{Lorentzian_higher_spin_HKLL} integrated over the geodesic $\gamma_{12}$ in AdS.
In the scalar case \eqref{Scalar_HKLL_field} the expression is known as the Poincar\'e smearing of an AdS scalar field where the kernel has support on the entire Poincar\'e boundary \cite{Hamilton:2006az}.
On the other hand there is an alternative construction of the same field in the AdS-Rindler coordinates
known as the AdS-Rindler smearing  \cite{Hamilton:2006az} where the boundary operator is smeared in the (complexified) boundary region space-like to the bulk point $X$.
This expression is more illuminating than the global one as it allows us to reconstruct a bulk field in the AdS causal wedge anchored on a causal diamond by smearing the CFT operators only on the causal diamond \cite{Morrison:2014jha}.
The region associated with a pair of time-like separated points $x_1$ and $x_2$ is the causal diamond $\diamondsuit_{12}$, which can be mapped to $\BR\times \BH^{d-1}$ by a conformal map \cite{Casini:2011kv}.
Since $\BR\times \BH^{d-1}$ is the conformal boundary of the AdS-Rindler coordinates it is reasonable to expect the AdS-Rindler smearing appears in the time-like OPE block when represented in the $\BR\times \BH^{d-1}$ coordinates.
We hope to investigate this interesting problem elsewhere.

The bulk field $\Phi^{(\rL)}_{\Delta, J}$ we defined by \eqref{Lorentzian_higher_spin_HKLL_diff} supposedly describes a massive higher spin in the AdS space-time as its Euclidean counterpart $\Phi^{(\rE)}_{\Delta, J}$ given by \eqref{Diff-HKLL-Spinning} and its equivalent form \eqref{Euclidean_higher_spin_HKLL_in_Embedding} clearly satisfies the equation of motion.
Indeed $\Phi^{(\rL)}_{\Delta, J}$ correctly reduces to the massless higher spin field \eqref{Sarkar-Xiao_field} when the dual operator is conserved as shown in section \ref{ss:conserved}.
In the spin-$1$ case, we found a discrepancy between our derived bulk field and the HKLL representation of the massive spin-$1$ field existing in the literature \cite{Kabat:2012hp}, as is mentioned below \eqref{Spin1_OPE_Block}.
While we do not have a complete resolution of this discrepancy, we suggest two possibilities to reconcile them.
First one is that since the bulk field appears in the OPE block in the integrated form the difference between the two fields vanishes when integrated over the geodesic.
Second one is that the difference originates from the ``gauge" choice in deriving the HKLL representation as there is a freedom for the choice of the integral kernel in constructing the same field \cite{Hamilton:2005ju,Kabat:2012hp}.
It is desirable to resolve this discrepancy completely in the future.

It is worthwhile emphasizing that the Euclidean OPE block \eqref{Spinning-OPE block} (or its differential representation \eqref{OPE-block-Diff}) has the same structure as the Lorentzian block \eqref{space-like_OPEB_HS} with the common coefficient $\CN_{12, [\Delta, J]}$.
It means that the location of the physical poles in the complex $\Delta$ plane is the same as it is determined by $\CN_{12, [\Delta, J]}$, so their pole structures also agree while the Euclidean block includes the shadow poles in addition to the physical poles.

Finally let us end with a couple of future research directions.
\begin{itemize}
    \item In this work, we have specifically constructed the holographic description of the OPE block by considering the shadow projector. However in the Lorentzian CFT's, it is known that we can also have spin-shadow and light ray transformations \cite{Isachenkov:2017qgn, Kravchuk:2018htv}, such that the full projector should include these transformations as they are also part of the $D_8$ Weyl group of the Lorentzian conformal symmetry group $\SO(2, d)$. It would be very interesting to generalize our construction to consider the OPE blocks involving these non-local exchange operators, and how causality conditions can modify the definitions of HKLL-type representation of their bulk dual fields. These additional generalizations will naturally lead us to constructing the holographic dual configurations for various Lorentzian conformal blocks, extending the relatively straightforward Euclidean construction.
    \item The bulk field we defined in \eqref{Euclidean_higher_spin_HKLL_in_Embedding} clearly satisfies the equation of motion of a massive higher spin field in the Euclidean AdS space.
    When $J=0$ and $\Delta_1 = \Delta_2 = 0$ \eqref{Spinning-OPE block} can be seen as the Radon transform that intertwines between the bulk scalar equation of motion and the Casimir equation of the scalar OPE block \cite{Czech:2016xec} (see also \cite{Das:2017wae,Raben:2018rbn}).
    Based on this observation we speculate that \eqref{Spinning-OPE block} is a variant of the Radon transform between functions on $\BH^{d+1}$ and on the moduli space of a pair of points, and
    the equation of motion follows from the quadratic Casimir equation for the OPE block.
    It would also be intriguing to see if \eqref{space-like_OPEB_HS} can be regarded as a Lorentzian analogues of the Radon transform and a similar story holds.
\end{itemize}

\acknowledgments
We would like to thank Chung-I Tan and Volker Schomerus for useful discussions.
A part of this work has been done during East Asia Joint Workshop on Fields and Strings 2019 and 12th Taiwan String Theory Workshop at National Center for Theoretical Sciences in Taiwan.
The work of H.\,Y.\,C. was supported in part by Ministry of Science and Technology (MOST) through the grant
107 -2112-M-002-008-.
The work of N.\,K. was supported in part by the Program for Leading Graduate Schools, MEXT, Japan and also supported by World Premier International Research Center Initiative (WPI Initiative), MEXT, Japan.
The work of T.\,N. was supported in part by the JSPS Grant-in-Aid for Scientific Research (C) No.19K03863 and the JSPS Grant-in-Aid for Scientific Research (A) No.16H02182.

\appendix
\section{Miscellaneous}\label{sec:Miscellaneous}
\subsection{Useful formulas}
Here we collect some of the useful mathematical formulae and identities used in the main text.

\paragraph{Schwinger parametrization}
\begin{align}\label{Schwinger_parametrization}
	\frac{1}{x^\Delta} = \frac{1}{\Gamma (\Delta)}\, \int_0^\infty\, \frac{\d t}{t}\,t^\Delta\, e^{-t x} \ .
\end{align}
Alternatively, \cite{Smirnov:2006ry}
\begin{align}
	\frac{1}{(x - \i\,0)^\Delta} = \frac{\i^\Delta}{\Gamma (\Delta)}\, \int_0^\infty\, \frac{\d t}{t}\,t^\Delta\, e^{-\i\,t x} \ .
\end{align}

\paragraph{A useful identity}
Here we will derive the following identity which is used for deriving the momentum space representation of the three-point function:
\begin{align}\label{useful_formula}
    \begin{aligned}
    \int [\d^d x_3]_\rE \, &e^{\i\, p\cdot x}\, (x_{13}^2)^{-\delta_1} (x_{23}^2)^{-\delta_2} 
        = \frac{2 \pi^{h}}{\Gamma(\delta_1)\, \Gamma(\delta_2)} \left(\frac{p^2}{4x_{12}^2} \right)^{\frac{\delta_1 + \delta_2 - h}{2}}\\
        &\times \int_0^1 \d u \, u^{\frac{\delta_1 - \delta_2 + h}{2}-1} (1-u)^{\frac{\delta_2 - \delta_1 + h}{2}-1}  e^{\i\,p\cdot \left( u x_1 + (1-u)  x_2\right)} \, K_{\delta_1 + \delta_2 - h}\left(\sqrt{u(1-u)\,p^2\, x_{12}^2}\right) \ .
     \end{aligned}
\end{align}
To show it we adopt the Schwinger parametrization
\begin{align}
    \int [\d^d x_3]_\rE \, e^{\i\, p\cdot x}\, (x_{13}^2)^{-\delta_1} (x_{23}^2)^{-\delta_2} = \frac{1}{\Gamma(\delta_1)\, \Gamma(\delta_2)} \int_0^\infty \frac{\d \alpha}{\alpha} \int_0^\infty \frac{ \d \beta}{\beta} \alpha^{\delta_1} \beta^{\delta_2} \, \int \d^d x_3\, e^{-\alpha x_{13}^2 - \beta x_{23}^2 + \i\, p\cdot x_3} \ ,
\end{align}
Using the translation invariance, the $x_3$-integral can be done as
\begin{align}
 \int [\d^d x_3]_\rE\, e^{-\alpha x_{13}^2 - \beta x_{23}^2 + \i\, p\cdot x_3} 
 &= \ \int [\d^d y]_\rE \,  e^{-\alpha y^2 - \beta (x_{12} +y)^2 + \i\, p\cdot (y + x_1)}\nn \\
 &= e^{\i\, p\cdot x_1- \beta x_{12}^2}\, \int [\d^d y]_\rE \, e^{-(\alpha + \beta) y^2 - (2 \beta x_{12} - \i\, p )\cdot y}\nn \\
 &= \frac{\pi^{h} e^{\i\, p\cdot x_1- \beta x_{12}^2}}{ (\alpha + \beta)^{h}} \, e^{\frac{(2 \beta x_{12} - \i\, p )^2}{4 (\alpha + \beta)}}\ .
\end{align}
Then introducing the change of variables $\alpha = y u$, $\beta = y (1-u)$ we obtain the expression we wanted to derive:
\begin{align}
    &\frac{\pi^{h} e^{\i\, p\cdot x_1} }{\Gamma(\delta_1)\, \Gamma(\delta_2)} \int_0^1 \d u \, \int_0^\infty \d y \, u^{\delta_1 -1} (1-u)^{\delta_2 - 1} y^{\delta_1 + \delta_2 - h - 1} \, e^{- y u (1-u)x_{12}^2 - \frac{p^2}{4 y} - \i\, (1-u)p\cdot x_{12}} \nonumber \\
    &= \frac{2 \pi^{h}}{\Gamma(\delta_1)\, \Gamma(\delta_2)} \left(\frac{p^2}{4x_{12}^2} \right)^{\frac{\delta_1 + \delta_2 - h}{2}} \int_0^1 \d u \, u^{\frac{\delta_1 - \delta_2 + h}{2}-1} (1-u)^{\frac{\delta_2 - \delta_1 + h}{2}-1}  e^{\i\, p\cdot \left( u x_1 + (1-u)  x_2\right)} \nonumber \\
     &\times \, K_{\delta_1 + \delta_2 - h}\left(\sqrt{u(1-u)\,p^2\, x_{12}^2}\right) \ ,
\end{align}
where we used the integral formula:
\begin{align}
    \int_0^\infty \d x \, x^{\nu-1} e^{-\frac{\beta}{x} - \gamma x} = 2 \left(\frac{\beta}{\gamma}\right)^{\nu/2} K_{-\nu} (2 \sqrt{\beta \gamma}) \ .
\end{align}

\subsection{Embedding formalism in $\BR^{1,d+1}$ and $\BR^{2,d}$}\label{app:Embedding}
In this appendix, we give the necessary details about the embedding space for both Euclidean $\BR^{1,d+1}$ and Lorentzian $\BR^{2,d}$, following \cite{Kravchuk:2018htv}, which are relevant for the discussions of OPE blocks discussed in the main text.
 
For the Euclidean embedding space $\BR^{1,d+1}$, the so-called Poincar\'{e} section relating it coordinates and the ones for the physical space $\BR^d$ and $(d+1)$-dimensional Euclidean AdS space is well-known \cite{Costa:2011dw, Costa:2014kfa}:
\begin{align}\label{EmbeddingCoorPola}
    \begin{aligned}
    (P^+, P^-, P^i) &= (1,x^2, x^i)\ ,\qquad &
    (X^+, X^-, X^i) &= \frac{1}{\eta}(1, y^2+\eta^2, y^i)\ , \\
    (Z^+, Z^-, Z^i) &= (0,2x\cdot z, z^i)\ , \quad &
    (W^+, W^-, W^i) &= (0, 2y\cdot w, w^i)\ ,\quad i=1, \dots, d\ .
    \end{aligned}
\end{align}
where $\eta$ is the radial coordinate in AdS-Poincar\'{e} metric \cite{Costa:2011dw,Costa:2014kfa}.
Here we have also introduced both boundary and bulk polarization vectors which satisfy the transverse conditions
\begin{align}
    P^A Z_A = X^A W_A = 0 \ .
\end{align}

For the Lorentzian embedding space $\BR^{2,d}$ we can do similar, let us denote its coordinates as $P^{-1}, P^{0}, \dots, P^{d}$ with the metric:
\begin{align}\label{Rd2-metric}
P^2 = -(P^{-1})^2-(P^{0})^2 + (P^{1})^2 + \dots + (P^{d})^2\ ,
\end{align}
they transform as the vector representation under $\SO(2,d)$ group.
Similar to the embedding formalism for Euclidean space, we can consider the null cone in $\BR^{2,d}$:
\begin{align}\label{Rd2-null}
(P^{-1})^2+(P^{0})^2 = (P^{1})^2 + \dots + (P^{d})^2. 
\end{align}
We can embed $d$-dimensional Minkowski space $\CM_d$ in the null cone \eqref{Rd2-null} by introducing the lightcone coordinates:
\begin{align}\label{LightCone-Lorentz}
P^{\pm} = P^{-1} \pm P^{d}\ , 
\end{align}
and restricting along the $P^+ \neq 0$ locus. This allows us to use $\BR$-rescaling to set $P^+$ to 1, i.\,e., choosing the Lorentzian analogue of Poincar\'{e} patch.
We can now further identify the Minkowski coordinates to be $P^{\mu} = x^\mu$, $\mu =0, 1, \dots, d-1$, and have the Lorentzian analogue of Poincar\'{e} section:
\begin{align}\label{Def:Canonical}
(P^+, P^-, P^{\mu}) = (1, x^2, x^\mu)\ , \quad (Z^+, Z^-, Z^\mu) = (0, 2z^2, z^\mu )
\end{align}
where $x^2 = \eta_{\mu\nu} x^\mu x^\nu$ and $z^2= \eta_{\mu\nu} z^\mu z^\nu$ evaluated with respect to the Minkowski metric $\eta_{\mu\nu}$.
Similarly we can also consider embedding $(d+1)$-dimensional Lorentzian Anti-de Sitter space-time  in $\BR^{2, d}$ by the following hyperboloid:
\begin{align}\label{Embed-Lorentz-AdS}
 -X^+ X^- - (X^{0})^2  +(X^{1})^2 + \dots + (X^{d-1})^2 = -1\ .
\end{align}
where we have again defined the lightcone coordinates $X^\pm = X^{-1}\pm X^d$ as in \eqref{LightCone-Lorentz}, and make the following identifications:
\begin{align}\label{Def:Lorentzian AdS}
 (X^+, X^-, X^{\mu}) = \frac{1}{\eta}(1, y^2+\eta^2, y^\mu) \ ,
\end{align}
which satisfy \eqref{Embed-Lorentz-AdS}.
Finally, we can also consider embedding the following hypersurface:
\begin{align}\label{deSitter_embedding1}
-Y^+Y^- - (Y^{0})^2 + (Y^{1})^2 + \dots + (Y^{d-1})^2  = +1 \ ,
\end{align}
which can be satisfied by \eqref{deSitter_embedding_coord}, and it can be related to de Sitter space-time by Wick rotating the Lorentzian time coordinate $x^0$.

\section{Wightman and time-ordered functions} \label{Appendix:Wightman_and_Feynman}
 
In this appendix, we summarize our conventions for Wightman and time-ordered correlation functions obtained from the Euclidean correlation function by the analytic continuation.

\subsection{Scalar primary}
 
Let us start with the Euclidean two-point correlation function of a scalar primary operator $\CO_\Delta(x)$ is fixed by conformal symmetry to be
\begin{align}
     \langle\, \CO_\Del(x)\,\CO_\Del(0)\,\rangle_{\rE} = \frac{1}{(x^2)^\Delta}  = E_\Delta (x) \ ,
\end{align}
where $x = (x^1, \cdots, x^d)$ and we have introduced the shorthand notation $E_{\Del}(x)$ to denote Euclidean correlation function.
The Fourier transform of the two-point function is given by:
\begin{align}\label{Euclid_Fourier}
    E_\Delta(x) = \alpha_{\Del, 0} \, \int \frac{[\d^d p]_\rE}{(2\pi)^d}\,e^{\i\,p\cdot x}\, (p^2)^{\Delta - h}  \ ,
\end{align}
with $\alpha_{\Delta, J}$ defined by \eqref{Def:alpha}.

\paragraph{Wightman correlator}
The analytic continuation 
\begin{align}\label{Analytic-cont}
    x^d = \i\, x^0 + \epsilon \ , \qquad \epsilon >0 
\end{align}
in $E_\Delta(x)$ yields the Wightman two-point function:
\begin{align}
    W_\Delta (x) &\equiv \langle 0 |\, \CO(x)\,\CO(0)\,|0\rangle = \frac{1}{\left[-(x^0 - \i\,\epsilon)^2 + ({\bf x})^2\right]^\Delta} \ ,
\end{align}
for the Lorentzian coordinates $x^\mu = (x^0, {\bf x}^i),~(i=1,\cdots, d-1)$.
The Fourier transform is derivable from \eqref{Euclid_Fourier} as follows:
\begin{align}
    \begin{aligned}
        W_\Delta (x) 
            &= \alpha_{\Del, 0} \, \int \frac{\d^{d-1} {\bf p}}{(2\pi)^{d-1}}\,e^{\i {\bf p}\cdot {\bf x}}\, \int_{-\infty}^\infty\frac{\d p^d}{2\pi}\,e^{-p^d (x^0 - \i\,\epsilon)}\,\left( {\bf p}^2 + (p^d)^2\right)^{\Delta - h}\\
             &= \alpha_{\Del, 0} \, \int \frac{\d^{d-1} {\bf p}}{(2\pi)^{d-1}}\,e^{\i {\bf p}\cdot {\bf x}}\, \i\,\int_{0}^\infty\frac{\d p^0}{2\pi}\,e^{-\i\,p^0 x^0 - \epsilon p^0}\,\left[ \left( {\bf p}^2 - (p^0 - \i\,\delta)^2\right)^{\Delta - h} - \left( {\bf p}^2 - (p^0 + \i\,\delta)^2\right)^{\Delta - h}\right] \ ,
    \end{aligned}
\end{align}
where we defined $p^d = \i\,p^0$ and introduced a small cutoff $\delta >0$ to deform the contour of integration for $p^d$ from the real axis to the imaginary axis. Using a formula 
\begin{align}
    (x + \i \,\delta)^a - (x - \i \,\delta)^a = 2\,\i\,\sin( \pi a) \,\Theta(-x)\,(-x)^a \ ,
\end{align}
for the discontinuity in the integrand, we obtain 
\begin{align}\label{Wightman_2pt_position}
    W_\Delta (x) &=  \frac{\pi^{h +1}}{2^{2\Delta - d - 1}\Gamma(\Delta)\,\Gamma(\Delta - h +1)}\, \int \frac{[\d^d p]_\rL}{(2\pi)^d}\,e^{\i\,p\cdot x}\,\Theta(p^0)\,\Theta(-p^2)\,(-p^2)^{\Delta - h} \ ,
\end{align}
which is the momentum space two point function we used in the main text.

\paragraph{Time-ordered correlation function}
Here we also consider the time-ordered two point correlation function constructed from the Wightman function:
\begin{align}
    \begin{aligned}
        F_\Delta (x) &\equiv \langle 0 |\, {\rm T}\left[\CO(x)\,\CO(0)\right]\,|0\rangle \\
            &= \Theta(x^0)\,W_\Delta(x) + \Theta(-x^0)\,W_\Delta(-x) \\
            &= \frac{1}{(x^2 +\i\,\epsilon)^\Delta}.
    \end{aligned}
\end{align}
It is obtained from the Euclidean correlation function by the Wick rotation instead (c.\,f.\, \eqref{Analytic-cont} ):
\begin{align}\label{Wick-Rot}
    x^d = (\i + \epsilon) x^0 \ ,
\end{align}
thus has the Fourier transform
\begin{align}
    F_\Delta (x) = -\i\, \alpha_{\Del, 0} \, \int \frac{[\d^d p]_\rL}{(2\pi)^d}\,e^{\i\,p\cdot x}\, (p^2-\i\,\epsilon)^{\Delta - h} \ .
\end{align}

\subsection{Spin-$J$ primaries}
 
In this appendix, we would like to perform similar analytic continuation to obtain Spinning Wightman function.
To write down the two-point function of a spin-$J$ primary operator we introduce the index-free notation
\begin{align}
    \CO_{\Delta, J}(x, z) \equiv \CO_{\Delta, \mu_1\cdots\mu_J}(x)\,z^{\mu_1}\cdots z^{\mu_J} \ ,
\end{align}
where $z^\mu$ is a null polarization vector $z^2 = 0$.
In the index-free notation the normalized Euclidean two-point function can be written as
\begin{align}
    \begin{aligned}
    E_{[\Delta, J]}(x; z_1, z_2) 
        &= \langle\,\CO_{\Delta,J}(x, z_1)\,\CO_{\Delta,J}(0, z_2)\,\rangle\\
        &= 
        \frac{[x^2(z_1\cdot z_2) - 2 (x\cdot z_1)(x\cdot z_2)]^J}{(x^2)^{\Delta+J}} \ .
    \end{aligned}
\end{align}
The Fourier transform becomes (see e.g.\,\cite{Isono:2018rrb})
\begin{align}
    \begin{aligned}
    E_{[\Delta, J]}&(p; z_1, z_2) \\
        &= \frac{\pi^J\,J!\,\Gamma(h-\Delta)}{2^{2\Delta - d - J}\Gamma(\Delta +J)}\, (p^2)^{\Delta -h}\,\left( \frac{(p\cdot z_1)(p\cdot z_2)}{-p^2} \right)^J\,P^{(\Delta - h - J,\, h-2)}_J \left( 1- \frac{p^2\,(z_1\cdot z_2)}{(p\cdot z_1)(p\cdot z_2)} \right) \\
        &= \alpha_{\Delta, J}\,(p^2)^{\Delta -h}\,\sum_{r=0}^J 2^r\,\binom{J}{r}\,\frac{(h-\Delta)_r}{(2-\Delta - J)_r}\,(z_1\cdot z_2)^{J-r}\,\left( \frac{(p\cdot z_1)(p\cdot z_2)}{-p^2} \right)^r \ ,
    \end{aligned}
\end{align}
with $\alpha_{\Delta, J}$ defined by \eqref{Def:alpha}.

The Wightman two-point function can be obtained by the analytic continuation as in the same way as the scalar case \eqref{Wightman_2pt_position} (see e.g.\,section 5.D in \cite{Dobrev:1977qv} for details) or solving the conformal Ward identity \cite{Gillioz:2016jnn}:
\begin{align}
    W_{[\Delta, J]}(p; z_1, z_2) =
    C_{\Delta, J}\,\Theta(p^0)\,\Theta(-p^2)\, (-p^2)^{\Delta -h}\,\sum_{r=0}^J 2^r\,\binom{J}{r}\,\frac{(h-\Delta)_r}{(2-\Delta - J)_r}\,(z_1\cdot z_2)^{J-r}\,\left( \frac{(p\cdot z_1)(p\cdot z_2)}{-p^2} \right)^r \ ,
\end{align}
where the normalization constant is given by:
\begin{align}\label{Wightman_constant}
    \begin{aligned}
    C_{\Delta, J} 
        &= 2\sin\left(\pi(\Delta-h)\right)\,\alpha_{\Delta, J} \\
        &= \frac{\pi^{h+1}}{2^{2\Delta - d -1}\,(\Delta + J -1)\,\Gamma(\Delta - 1)\,\Gamma(\Delta -h + 1)} \ .
    \end{aligned}
\end{align}
With this normalization, the Wightman function satisfies the identity:
\begin{align}\label{Wightman_inverse}
    W_{[\Delta, J]}(p; z_1, d_z)\,  W_{[\bar \Delta, J]}(p; z, z_2) = C_{\Delta, J}\,C_{\bDel, J}\,\Theta(p^0)\,\Theta (-p^2)\,(z_1\cdot d_z)^J\, (z\cdot z_2)^J \ , 
\end{align}
where $d_z$ is the Todorov operator
\begin{align}\label{Def:Todorov-op}
    (d_z)_\mu \equiv \left( h-1 + z\cdot \frac{\partial}{\partial z}\right) \frac{\partial}{\partial z^\mu} - \frac{1}{2}\,z_\mu\,\frac{\partial^2}{\partial z\cdot \partial z} \ .
\end{align}
Written in the physical space, it implies that $W_{[\bar\Delta,J]}$ is the inverse of $W_{[\Delta,J]}$ in the following sense:
\begin{align}
    W_{[\Delta, J]}^{\mu_1\cdots\mu_J, \rho_1\cdots \rho_J}(p)\,  W_{[\bar \Delta, J], \rho_1\cdots \rho_J, \nu_1\cdots\nu_J}(p) = C_{\Delta, J}\,C_{\bDel, J}\,\Theta(p^0)\,\Theta (-p^2)\,\left[ \delta_{(\nu_1}^{\mu_1}\cdots \delta_{\nu_J)}^{\mu_J} - \text{traces} \right] \ .
\end{align}
For completeness, here we also list the embedding space lift of the Todorov operator \eqref{Def:Todorov-op}:
\begin{align}\label{Def:DZ}
    D_{Z}^A =  \left( h-1 + Z\cdot \frac{\partial}{\partial Z}\right) \frac{\partial}{\partial Z^A} - \frac{1}{2}Z_A\,\frac{\partial^2}{\partial Z\cdot \partial Z}\ ,
\end{align}
and its $(d+1)$-dimensional AdS space counterpart:
\begin{align}\label{Def:AdS-Todorov-op}
    \begin{aligned}
        K_A &= \frac{d-1}{2}\left[\frac{\partial}{\partial W^A}+X_A \left(X\cdot \frac{\partial}{\partial W}\right)\right] + \left(W\cdot \frac{\partial}{\partial W}\right)\frac{\partial }{\partial W^A} \\
        &\quad + X_A \left(W\cdot \frac{\partial}{\partial W}\right) \left(X\cdot \frac{\partial}{\partial W}\right)-\frac{1}{2}W_A \left[\frac{\partial^2}{\partial W \cdot \partial W} +\left(X\cdot \frac{\partial}{\partial W}\right) \left(X\cdot \frac{\partial}{\partial W}\right) \right]\ .
    \end{aligned}
\end{align}

\section{Integral representation of Bessel functions}\label{Appendix:Integral_reps_of_Bessel}
 
In the main text, we used integral representations of Bessel functions to rewrite the OPE block into the form allowing for the holographic interpretation.
In this appendix we will give the details of the derivations.
To this end we start with proving the identity,
\begin{align}\label{Spherical_Integral_Bessel}
    \int [\d^d \hat{x}]_{\rE}\, e^{\i\,p \cdot x} = 2 \pi^h \left( \frac{p x}{2} \right)^{1-h} J_{h-1} (p x)\ ,
\end{align}
where $\hat{x}$ denote the angular variables in (Euclidean) $d$-dimensional space and $h = d/2$, $p = |p|$, $x = |x|$.
This formula can be understood by considering the following expansion of the plane wave,
\begin{align}
    e^{\i\,p \cdot x} =  \Gamma(h-1) \left( \frac{p x}{2} \right)^{1-h} \, \sum_{l=0}^\infty\, \i^l\, (h-1 + l)\, J_{h-1+l}(p x)\, C_{l}^{h-1} (\hat{p}\cdot \hat{x}) \, ,
\end{align}
and the orthogonality relation of the Gegenbauer polynomials \cite{Chetyrkin:1980pr}
\begin{align}
    \int [\d^d \hat{x}_2]_{\rE}\, C_{m}^\nu (\hat{x}_1 \cdot \hat{x}_2 ) \, C_n^{\nu} (\hat{x}_2 \cdot \hat{x}_3 )  = \delta_{nm}\,\frac{2 \pi^{\nu+1} \nu }{(n+\nu) \Gamma(\nu +1)}\, C_n^{\nu} (\hat{x}_1 \cdot \hat{x}_3) \ ,
\end{align}
for $n=0$ with $C_0^\nu(x)=1$.

\subsection{Modified Bessel function of the second kind $K_\nu$}
Let us consider the following integral
\begin{align}
    \int [\d^d  y]_{\rE} \, (x^2 + y^2)^{\nu -h}\,e^{\i\,p\cdot y} \ .
\end{align}
Using the identity \eqref{Spherical_Integral_Bessel}, it is written as
\begin{align}
    \begin{aligned}
        \int [\d^d y]_{\rE} \, (x^2 + y^2)^{\nu -h}\,e^{\i\,p\cdot y}
        &= \int_0^\infty \d y\,y^{d-1}\int [\d^d \hat y]_\rE\,(x^2 + y^2)^{\nu -h}\,e^{\i\,p\cdot y} \\
        &= (2 \pi)^h\, p^{1-h} \int_0^\infty \d y\,y^{h}\,(x^2 + y^2)^{\nu -h}\,J_{h-1} (p y)\\
        &= \frac{2^{\nu +1}\pi^h}{\Gamma(h-\nu )}\, \left( \frac{x}{p}\right)^\nu\,K_\nu (px) \ ,
    \end{aligned}
\end{align}
where we used the formula (6.565.4) of \cite{gradshteyn2014table}
\begin{align}
    \int_0^\infty\,\d y\,y^{\alpha+1}\,(y^2 + x^2)^{-(\beta +1)}\,J_\alpha(py) = \frac{x^{\alpha-\beta} p^\beta}{2^\beta\, \Gamma(\beta + 1)}\,K_{\alpha - \beta}(px) \ ,
\end{align}
which holds for $-1 < \text{Re}\, \alpha < \text{Re} (2\beta + 3/2)$, $x>0$ and $p>0$, in the last equality.
 
It is also instructive to give an alternative derivation using the Schwinger parametrization \eqref{Schwinger_parametrization}
\begin{align}
    \begin{aligned}
        \int [\d^d y]_{\rE} \, (x^2 + y^2)^{\nu-h}\,e^{\i\,p\cdot y} 
        &= \frac{1}{\Gamma(h-\nu)}\int_0^\infty \frac{\d t}{t}\,t^{h-\nu}\,e^{-t x^2}\int [\d^d y]_\rE\,e^{\i\,p\cdot y - t y^2}\\
        &= \frac{\pi^h}{\Gamma(h-\nu )}\int_0^\infty \frac{\d t}{t}\,t^{-\nu}\,e^{-t x^2- \frac{p^2}{4t}}\\
        &= \frac{2^{\nu +1}\pi^h}{\Gamma(h-\nu )}\, \left( \frac{x}{p}\right)^\nu\,K_\nu (px) \ .
    \end{aligned}
\end{align}
Hence we find the following integral representation of the modified Bessel function of the second kind,
\begin{align}\label{Integral_rep_K}
    K_\nu (px) = \frac{\Gamma(h-\nu )}{2^{\nu +1}\pi^h}\,\left( \frac{p}{x}\right)^\nu\,\int [\d^d y]_{\rE} \, (x^2 + y^2)^{\nu-h}\,e^{\i\,p\cdot y} \ .
\end{align}

\subsection{Bessel function of the first kind $J_\nu$}\label{ss:besselJ}
 
Next we show the following integral representation of the Bessel function of the first kind:
\begin{align}
    J_{\nu} (p x) = \frac{1}{2^\nu \pi^h\Gamma(\nu - h + 1)}\left(\frac{p}{x}\right)^\nu\,\int_{y \le x}[\d^d y]_{\rE}\,(x^2 - y^2)^{\nu -h}\,e^{\i\,p\cdot y} \ .
    \label{integral_rep_of_Bessel}
\end{align}
We prove it by rewriting the integral in the right hand side,
\begin{align}
    \begin{aligned}
        \int_{y \le x}[\d^d y]_{\rE}\,(x^2 - y^2)^{\nu -h}\,e^{\i\,p\cdot y} 
            &
            = \int_0^x \d y\,y^{d-1}\,\int \d^d \hat y\, (x^2 - y^2)^{\nu - h}\,e^{\i\,p\cdot y} \\
            &= \int_0^x \d y\,y^{d-1}\,(x^2 - y^2)^{\nu - h}\, 2\pi^h \left(\frac{py}{2}\right)^{1-h}\,J_{h-1}(py)\\
            &= 2\pi^h \left(\frac{p}{2}\right)^{1-h}\,x^{h+2\mu+1}\, \int_0^1\d r\, r^h\,(1-r^2)^{\nu -h}\,J_{h-1}(px r) \\
            &= 2^{\nu}\pi^h\,\Gamma(\nu - h + 1)\,\left(\frac{x}{p}\right)^\nu\,J_{\nu} (p x)\ ,
    \end{aligned}
\end{align}
where we used the formula (6.567.1) of \cite{gradshteyn2014table}
\begin{align}
    \int_0^1 \d x\, x^{\alpha+1}\,(1-x^2)^\beta\, J_\alpha (b x) = 2^\beta\,\Gamma (\beta+1)\,b^{-(\beta +1)}\,J_{\alpha +\beta +1}(b x) \ ,
\end{align}
which holds for $b>0$, $\text{Re}\,\alpha >-1$ and $\text{Re}\,\beta > -1$, 
in the last equality.

\subsection{Modified Bessel function of the first kind $I_\nu$}\label{ss:besselI}

Consider the following integral:
\begin{align}
    \begin{aligned}
    \int_{y\le x}[\d^d y]_{\rE} \,(x^2 - y^2)^{\nu-h}\,e^{-p\cdot y} 
        &= \text{Vol}(\BS^{d-2})\,\int_0^x \d y\,y^{2h-1}\,(x^2 - y^2)^{\nu-h}\,\int_0^\pi \d \theta \sin^{2(h-1)}\theta\,e^{-p y \cos\theta} \\
        &= (2\pi)^h\,p^{1-h}\,\int_0^x \d y\,y^{2h-1}\,(x^2 - y^2)^{\nu-h}\,I_{h-1}(py)\\
        &= 2^\nu \pi^h\,\Gamma(\nu-h+1)\,\left( \frac{x}{p}\right)^\nu\,I_\nu (px) \ ,
    \end{aligned}
\end{align}
where we used in the last equality the identity (19.5.5) in \cite{bateman1954tables})
\begin{align}
    \int_0^x \d y\,y^{\alpha + 1}\,(x^2 - y^2)^{\beta -1}\,I_\alpha (y) = 2^{\beta-1}x^{\alpha + \beta}\,\Gamma(\beta)\,I_{\alpha + \beta}(x) \ ,
\end{align}
which holds for $\text{Re}\,\alpha >-1$ and $\text{Re}\,\beta >0$.
Hence we find the integral representation of $I_\nu$:
\begin{align}
    I_\nu (px) = \frac{1}{2^\nu \pi^h\,\Gamma(\nu-h+1)}\,\,\left( \frac{p}{x}\right)^\nu\,\int_{y\le x}[\d^d y]_{\rE} \,(x^2 - y^2)^{\nu-h}\,e^{-p\cdot y} \ ,
\end{align}
which can be seen as an alternative derivation of \eqref{integral_rep_of_Bessel} using the relation \eqref{Inu_to_Jnu}.

\section{Calculating three-point function by surface Witten diagram}\label{app:SurfaceWitten}
 
In this section, we check whether the surface Witten diagram \eqref{Surface_Witten} is a correct holographic description of the time-like OPE block.
We will compute the two-point function of the surface Witten diagram and a CFT operator to see if it correctly reproduces the structure of the scalar three-point function.
We can choose $x_1 = (R/2, \vec 0)$ and $x_2 = (-R/2, \vec 0)$, (i.\,e., a pair of time-like separated points centered at the origin) without loss of generality as a general configuration is easily achieved by performing Lorentz transformation.
The dual defect $\Sigma_{12}$ is a codimension-two sphere of radius $r$ located at the origin.
$\Sigma_{12}$ extends to the bulk surface $\Xi_{12}$ whose worldvolume is specified in the Poincar\'e coordinates by:
\begin{align}\label{Bulk_surface}
    t = 0 \ , \quad \rho^2 + z^2 = R^2 \ ,
\end{align}
where we use the polar coordinates to parameterize the boundary space-time
\begin{align}
\d s^2_{\BR^{1,d-1}} = -\d t^2 + \d \rho^2 + \rho^2 \, \d \Omega_{d-2}^2 \ .
\end{align}
The bulk surface $\Xi_{12}$ defined by \eqref{Bulk_surface} inherits a spherically symmetry $\SO(d-1)$ from the spherical defect $\Sigma_{12}$.
We choose the worldvolume coordinates $\xi$ to be $(u,\theta_i)~(i=1,\cdots\, d-2)$ where $\theta_i$ are the spherical coordinates and $u$ is a coordinate parameterizing the semi-circle
\begin{align}
    \rho = R\,(2u -1) \ , \quad z = 2R\,\sqrt{u(1-u)} \ .
\end{align}
The spherical symmetry allows us to reduce \eqref{Surface_Witten} to an integral on the geodesic from a point $x_1'$ on $\Sigma_{12}$ to the antipodal point $x_2'$, which is the great circle parameterized by $u$ $(1/2\le u\le 1)$, times the volume of the sphere.
With these parametrization, \eqref{Surface_Witten} becomes
\begin{align}\label{Surface_Witten_higher_dim}
    \CB_{\Delta}(x_1, x_2) \sim \frac{1}{|x_{12}^2|^\frac{\Delta_{12}^+}{2}}\,\frac{\text{Vol}(\BS^{d-2})}{2}\,\int_0^1\frac{\d u}{u(1-u)}\,\frac{1}{2^{d-1}}\left(\frac{(2u-1)^2}{u(1-u)}\right)^{\frac{d-2}{2}}\,\Phi_0( x'(u)\, z'(u))\ , 
\end{align}
where $x'(u)$ and $z'(u)$ are the AdS coordinates on the geodesic from $x_1'$ to $x_2'$.

Let us confirm that \eqref{Surface_Witten_higher_dim} reproduces the CFT three-point function. 
We consider the following two-point function,
\begin{align} \label{Two-point of OPE block and primary}
    \langle\, \CB_{\Delta}(x_1, x_2)\,  \CO_\Delta (x_3 )\, \rangle \sim  \frac{1}{|x_{12}^2|^\frac{\Delta_{12}^+}{2}}\,\frac{\text{Vol}(\BS^{d-2})}{2}\,\int_0^1\frac{\d u}{u(1-u)}\,\frac{1}{2^{d-1}}\left(\frac{(2u-1)^2}{u(1-u)}\right)^{\frac{d-2}{2}}\, \langle\, \Phi_0( x'(u)\, z'(u))\, \CO_\Delta (x_3)\, \rangle\ .
\end{align}
The HKLL construction of the bulk scalar field lets $\langle\, \Phi_0( x'(u)\, z'(u))\, \CO_\Delta (x_3)\, \rangle$ be proportional to the bulk-to-boundary propagator $K_\Delta (x'(u), z'(u),x_3)$.
Let us further choose $x_3 = 0$ so that the propagator simplifies to
\begin{align}
    \begin{aligned}
    \langle\, \Phi_0( x'(u), z'(u))\, \CO_\Delta (0)\, \rangle &= K_\Delta \left(x'(u), z'(u),0\right) \\
    &= \left( \frac{2\sqrt{u(1-u)}}{R} \right)^\Delta \, .
    \end{aligned}
\end{align}
Substituting it to \eqref{Two-point of OPE block and primary}, we end up with
\begin{align}\label{Two-point of OPE block and primary at origin}
    \langle\, \CB_{\Delta}(x_1, x_2) \, \CO_\Delta (x_3 )\, \rangle \sim  \frac{\text{Vol}(\BS^{d-2})}{2^{d-\Delta}}\, \frac{1}{R^{2\Delta_1 + \Delta} } \,\int_0^1\frac{\d u}{u(1-u)}\, \left(\frac{(2u-1)^2}{u(1-u)}\right)^{\frac{d-2}{2}} (u(1-u))^\frac{\Delta}{2} \ ,
\end{align}
where we substitute $\Delta_{12}^+ = \Delta_1 + \Delta_2 = 2 \Delta_1$.
The last integral can be done analytically,
\begin{align}
& \int_0^1\frac{\d u}{u(1-u)}\,  \left(\frac{(2u-1)^2}{u(1-u)}\right)^{\frac{d-2}{2}} (u(1-u))^\frac{\Delta}{2} \nonumber \\
&= (-1)^{-d}\, 2^{\frac{d}{2}-\Delta -1} \Gamma
   \left(\frac{-d+\Delta +2}{2}\right)  
   \left[2^{d/2}\,\frac{\sqrt{\pi }  \,
   _2F_1\left(2-d,\frac{-d+\Delta +2}{2};-d+\Delta +2;2\right)}{\Gamma
   \left(\frac{-d+\Delta +3}{2}\right)} \right. \nonumber \\  
     & \qquad + \left.\frac{\left((-1)^d-1\right) 2^{\Delta /2}\,
   \Gamma (d-1) \, _2F_1\left(\frac{d-\Delta }{2},\frac{-d+\Delta
   +2}{2};\frac{d+\Delta }{2};\frac{1}{2}\right)}{\Gamma \left(\frac{d+\Delta
   }{2}\right)} \right] \ , 
\end{align}
which gives the overall coefficient.
This reproduces the correct $R$-dependence of the three-point structure when $x_1 = (R/2, 0)$ , $x_2 = (-R/2, 0)$ and $x_3 = 0$.

\bibliographystyle{JHEP}
\bibliography{Lorentzian_split_rep}

\end{document}